\documentclass[twocolumn,english,aps,superscriptaddress,floatfix,longbibliography,nobibnotes,nofootinbib]{revtex4-1}
\usepackage[latin9]{inputenc}
\usepackage{color}
\usepackage{babel}
\usepackage{amsmath}
\usepackage{amssymb}
\usepackage{graphicx}
\usepackage{wasysym}
\usepackage{esint}
\usepackage[unicode=true,pdfusetitle,
 bookmarks=true,bookmarksnumbered=false,bookmarksopen=false,
 breaklinks=true,pdfborder={0 0 0},pdfborderstyle={},backref=false,colorlinks=true]
 {hyperref}
\hypersetup{
 linkcolor=blue,citecolor=blue,urlcolor=blue}
\usepackage{breakurl}

\makeatletter

\usepackage{color}
\usepackage{babel}

\usepackage{color}

\renewcommand{\textendash}{--}

\makeatother

\begin{document}

\title{The phase diagram of a frustrated Heisenberg model: from disorder
to order and back again}

\author{Michel M. J. Miranda}

\affiliation{Instituto de F\'isica de S\~ao Carlos, Universidade de S\~ao Paulo,
C.P. 369, 13560-970, S\~ao Carlos, SP, Brazil}

\affiliation{Institut f\"ur Theoretische Physik and W\"urzburg-Dresden Cluster
of Excellence ct.qmat, Technische Universit\"at Dresden, 01062 Dresden,
Germany }

\affiliation{Max-Planck-Institut für Chemische Physik fester Stoffe, N\"othnitzer
Stra{\ss}e 40, 01187 Dresden, Germany}

\author{Igor C. Almeida}

\affiliation{Instituto de F\'isica de S\~ao Carlos, Universidade de S\~ao Paulo,
C.P. 369, 13560-970, S\~ao Carlos, SP, Brazil}

\author{Eric C. Andrade}

\affiliation{Instituto de F\'isica de S\~ao Carlos, Universidade de S\~ao Paulo,
C.P. 369, 13560-970, S\~ao Carlos, SP, Brazil}

\author{Jos\'e A. Hoyos}

\affiliation{Instituto de F\'isica de S\~ao Carlos, Universidade de S\~ao Paulo,
C.P. 369, 13560-970, S\~ao Carlos, SP, Brazil}
\begin{abstract}
We study the effects of bond and site disorder in the classical $J_{1}$-$J_{2}$
Heisenberg model on a square lattice in the order-by-disorder frustrated
regime $2J_{2}>\left|J_{1}\right|$. Combining symmetry arguments,
numerical energy minimization and large scale Monte Carlo simulations,
we establish that the finite-temperature Ising-type transition of
the clean system is destroyed in the presence of any finite concentration
of impurities. We explain this finding via a random-field mechanism
which generically emerges in systems where disorder locally breaks
the same real-space symmetry spontaneously globally broken by the
associated order parameter. We also determine that the phase replacing
the clean one is a paramagnet polarized in the nematic glass order
with non-trivial magnetic response. This is because disorder also
induces non-collinear spin-vortex-crystal order and produces a conjugated
transverse dipolar random field. As a result of these many competing
effects, the associated magnetic susceptibilities are non-monotonic
functions of the temperature. As a further application of our methods,
we show the generation of random axes in other frustrated magnets
with broken SU(2) symmetry. We also discuss the generality of our
findings and their relevance to experiments.\\
\\
Published in \href{https://journals.aps.org/prb/abstract/10.1103/PhysRevB.104.054201}{Phys. Rev. B {\bf 104}, 054201 (2021)};
DOI: \href{https://doi.org/10.1103/PhysRevB.104.054201}{10.1103/PhysRevB.104.054201}
\end{abstract}

\date{August 4, 2021}

\keywords{doidsd}
\maketitle

\section{Introduction}

Understanding the effects of quenched disorder is a long-standing
and fundamental problem in condensed matter systems. It has long been
recognized that it can qualitatively modify the properties of the
phases near a phase transition~\citep{griffiths-prl69,mccoy-prl69},
change the critical behavior of continuous phase transitions~\citep{harris74},
turn a first-order phase transition into a continuous one~\citep{imry-wortis-prb79},
and even destroy the long-range order of the phase itself~\citep{imry_ma},
among other effects (for a review, see, e.g., Ref.~\citealp{thomas19}). 

All these effects can be understood in the context of single-order-parameter
field theory in which disorder adds a random component to either the
conjugate field or to the mass terms. In non-frustrated magnetic systems
(such as those described by the Ising or Heisenberg models), the random
mass term is generated by any inhomogeneities preserving the symmetry
of the Hamiltonian interactions such as site dilution or bond defects.
This term is particularly important near phase transitions yielding
to the so-called Griffiths singularities~\citep{griffiths-prl69,mccoy-prl69}
in addition possibly changing the universality class of a continuous
phase transition~\citep{harris74}. (For more recent developments,
see, e.g., Ref.~\citealp{neto14} and references therein.) 

The effects of random conjugate fields can be even more dramatic.
At sufficiently low dimensions, they produce non-perturbative effects
completely destroying the phase transition~\citep{imry_ma}. However,
random fields are rarely realized in non-frustrated systems because
site and bond disorder generally preserve the symmetry broken by the
resulting magnetic order. In frustrated systems, conversely, the resulting
order commonly breaks a real-space lattice symmetry which is usually
locally broken by the impurities. Therefore, a direct coupling between
disorder and the order parameter field is expected. How this coupling
manifests itself in the associated field theory depends on how the
order-parameter degeneracy is lifted by disorder. For a simple $\text{Z}_{2}$-symmetric
field, only a random field can be generated. For a $\left(\text{Z}_{n}\otimes\text{Z}_{2}\right)$-symmetric
field with disorder lifting the associated degeneracy down to a $\text{Z}_{2}$
one, the generated term is then a random easy axis. 

In the frustrated $J_{1}$-$J_{2}$ Ising model, bond and site disorder
generate a random field term precluding any long-range stripe order
in dimensions $d\leq2$~\citep{fernandez-epl88}. Later, it was shown
that part of this result also applies to the $J_{1}$-$J_{2}$ Heisenberg
model: bond disorder generates a random nematic field term~\citep{fyodorov-shender-jpc91}.
However, it became unclear whether site vacancies produce the same
outcome because (i) the random-field term is not generated in the
one-loop approximation~\citep{fyodorov-shender-jpc91} and (ii) there
is another effect taking place: due to the continuous symmetry character
of the interactions, site vacancies nucleate a non-collinear spin-vortex-crystal
order via the order-by-disorder mechanism~\citep{henley89}. In addition,
Monte Carlo numerical simulations were interpreted as supporting nematic
long-range order in the regime of low dilution~\citep{weber12}.\footnote{The nematic and spin-vortex-crystal orders are commonly designated
by collinear (or columnar antiferromagnetic) and anti-collinear order,
respectively.}

Clearly, from the symmetry perspective, site and bond disorder should
produce qualitatively the same effects in this scenario. Thus, the
aforementioned reason (i) is refuted. A random-field term is expected
in higher orders of approximation. Reason (ii) is less clear, but
the symmetry argument can be played in the other way around: the spin-vortex-crystal
order is also nucleated by bond disorder, an effect not explored in
the literature so far.

Finally, we point out to a recent development which makes this problem
even less clear. It was shown that more subtle situations can also
generate a coupling between disorder and the order-parameter field.
When the ordered state breaks inversion symmetry (such as the spin-vortex-crystal
one), a single-bond defect (which does not break this symmetry) generates
a slowly decaying transverse dipolar field~\citep{santanu20}. In
the presence of a finite concentration of bond defects, the non-collinear
ordered state is destroyed by these random dipolar fields.

Having enumerated all these uncertainties, a deeper understanding
of the effects of disorder in frustrated magnets is desirable. In
this work, we revisit the effects of site and bond disorder on the
$J_{1}$-$J_{2}$ classical Heisenberg model and show that any finite
amount of disorder (either site or bond impurities) precludes the
nematic-paramagnet phase transition due to generation of random fields.
Although the effect of a single-site vacancy is different from that
of a single-bond defect, the effects of any finite fraction of these
impurities are equivalent because it is possible to find configurations
of these impurities breaking the same real-space symmetries. In addition,
we show that the resulting paramagnet is polarized in the nematic
spin-cluster glass order as the system is broken into domains exhibiting
local nematic order.\footnote{We adopt the terminology of Ref.~\citealp{andrade18} in which the
spin-cluster glass differs from the usual Edwards-Anderson spin glass
(or other glasses) as the magnetic correlation length (the nematic
domain) in the former can be arbitrarily large.} The domain walls exhibit non-collinear competing order with temperature-dependent
thickness, yielding a non-trivial behavior to the susceptibilities.
At zero temperature, the spin configuration remains coplanar for weak
disorder. The spin-vortex-crystal order is destroyed by the transverse
dipolar random fields resulting in a spin-vortex-crystal glass. However,
the associated vestigial spin-vorticity density-wave order is perturbatively
stable against these fields. Our conclusions are based on a simple
symmetry analysis for determining how disorder locally lifts the degeneracy
of the order-parameter manifold and are confirmed by numerical energy
minimization and Monte Carlo simulations. 

The remainder of this paper is organized as follows. In Sec.~\ref{sec:CLEAN}
we introduce the model and review the order-by-disorder mechanism
which is responsible for stabilizing the long-range nematic (collinear)
order. In Sec.~\ref{sec:Disorder} we study the effects of bond and
site disorder. We review the known effects of single impurities and
establish the equivalence of their effects when a finite density of
them is present. We also show that random dipolar fields conjugate
to the spin-vortex-crystal order is generated. Finally, we determine
the phase diagram and characterize the resulting thermal state. In
Sec.~\ref{sec:MC} we present extensive numerical simulations to
support our scenario. As an example of our method, in the \hyperref[sec:easy-axes]{Appendix},
we analyze a class of XY frustrated magnets with off-diagonal (spin-orbit-induced)
couplings and show the disorder-induced generation of a random easy-axis
term. In $d=2$, this term destroys long-range order and in $d=3$,
it stabilizes a cluster-spin-glass phase for sufficiently strong disorder.
Finally, in Sec.~\ref{sec:conclusion} we present our concluding
remarks and briefly discuss the implication of our results to the
case of quantum spins and materials compounds.

\section{The model and the entropic selection\label{sec:CLEAN}}

We revisit the random $J_{1}$-$J_{2}$ classical Heisenberg model,
\begin{equation}
\mathcal{H}=\sum_{\left\langle i,j\right\rangle }J_{1,ij}\mathbf{S}_{i}\cdot\mathbf{S}_{j}+\sum_{\left\langle \left\langle i,j\right\rangle \right\rangle }J_{2,ij}\mathbf{S}_{i}\cdot\mathbf{S}_{j},\label{eq:H}
\end{equation}
 on a square lattice of $N=L\times L$ sites with periodic boundary
conditions. The classical spins $\mathbf{S}_{i}$ are three-component
unity vectors. The exchange interactions $J_{1\left(2\right),ij}$
are between nearest- (next-nearest-) $\left(i,j\right)$ neighbor
sites quantified by $J_{\alpha,ij}=J_{\alpha}+\delta J_{\alpha,ij}$
($\alpha=1$ or $2$). Quenched disorder is parametrized by the set
$\{\delta J_{\alpha,ij}\}$ of random variables. As we are interested
in the effects of geometric frustration, we consider antiferromagnetic
next-nearest neighbors interactions $J_{2}>0$. As will become clear,
our results do not depend on whether the nearest-neighbor interactions
are ferromagnetic or antiferromagnetic and, thus, for concreteness,
we consider $J_{1}>0$ from now on. 

\begin{figure}
\begin{centering}
\includegraphics[width=1\columnwidth]{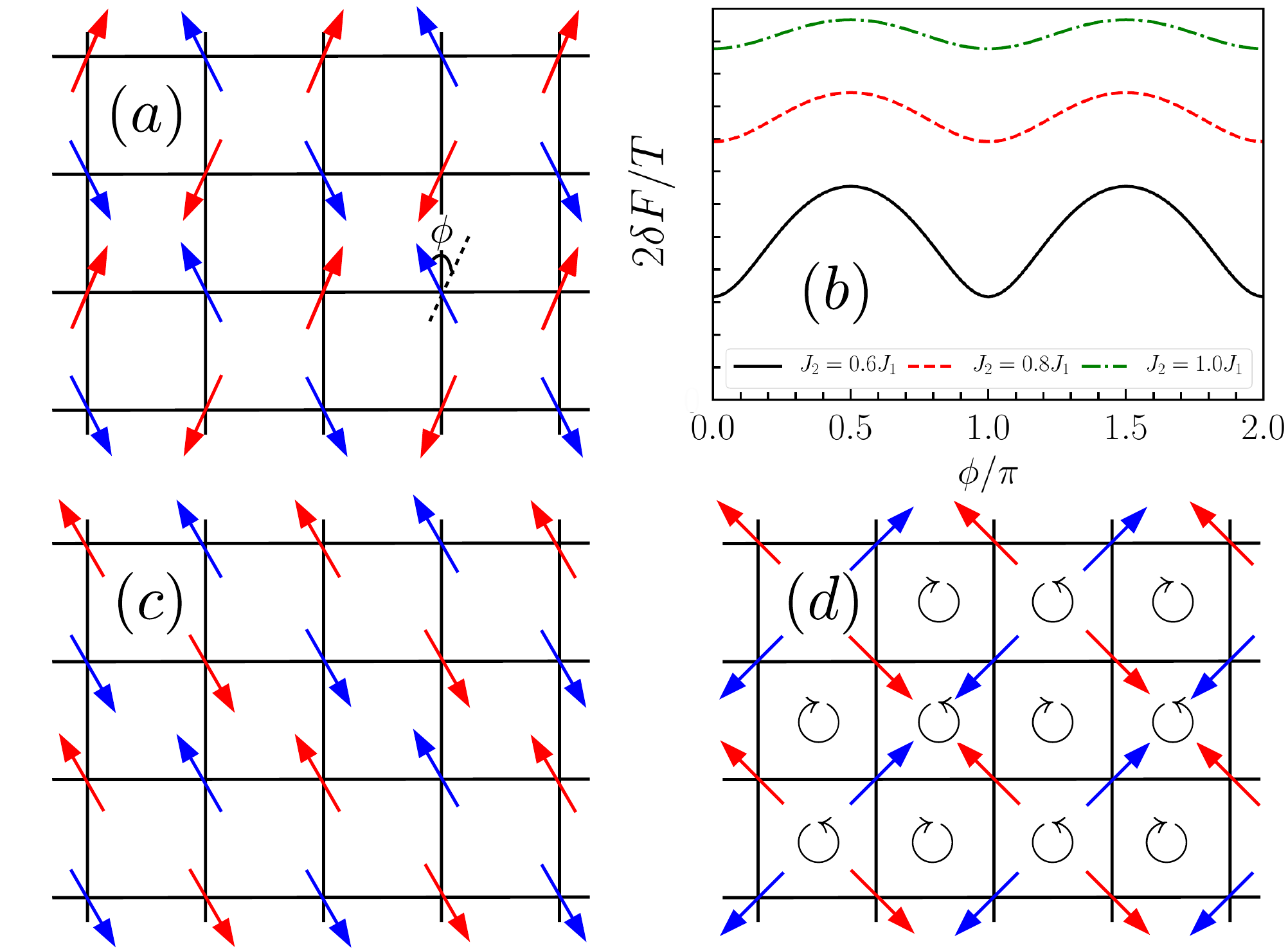}
\par\end{centering}
\caption{\label{fig:selection} (a) The clean ($\delta J_{\alpha,ij}=0$) classical
ground state of the Hamiltonian \eqref{eq:H} in the regime $J_{1}<2J_{2}$:
two decoupled antiferromagnetic sublattices (red and blue arrows)
with the polar $\phi$ and azimuthal $\varphi$ (not shown) angles
parameterizing a nontrivial O(3) degeneracy. (b) The low-$T$ corrections
to the free energy $\delta F$ as a function of the polar angle $\phi$,
indicating the selection of the states $\phi=0$ or $\pi$. (c) The
collinear (stripe) state $\phi=0$ {[}ordering wave vector $\mathbf{Q}_{+}=\left(0,\pi\right)${]}.
(d) The non-collinear (spin-vortex-crystal) state $\phi=\frac{\pi}{2}$,
$\varphi=0$ with staggered ``handness'' pattern.}
 
\end{figure}

Let us start our analysis reviewing the well-established clean-system
($\delta J_{\alpha,ij}=0$) physics~\citep{henley89,chandra90,weber03}
(see also Ref.~\citealp{lmm_frusmag} for a review). For $J_{1}>2J_{2}$,
the classical ground state is a conventional Néel ground state with
ordering vector $\mathbf{Q}_{\text{AF}}=\left(\pi,\pi\right)$ and
ground-state energy $E_{N}=-2N\left(J_{1}-J_{2}\right)$. On the other
hand for $J_{1}<2J_{2}$, it breaks up into two interpenetrating square
sublattices (of unity cell $\sqrt{2}\times\sqrt{2}$), each one ordered
antiferromagnetically in its own Néel state minimizing the next-nearest-neighbor
interaction $J_{2}$ as sketched in Fig.~\hyperref[fig:selection]{\ref{fig:selection}(a)}.
The corresponding energy is $E_{S}=-2J_{2}N$ independent of $J_{1}$
and on the polar $\phi$ and azimuthal $\varphi$ angles between two
Néel states, and thus, the ground state has an \emph{additional} O(3)
accidental degeneracy (parametrized by a polar $\phi$ and an azimuthal
$\varphi$ angle) known to be lifted by the many sorts of fluctuations:
a mechanism known as order by disorder (ObD)~\citep{villain80,shender82}.

The effects of thermal fluctuations are understood in the following
way. Expanding the free energy up to Gaussian fluctuations above the
highly degenerate ground state, the resulting $\left(\phi,\varphi\right)$-dependent
correction to the free energy is 
\begin{equation}
\delta F\propto-T\cos^{2}\phi\label{eq:F-ObtD}
\end{equation}
 {[}see Fig.~\hyperref[fig:selection]{\ref{fig:selection}(b)}{]}
and, therefore, a stripe (collinear) state ($\phi=0$ or $\pi$) is
entropically selected {[}see Fig.~\hyperref[fig:selection]{\ref{fig:selection}(c)}{]}.
In other words, the accidental ground-state O(3) degeneracy is partially
lifted by thermal fluctuations remaining a discrete $\text{Z}{}_{2}$
one: this is the order-by-(thermal)disorder mechanism.

Is the stripe phase stable at low temperatures? If that was the case,
then both the continuous O(3) spin symmetry and the discrete $\text{Z}{}_{2}$
symmetry would be spontaneously broken. However, the Mermin-Wagner
theorem~\citep{mermin-wagner-theorem} dictates that no continuous
symmetry can be spontaneously broken at and below the lower critical
dimension $d_{c,\text{MW}}^{-}=2$ at finite $T$ in systems with
short-ranged interactions. Interestingly, the vestigial (or composite)
nematic order associated to the $\text{Z}{}_{2}$ symmetry breaking
still takes place at sufficiently low temperatures. Consequently,
a continuous finite-$T$ phase transition in the two-dimensional (2D)
Ising universality class occurs in the ObD regime $J_{1}<2J_{2}$.
Here, the nematic order parameter is quantified by 
\begin{eqnarray}
\left\langle m_{\parallel}\right\rangle  & = & \left\langle N^{-1}\sum_{i}m_{\parallel,i}\right\rangle ,\mbox{ with }\nonumber \\
m_{\parallel,i} & = & 4^{-1}\left(\mathbf{S}_{i}-\mathbf{S}_{j}\right)\cdot\left(\mathbf{S}_{k}-\mathbf{S}_{l}\right)\label{eq:Nematic-OP}
\end{eqnarray}
 being the local nematic order parameter involving the four spins
in the $i$th plaquette. {[}Plaquette sites $ikjl$ are arranged counterclockwise
as shown in Fig.~\hyperref[fig:dipole]{\ref{fig:dipole}(a)}{]}.
Here, $\left\langle \cdots\right\rangle $ represents the standard
thermal average. The order parameter being $\left\langle m_{\parallel}\right\rangle >0$
{[}$\left\langle m_{\parallel}\right\rangle <0${]} means a nematic
state with ordering vector $\mathbf{Q}_{+}=\left(0,\pi\right)$ {[}$\mathbf{Q}_{-}=\left(\pi,0\right)${]}
and polar angle $\phi=0$ {[}$\phi=\pi${]}. 

In Ref.~\citealp{weber03}, the critical temperature was found to
be $T_{c}\approx0.55J_{2}$, for $J_{2}\apprge0.9J_{1}$. Thus, the
effective nematic coupling constant is $J_{\parallel}\propto J_{2}$
in this regime which is in agreement with the field-theory predictions~\citep{chandra90}.

\section{The effects of quenched disorder\label{sec:Disorder}}

In this section, we describe the effects of quenched disorder (namely,
vacancies or bond defects) on the model Hamiltonian \eqref{eq:H}
in the ObD regime ($J_{1}<2J_{2}$) which will guide our interpretation
of the numerical data (see Sec.~\ref{sec:MC}).

We start analyzing how a single and a impurity pair lift the O(3)
ground-state accidental degeneracy. In the following, we show that
any finite concentration of impurities yields to non-perturbative
effects hindering any paramagnet-nematic phase transition. Finally,
combining all these effects in addition to the thermal fluctuations,
we characterize the resulting paramagnet.

\subsection{Selection by a single impurity\label{subsec:single-impurity}}

In the ObD regime $J_{1}<2J_{2}$, perturbative approaches~\citep{henley89,fyodorov-shender-jpc91}
confirmed by numerical studies~\citep{weber12} predict that a site
vacancy selects a set of non-collinear states out of the highly degenerate
ground-state manifold, namely the spin-vortex-crystal (SVC) states
{[}see Fig.~\hyperref[fig:selection]{\ref{fig:selection}(d)}{]}.
The corresponding correction to the ground-state energy due to a density
$x\ll1$ of vacancies is 
\begin{equation}
\delta E\propto x\cos^{2}\phi,\label{eq:E-ObsD}
\end{equation}
 which favors $\phi=\frac{\pi}{2}$. (Notice that this order-by-(quenched)disorder
selection is an energetic one.) Therefore, isolated vacancies lift
the accidental O(3) ground-state degeneracy leaving a remaining O(2)
(associated with the azimuthal angle $\varphi$ between the two Néel
states). The related vestigial order, named spin-vorticity density-wave
(SVDW) order~\citep{rafael16}, is quantified by the axial vector
\begin{eqnarray}
\left\langle \mathbf{m}_{\perp}\right\rangle  & = & \left\langle N^{-1}\sum_{i}\mathbf{m}_{\perp,i}\right\rangle ,\mbox{ with }\nonumber \\
\mathbf{m}_{\perp,i} & = & \left(-1\right)^{i_{x}+i_{y}}4^{-1}\left(\mathbf{S}_{i}-\mathbf{S}_{j}\right)\times\left(\mathbf{S}_{k}-\mathbf{S}_{l}\right)\label{eq:SVDW-OP}
\end{eqnarray}
 being the local SVDW order parameter which involves the four spins
of the $i$th plaquette {[}as in Eq.~\eqref{eq:Nematic-OP}{]}, and
$i_{x}$ and $i_{y}$ being the coordinates of site $i$.

On the other hand, the selection by a single $J_{1}$-bond defect
is quite different. A vertical bond defect $J_{1}+\delta J$, for
instance, provides an energy correction to the ground state equal
to $\delta E=-\delta J\cos\phi$, and thus, selects the stripe state
$\mathbf{Q}_{+}$ ($\mathbf{Q}_{-}$) if $\delta J$ is positive (negative).
Interestingly, the two stripe states extremize the energy correction.
This is of no surprise since the stripe states $\mathbf{Q}_{\pm}$
break the vertical/horizontal real-space symmetry. As a result, $J_{1}$-bond
disorder acts like a local conjugate field breaking the symmetry between
the two stripe states~\citep{fyodorov-shender-jpc91}.

Finally, let us discuss the effect of a weak $J_{2}$-bond defect.
Clearly, it does not lift the ground-state degeneracy since it does
not affect the collinear Néel states in each sublattice. We have numerically
verified (see Sec.~\ref{subsec:J2-disorder}) that a finite concentration
of $J_{2}$-bond impurities induces only random-mass disorder which
produces only mild effects far from the $J_{1}=2J_{2}$ transition. 

\subsection{Equivalence between site and bond disorder\label{subsec:Equivalence}}

The effects of a single site vacancy and a single $J_{1}$-bond impurity
are quite different as discussed in Sec.~\ref{subsec:single-impurity}.
The main reason is because they break different real-space symmetries
which selects different states out of the clean ground-state manifold.
However, for a finite concentration of impurities, site and bond disorder
become equivalent since the symmetries broken by them are the same.

\begin{figure}
\begin{centering}
\includegraphics[clip,width=0.8\columnwidth]{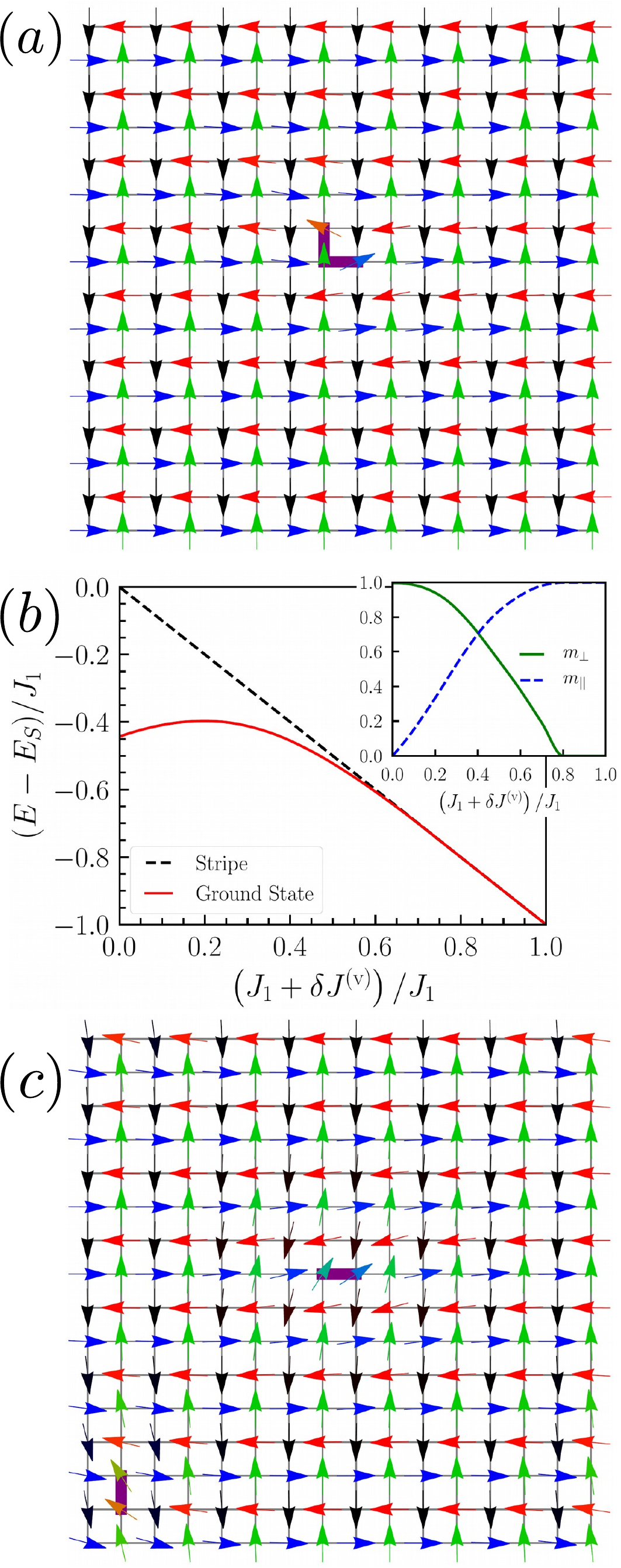}
\par\end{centering}
\caption{(a) Ground-state configuration obtained by the iterative energy minimization
method of the spin configurations. The strong links represent missing
nearest-neighbor $J_{1}$ bonds ($\delta J^{\text{(v)}}=\delta J^{\text{(h)}}=-J_{1}$).
(b) The ground-state energy (measured with respect to the clean one
$E_{S}$) (solid line) and the perfect-stripe-state energy {[}$E_{S}-\left(J_{1}+\delta J^{\text{(v)}}\right)${]}
(dashed line) as a function of the bond magnitude $J_{1}+\delta J^{\text{(v)}}$
(the magnitude of the other bond is fixed at $\delta J^{\text{(h)}}=-J_{1}$).
Inset: the corresponding nematic $m_{\parallel}$ and SVDW $m_{\perp}$
order-parameter magnitudes. (c) Same as (a) but with far-apart defects.
The system size is $L=16$ and periodic boundary conditions are considered.
The value $J_{2}=0.55J_{1}$.\label{fig:SVC-bond-defect}}
\end{figure}

For concreteness, consider for instance two $J_{1}$-bond impurities
of same magnitude, one vertical and the other horizontal, meeting
at the same site. As the vertical/horizontal symmetry is not locally
broken, the $\mathbf{Q}_{\pm}$ stripe states cannot be selected.
Performing numerical energy minimization~\citep{walker77},\footnote{We simply sweep the entire lattice and align the spins to their local
exchange field. This procedure is guaranteed to find a local energy
minima, but it usually struggles to find the global minima in the
case of disordered frustrated system, being therefore limited to small
system sizes~\citep{andreanov15,maryasin14,santanu20}. In order
to ensure convergence, we have used many different random initial
states, SVC states, and states fed from low-temperatures Monte Carlo
simulations.} we have verified that the selected ground state is the SVC, as expected
{[}see Fig.~\hyperref[fig:SVC-bond-defect]{\ref{fig:SVC-bond-defect}(a)}{]}.

We now investigate the robustness of this selection with respect to
(i) the anisotropy between vertical and horizontal defects, and with
respect to (ii) their distance. Using the energy minimization method,
we plot in Fig.~\hyperref[fig:SVC-bond-defect]{\ref{fig:SVC-bond-defect}(b)}
the ground-state energy (solid line) as a function of the defective
vertical coupling amplitude $J_{1}+\delta J^{\text{(v)}}$ while the
horizontal one is kept missing ($\delta J^{(\text{h)}}=-J_{1}$).
Clearly, the non-collinear SVC state is selected even for a fairly
large amount of anisotropy $\delta J^{\text{(v)}}\gtrapprox0.4\delta J^{\text{(h)}}$.
Beyond that, the SVC ground state smoothly evolves into the stripe
state. (The perfect stripe state energy is given by the dashed line.)
The inset shows the nematic and SVDW order parameters. Finally, we
have verified that the SVC state is energetically selected by any
two $J_{1}$-bond impurities regardless of their distance provided
that one is vertical and the other is horizontal (and their magnitude
are not sufficiently different) as shown in Fig.~\hyperref[fig:SVC-bond-defect]{\ref{fig:SVC-bond-defect}(c)}. 

Given that these two $J_{1}$-bond impurities separately select the
two different stripe states, one could suppose that the system would
break into two stripe domains (which is the case for Ising spins).
However, the domain-wall cost disfavors this configuration and the
system prefers the SVC state. Alternatively, we can say that, in order
to minimize energy, the domain wall thickness is unbounded (at $T=0$)
leaving no room for the stripe domains. More importantly, the domain
wall exhibits SVC order, the structure factor of which peaks at the
two stripe ordering vectors $\mathbf{Q}_{+}$ and $\mathbf{Q}_{-}$.
As will become clear later, this interpretation is helpful for interpreting
the numerical data.

In order to close the equivalence between site and bond disorder,
we now show that two site vacancies can also select one of the stripe
states. Consider for instance the case of two missing sites one on
top of each other. In this case, three vertical and four horizontal
$J_{1}$ bonds are missing, and thus, the $\mathbf{Q}_{+}$ ($\phi=0$)
state is energetically favored. Two nearest-neighbor site vacancies
are exactly what is needed to locally break the vertical/horizontal
symmetry.\footnote{Two distant site vacancies in the same row or column do break the
global vertical/horizontal real-space symmetry. However, they do not
select any of the stripe states as the same amount of vertical and
horizontal $J_{1}$ bonds is missing. Thus, the terminology of \emph{locally}
breaking a symmetry means that this is done in a plaquette.}

\subsection{Thermal fluctuations and screening\label{subsec:Screening}}

As shown in Secs.~\ref{subsec:single-impurity} and \ref{subsec:Equivalence},
a single or a pair of impurities either select the SVC or the stripe
states out of the clean ground-state manifold. 

At $T>0$, both orders are melted as they require the spontaneous
breaking of the O(3) (continuous) spin-rotation symmetry. The associated
nematic vestigial order, on the other hand, is stable at low temperatures
and is also entropically selected (see Sec.~\ref{sec:CLEAN}). The
associated SVDW vestigial order, however, is melted since it also
requires the spontaneous breaking of a continuous symmetry.\footnote{One could think that the associated O(2) symmetry could be quasi-broken
in a Berezinskii-Kosterlitz-Thouless transition. However, symmetry
considerations indicate that an O(3) symmetry {[}related to the axial
vector \eqref{eq:SVDW-OP}{]} must be broken before the O(2) one~\citep{rafael16,fernandes-orth-schmalian-arcmp19},
thus, precluding any quasi-long-range SVDW order at finite $T$. } 

In the remainder of this section, we discuss the resulting thermal
state in the presence of these impurities. Let us start with the trivial
case of a single $J_{1}$-bond defect or two nearest-neighbor site
vacancies which select, say, the $\mathbf{Q}_{+}$ nematic state.
As they act as a local nematic field, then at finite $T$ and for
finite system size $L$, the nematic order $\left\langle m_{\parallel}\right\rangle $
is always positive. It reaches $1$ as $T\rightarrow0$ and vanishes
in the $T\rightarrow\infty$ limit. Upon increasing the system size
at finite $T$, $\left\langle m_{\parallel}\right\rangle $$\rightarrow0$
as $L\rightarrow\infty$ because the free-energy difference between
the positive and negative nematic states is non-extensive. Thus, the
limits $T\rightarrow0$ and $L\rightarrow\infty$ do not commute.

Let us now discuss the more interesting case of a single-site vacancy
which selects the SVC order. There are two mechanisms confining that
order in a region of size $\xi_{\perp}$ around the impurity: the
competing nematic order and the thermal fluctuations. At finite temperatures
(below the clean critical temperature $T_{c}$), the impurity generates
a perturbation inside the nematic state which cannot be propagated
beyond the nematic correlation length $\xi_{\parallel}$. Thus, $\xi_{\perp}$
is bounded by $\xi_{\parallel}$. Evidently, SVC and SVDW orders are
also bounded by the thermal correlation length $\xi_{\text{MW}}$
of a continuous symmetry order parameter which, at low temperatures,
is $\propto e^{\rho_{\perp}/T}$, with $\rho_{\perp}$ being a related
 stiffness. (The exponentially large $\xi_{\text{MW}}$ is characteristic
of the lower critical dimension of the problem $d_{c,\text{MW}}^{-}=2$.)
Therefore, $\xi_{\perp}\propto\min\left\{ \xi_{\text{MW}},\xi_{\parallel}\right\} $.

An analogous effect would occur for two bond defects provided that
their distance is smaller than $\xi_{\perp}$, that one is vertical
and the other horizontal, and that their magnitudes are similar. If
their distance is greater than $\xi_{\perp}$, the SVDW domain wall
between them is entropically disfavored and the system will be in
either one of the equally equivalent nematic states. 

\subsection{The effects of disorder-induced random nematic fields \label{subsec:Nematic-RF}}

In this section we show the impossibility of a phase transition into
a nematic phase in the presence of any finite concentration of impurities
in $d\leq2$. The proof is by \emph{reductio ad absurdum}. We assume
the existence of the transition and arrive in a contradiction.

Assuming that the transition exists, the symmetry of nematic order
parameter \eqref{eq:Nematic-OP} dictates that a spontaneous symmetry-breaking
transition is in the unfrustrated 2D Ising universality class~\citep{chandra90,weber03,weber12,rafael16}.
Therefore, the nematic transition can be described by the 2D Ising
model with generic short-range ferromagnetic interactions with the
Ising variables representing the two nematic states $\mathbf{Q}_{\pm}$.
As discussed in Secs.~\ref{subsec:single-impurity} and \ref{subsec:Equivalence},
$J_{1}$-bond defects and nearest-neighbor vacancy pairs act as nematic
fields locally breaking the symmetry between the $\mathbf{Q}_{\pm}$
states. Therefore, in the presence of a finite concentration of impurities
we must furnish the effective Ising model with additional random fields
and random couplings. The lower critical dimension of this model is
known to be $d_{c,\text{\ensuremath{\parallel}RF}}^{-}=2$~\citep{imry_ma,grinstein-ma-prl82,grinstein-ma-prb83,villain-jpa82}.
Consequently, there is no phase transition in $d\leq2$. This ends
our proof.

We recall that, for $J_{1}$-bond disorder, these conclusions were
previously obtained in Ref.~\citealp{fyodorov-shender-jpc91} using
conventional perturbative field-theory methods. For site disorder,
it was suggested that the transition exists (see also Ref.~\citealp{weber12}).

In the case of anisotropic disorder,\footnote{By anisotropic disorder we mean those situations in which the global
vertical/horizontal isotropy is broken such as, for instance, the
cases in which the disorder averages of $\delta J_{1}^{\text{(v)}}$
and of $\delta J_{1}^{\text{(h)}}$ are different, or when the density
of horizontal and vertical vacancy pairs are different. } the mean value of the random fields is finite and hence the corresponding
nematic state is globally favored. In this case, $\left\langle m_{\parallel}\right\rangle $
is finite at any temperature. 

In the case of isotropic disorder, the random fields have zero mean
and, thus, the nematic order parameter is always vanishing. At low
temperatures, the system is broken into domains of different nematic
order of typical size $\xi_{\text{\ensuremath{\parallel}RF}}\propto e^{\tilde{J}^{2}/\sigma_{\tilde{h}}^{2}}$
(up to power-law corrections)~\citep{grinstein-ma-prl82,grinstein-ma-prb83,villain-jpa82}.
Here, $\tilde{J}$ is the mean value of the effective nematic coupling
constants $J_{\parallel}$ (proportional to $J_{2}$ in the regime
$J_{2}\apprge J_{1}$), and $\sigma_{\tilde{h}}^{2}$ is the variance
of the effective random fields $\tilde{h}$. In the case of low density
of bond defects $0<x\ll1$, $\sigma_{\tilde{h}}^{2}\propto x\left(1-x\right)$.
Likewise, for a low density $x$ of site vacancies, $\sigma_{\tilde{h}}^{2}\propto x^{2}\left(1-x^{2}\right)$.
The fact that $\xi_{\text{\ensuremath{\parallel}RF}}$ is exponentially
large in $1/\sigma_{\tilde{h}}^{2}$ is a consequence of the system
being at the lower critical dimension $d=d_{c,\text{\ensuremath{\parallel}RF}}^{-}$.
It also points out to the huge difference between the domain sizes
formed by bond and site defects in the limit of small defect concentrations
$x$.

In the \hyperref[sec:easy-axes]{Appendix}, we apply these simple
arguments to show the disorder-induced generation of random axes in
easy-plane pyrochlores.

\subsection{Disorder-induced dipolar random fields\label{subsec:Dipolar-RF}}

We now show that the spin-vortex-crystal state is unstable in the
presence of any finite concentration of impurities in $d\leq2$ even
at $T=0$.\footnote{This is a pertinent question even in the realistic case of quantum
spins. For a sufficiently high concentration of site vacancies, the
SVC state may percolate.} The reason is akin to the instability of the nematic order discussed
in Sec.~\eqref{subsec:Nematic-RF}: impurities-induced random fields.
As in the nematic case, it is also related to a real-space symmetry
broken by the ordered state: the inversion symmetry. However, the
random fields are non-local exhibiting a dipolar texture.

This result is a direct consequence of the theory developed in Ref.~\citealp{santanu20}
where it was shown that any amount of generic bond disorder destroys
any non-collinear order. As a consequence of the symmetry arguments
of Sec.~\ref{subsec:Equivalence}, we extend this result to site
disorder as well (see below). Finally, there is another important
difference with respect to the stripe/nematic case. The vestigial
SVDW is perturbatively stable against disorder at $T=0$. This does
not contradict the results of Ref.~\citealp{santanu20} since the
SVDW order is collinear.

\begin{figure}
\begin{centering}
\includegraphics[clip,width=0.65\columnwidth]{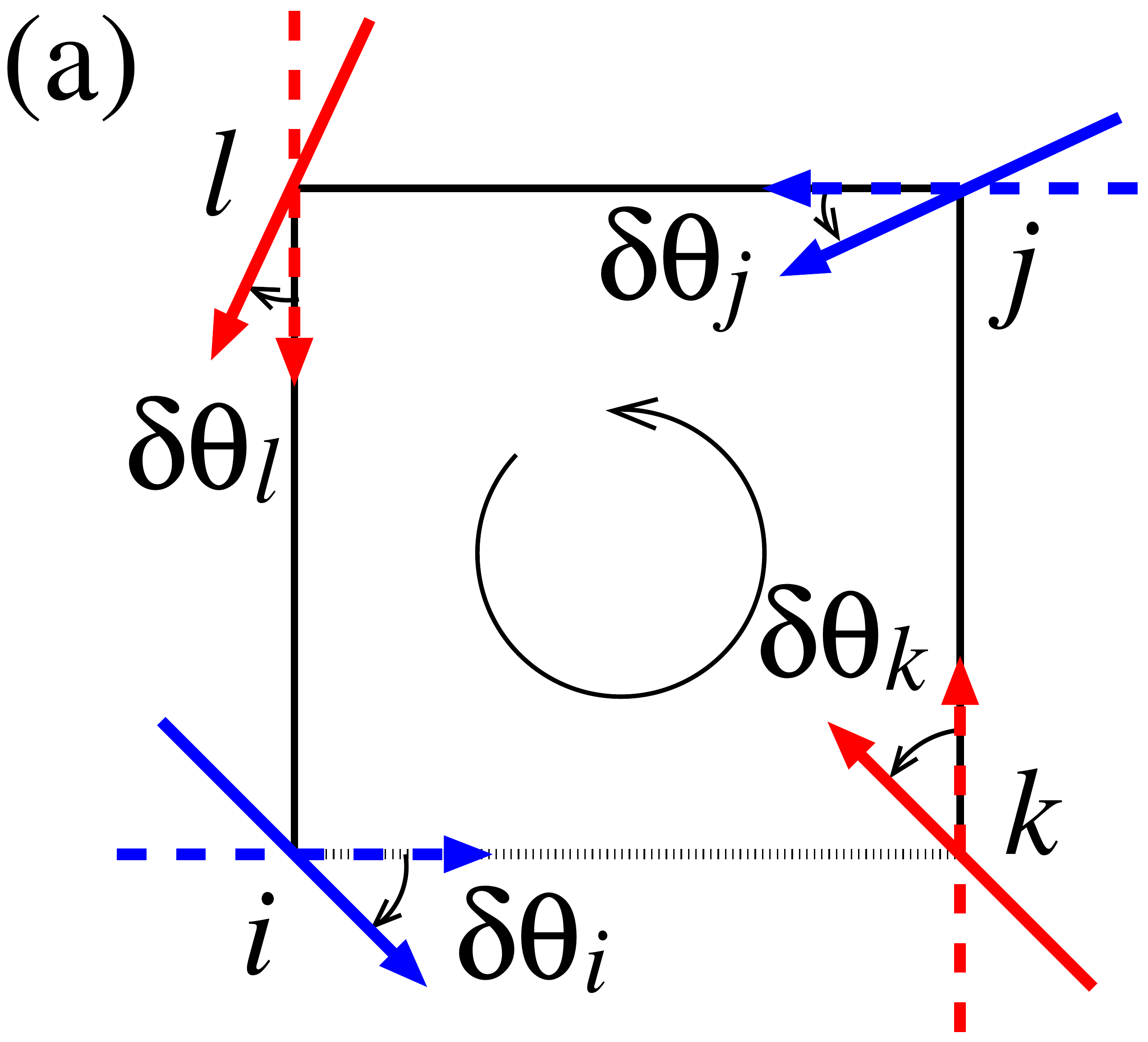}\\
\includegraphics[clip,width=0.8\columnwidth]{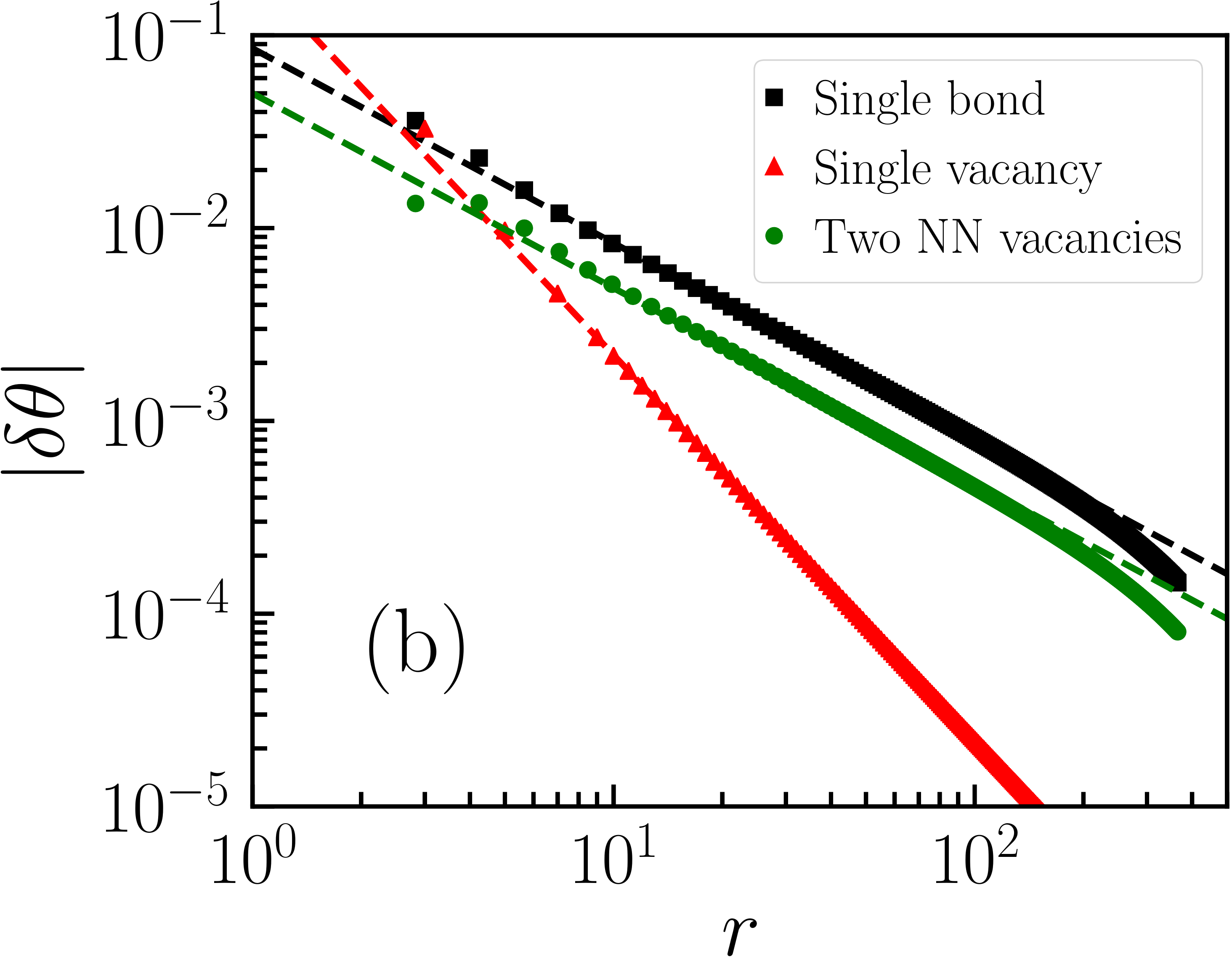}
\par\end{centering}
\caption{\label{fig:dipole}(a) The spin rotation field on a plaquette induced
by a strong ($\delta J>0$) $J_{1}$-bond impurity (dotted line).
(Dashed arrows represent the unperturbed state.) (b) The spin-rotation
field magnitude $\left|\delta\theta\right|$ plotted as a function
of distance $r$ from the defect (see text for details) for the cases
of a $J_{1}$-bond vacancy ($\delta J=-J_{1}$) (black squares), a
single vacancy (red triangles) and a pair of nearest-neighbor vacancies
(green circles). We have used $J_{2}=0.55J_{1}$ and fixed $K=0.05J_{1}$.
We used system sizes up to $L=2^{10}$ with periodic boundary conditions.
The dashed lines are power-law fits $\propto r^{-1}$ and $\propto r^{-2}$
in the range $10\lesssim r\lesssim10^{2}$. Deviations at large $r$
are due to finite-size effects. We have obtained similar results (not
shown) for system size $L=400$, $K=0.1J_{1}$, and $J_{2}/J_{1}=0.55$,
$0.6$ and $0.8$. }
\end{figure}

We start our analysis by numerically verifying that a single $J_{1}$-bond
defect produces a dipolar spin texture~\citep{santanu20} 
\begin{equation}
\delta\theta\left(\mathbf{r}\right)=\alpha\delta J\frac{\hat{e}\cdot\mathbf{r}}{r^{d}}\label{eq:spin-texture}
\end{equation}
 on the spin-vortex-crystal state {[}see Fig.~\hyperref[fig:dipole]{\ref{fig:dipole}(a)}{]}.
Here, $d=2$, $\delta\theta\left(\mathbf{r}\right)$ is the in-plane
impurity-induced angle deviation (with respect to the spin-vortex-crystal
state) of the spin at position $\mathbf{r}$ measured with respect
to the center of the defective bond, the unity vector $\hat{e}$ is
the direction of the defective bond ($\hat{x}$ or $\hat{y}$), and
$\alpha\equiv\alpha\left(J_{1},J_{2}\right)$ is a constant the sign
of which ensures that the two spins connected by the defective bond
align more (less) anti-parallel for $\delta J>0$ ($\delta J<0$).
(Notice therefore that the local random field lies within the SVC
plane and is perpendicular to the local spin order.)

In order to stabilize the non-collinear SVC state in a clean sample,
we supplement the Hamiltonian \eqref{eq:H} with a positive biquadratic
term $K\sum_{\left\langle ij\right\rangle }\left(\mathbf{S}_{i}\cdot\mathbf{S}_{j}\right)^{2}$~\citep{henley89,rafael16}.
The $T=0$ resulting spin configuration is obtained via numerical
energy minimization~\citep{walker77}. Convergence is confirmed by
starting from many different initial random states. In all cases the
final configuration is the same and remains coplanar. In Fig.~\hyperref[fig:dipole]{\ref{fig:dipole}(b)}
we plot $\left|\delta\theta\right|$ as a function of $r$ where $\mathbf{r}=r\hat{e}$.
We have verified the induced dipolar spin texture \eqref{eq:spin-texture}
by inspection of our data (not shown).\footnote{A careful reader will notice in Fig.~\hyperref[fig:SVC-bond-defect]{\ref{fig:SVC-bond-defect}(c)}
a non-dipolar spin texture near the impurities. We emphasize that
the dipolar character \eqref{eq:spin-texture} dominates only at larger
distances invisible to the naked eye.} This result can be derived analytically through linear response theory
in any dimension $d$ as shown in details in Ref.~\citealp{santanu20}.
Here, we only mention the reason for such. The fact that the SVC state
breaks the inversion symmetry makes the distortion field $\delta\theta\left(\mathbf{r}\right)$
an odd function with respect to the inversion $\mathbf{r}\rightarrow-\mathbf{r}$.
Therefore, the spin texture is $p$-wave like which is communicated
to the rest of the system by in-plane Goldstone modes and, thus, decays
as $\sim1/r$~\citep{villain79}.

The equivalence between bond and site disorder is confirmed in Fig.~\hyperref[fig:dipole]{\ref{fig:dipole}(b)}
where $\left|\delta\theta\right|$ is plotted for the case of a nearest-neighbor
vacancy pair. (Here, $\mathbf{r}$ is measured from the center of
the vacancy pair and $\hat{e}=\hat{y}$ or $\hat{x}$ is the direction
perpendicular to the vacant sites.) Again, we verify the induced dipolar
spin texture \eqref{eq:spin-texture}. (In contrast, a single vacancy
induces a quadrupolar texture which decays $\sim1/r^{2}$~\citep{weber12}.)

Having verified the dipolar spin texture \eqref{eq:spin-texture}
induced by a single-bond impurity or a nearest-neighbor vacancy pair,
it is now straightforward to show the instability of the SVC order
against any finite concentration of impurities in $d\leq2$. Averaging
over the disorder configurations (denoted by $\left[\cdots\right]$),
the mean angle deviation at site $j$ vanishes because disorder is
globally isotropic: $\left[\delta\theta_{j}\right]\propto\sum_{m}\delta J_{m}\frac{\hat{e}_{m}\cdot\mathbf{r}_{j,m}}{r_{j,m}^{d}}=0$,
where $\mathbf{r}_{j,m}$ is the position of site $j$ with respect
to the $m$th impurity. The variance, on the other hand, behaves quite
differently: $\left[\delta\theta_{j}^{2}\right]\propto\sum_{m,n}\delta J_{m}\delta J_{n}\frac{\hat{e}_{m}\cdot\mathbf{r}_{j,m}}{r_{j,m}^{d}}\frac{\hat{e}_{n}\cdot\mathbf{r}_{j,n}}{r_{j,n}^{d}}\sim\int r^{1-d}\text{d}r$
is infrared divergent for $d\le d_{c,\perp\text{RF}}^{-}=2$ precluding
any SVC order. At the lower critical dimension of the problem, it
was shown that SVC order is confined in a domain of temperature-independent
typical size $\xi_{\perp\text{RF}}\propto e^{\rho_{\parallel}^{2}/\sigma_{\delta J}^{2}}$,
where $\rho_{\parallel}$ is a related  stiffness and $\sigma_{\delta J}^{2}$
is the variance of $\delta J_{1}$~\citep{santanu20}. The resulting
ground state is a SVC spin-cluster glass. 

As expected, we verified that disorder on the $J_{2}$ bonds does
not induce dipolar random fields.

How about the stability of vestigial SVDW order? At $T=0$ in a coplanar
state, the local SVDW order \eqref{eq:SVDW-OP} reduces to a scalar
and therefore the result of Ref.~\citealp{santanu20} does not apply.
Let us consider the effects of the dipolar spin texture \eqref{eq:spin-texture}
on the SVDW order. Far from the impurity, the change in the local
order parameter is $\delta m_{\perp}(\mathbf{r})\propto-\left(1+d\left(d-2\right)\left(\hat{e}\cdot\hat{r}\right)^{2}\right)\left(\delta Jr^{-d}\right)^{2}$,
which vanishes sufficiently fast as $r\rightarrow\infty$. Consequently,
the mean deviation $\left[\delta m_{\perp}\right]\propto\left[\left(\delta J\right)^{2}\right]$
is finite and, more importantly, because $\delta m_{\perp}$ is isotropic
at $d=2$, the corresponding variance vanishes up to ${\cal O}\left(\delta J\right)^{4}$.
This strongly suggests that SVDW order is perturbatively stable at
$d=2$. Our numerics (see Sec.~\ref{subsec:GS}) support this scenario.
At weak disorder, the ground-state spin configuration remains coplanar
with vanishing SVC order and finite SVDW order. Upon increasing disorder,
the spin configuration becomes non-coplanar and the SVDW order is
destroyed.

\subsection{Phase diagram\label{subsec:PD}}

As discussed in the previous sections, in the presence of generic
disorder and in the ObD regime ($J_{1}<2J_{2}$) the model system
\eqref{eq:H} is a always a paramagnet at $T>0$ and a spin-vortex-crystal
cluster-spin glass at $T=0$ with finite spin-vorticity density-wave
vestigial order in the regime of weak disorder.

It is noteworthy that the paramagnet is not uninteresting. It is broken
into spin clusters locally exhibiting nematic order. Therefore, it
is polarized in the nematic spin-cluster-glass order. With respect
to the nematic order, this paramagnet is very much similar to that
of the Ising model in a random field which is not a glass~\citep{krzakala-etal-prl10}.
However, with respect to susceptibilities they are quite different
because (i) the nematic effective coupling constant is temperature
dependent (due to the entropic selection) and (ii) the domain walls
in our case have high SVDW susceptibility. As a consequence, the domain-wall
thickness is temperature dependent yielding nontrivial nematic and
SVDW susceptibilities. An explicit discussion on this issue involving
also the glassy properties is given in Sec.~\ref{subsec:Isotropic-disorder}.

\section{\label{sec:MC}Numerical simulations}

In this section we report our Monte Carlo (MC) simulations of the
system Hamiltonian \eqref{eq:H} and our numerical zero-temperature
energy minimization.

\subsection{Technical details}

In our MC code, we have used both the Heat Bath~\citep{miyatake-etal-jpc86}
and the Microcanonical~\citep{creutz-prd87} local update algorithms.
Precisely, we employ 5 Heat Bath steps followed by 5 Microcanonical
ones (1 MC step means 1 random sweep over the entire lattice), which
proved to be an optimal choice for equilibration. After these 10 MC
steps, we finally perform a parallel-tempering step. We define a temperature
grid such that the probability of swapping adjacent configurations
is independent of the temperature values and is sufficiently high
so as to ensure frequent exchanges~\citep{hukushima-nemoto-jpsj96}.
We typically perform $5\times10^{5}$ MC steps to reach equilibrium
and $M=5\times10^{5}$ MC steps to take averages.\footnote{We think it is worthy to mention that we have performed a MC study
without parallel tempering and obtained the same results. (Although
it needs $5\times10^{5}$ MC steps to reach equilibrium and $M=1.5\times10^{6}$
MC steps to take averages in order to obtain equivalent error bars.)
We interpret this result as the absence of a slowing down related
to any glass-like behavior.}

As usual in MC studies, the thermal average is replaced by the MC
average, i.e., 
\begin{equation}
\left\langle {\cal O}\right\rangle \rightarrow M^{-1}\sum_{t=1}^{M}{\cal O}(t),\label{eq:MC-average}
\end{equation}
 where ${\cal O}(t)$ is the value of the observable ${\cal O}$ at
the $t$th Monte Carlo step (after equilibration).

When studying disordered systems, we also average an observable over
$N_{d}=600$ different disorder realizations (denoted by $\left[\cdots\right]$),
namely, $\left[\left\langle {\cal O}\right\rangle \right]=N_{d}^{-1}\sum_{k=1}^{N_{d}}\left\langle {\cal O}\right\rangle _{k}$,
where $\left\langle {\cal O}\right\rangle _{k}$ is the MC average
of ${\cal O}$ in the $k$th disorder realization.

In addition to the nematic \eqref{eq:Nematic-OP} and SVDW \eqref{eq:SVDW-OP}
order parameters, we also study the corresponding susceptibilities,
the nematic Binder cumulant, and the specific heat, respectively given
by
\begin{eqnarray}
\chi_{a}^{\alpha} & = & \frac{N}{T}\left[\left\langle \left(m_{a}^{\alpha}\right)^{2}\right\rangle -\left\langle \left|m_{a}^{\alpha}\right|\right\rangle ^{2}\right],\label{eq:Chi}\\
U_{\parallel} & = & 1-\frac{1}{3}\left[\frac{\left\langle m_{\parallel}^{4}\right\rangle }{\left\langle m_{\parallel}^{2}\right\rangle ^{2}}\right],\label{eq:Binder}\\
c_{v} & = & \left(NT\right)^{-1}\left[\left\langle E^{2}\right\rangle -\left\langle E\right\rangle ^{2}\right],\label{eq:C}
\end{eqnarray}
 where $a=\parallel,\perp$, respectively, refers to the nematic and
SVDW orders, $\alpha,\beta=x,y,z$ are the order-parameter components
(not applicable to the nematic case), and $E$ is the spin configuration
energy \eqref{eq:H}.

In addition, we also have studied the nematic and SVDW Edwards-Anderson
order parameters defined as~\citep{fischer,pixley08} 
\begin{eqnarray}
\left[\left\langle m_{\text{EA},a}\right\rangle \right] & = & \left[\left\langle \sqrt{N^{-1}\sum_{\alpha,\beta}\left|\sum_{i}m_{a,i}^{\alpha(1)}m_{a,i}^{\beta(2)}\right|^{2}}\right\rangle \right],\label{eq:EA-OP}
\end{eqnarray}
 where the super-indices $(1)$ and $(2)$ denotes the two simulated
replicas, i.e., we independently simulate two copies of the system
(same disorder configuration), and thus, $\left\langle m_{a,i}^{\alpha(1)}m_{a,i}^{\beta(2)}\right\rangle =\left\langle m_{a,i}^{\alpha(1)}\right\rangle \left\langle m_{a,i}^{\beta(2)}\right\rangle $.
The corresponding spin-glass susceptibility $\chi_{\text{EA},a}$
is the corresponding thermal fluctuation magnitude of $m_{\text{EA},a}$
quantified by \eqref{eq:Chi} with $m_{a}^{\alpha}$ replaced by $m_{\text{EA},a}$.

\subsection{The clean system}

In order to benchmark our codes, we review the impurity-free nematic
phase transition in the ObD regime where we fix $J_{2}=0.55J_{1}$.

\begin{figure}[t]
\begin{centering}
\includegraphics[clip,width=0.5\columnwidth]{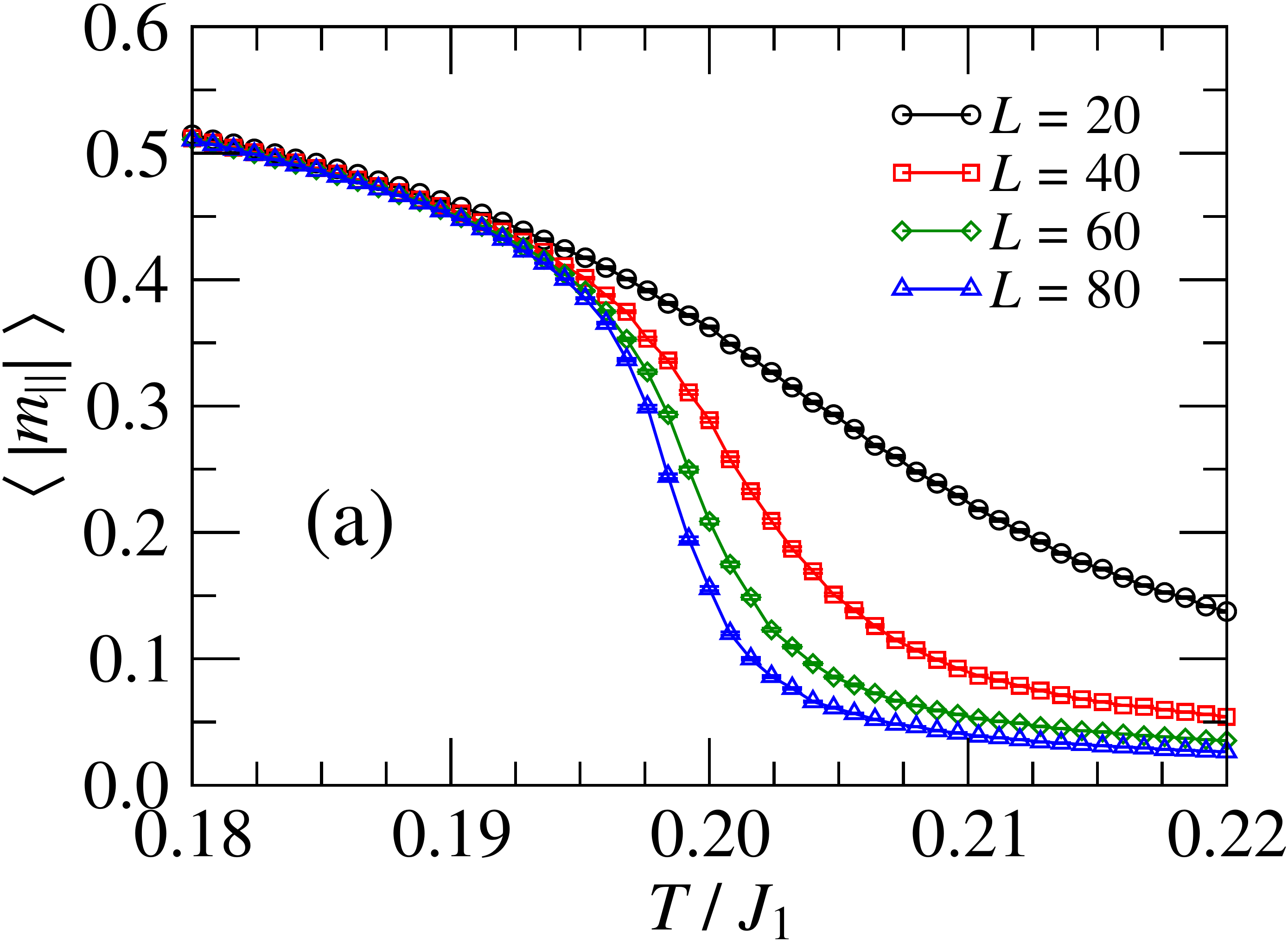}\includegraphics[clip,width=0.5\columnwidth]{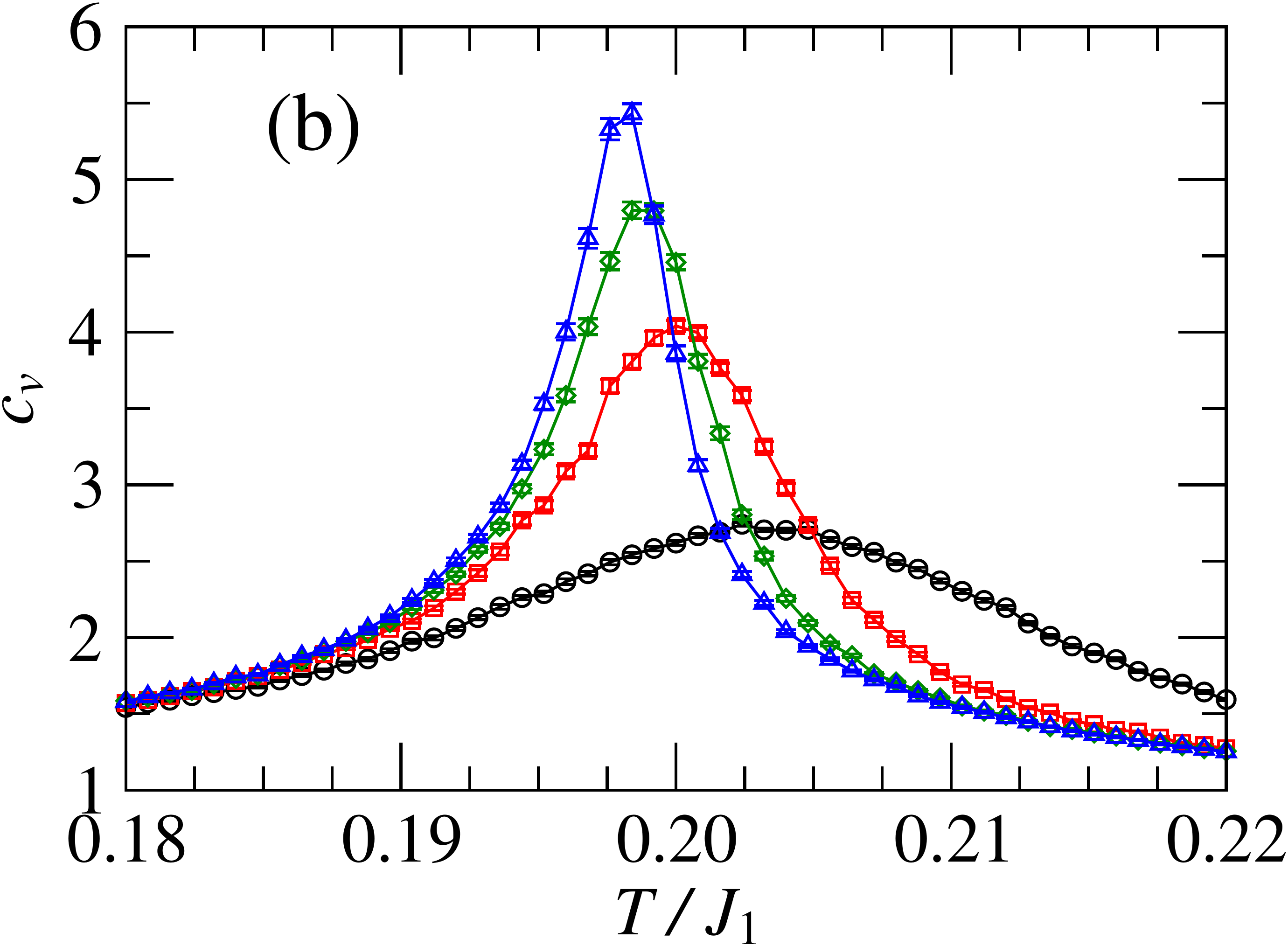}\\
\includegraphics[clip,width=0.5\columnwidth]{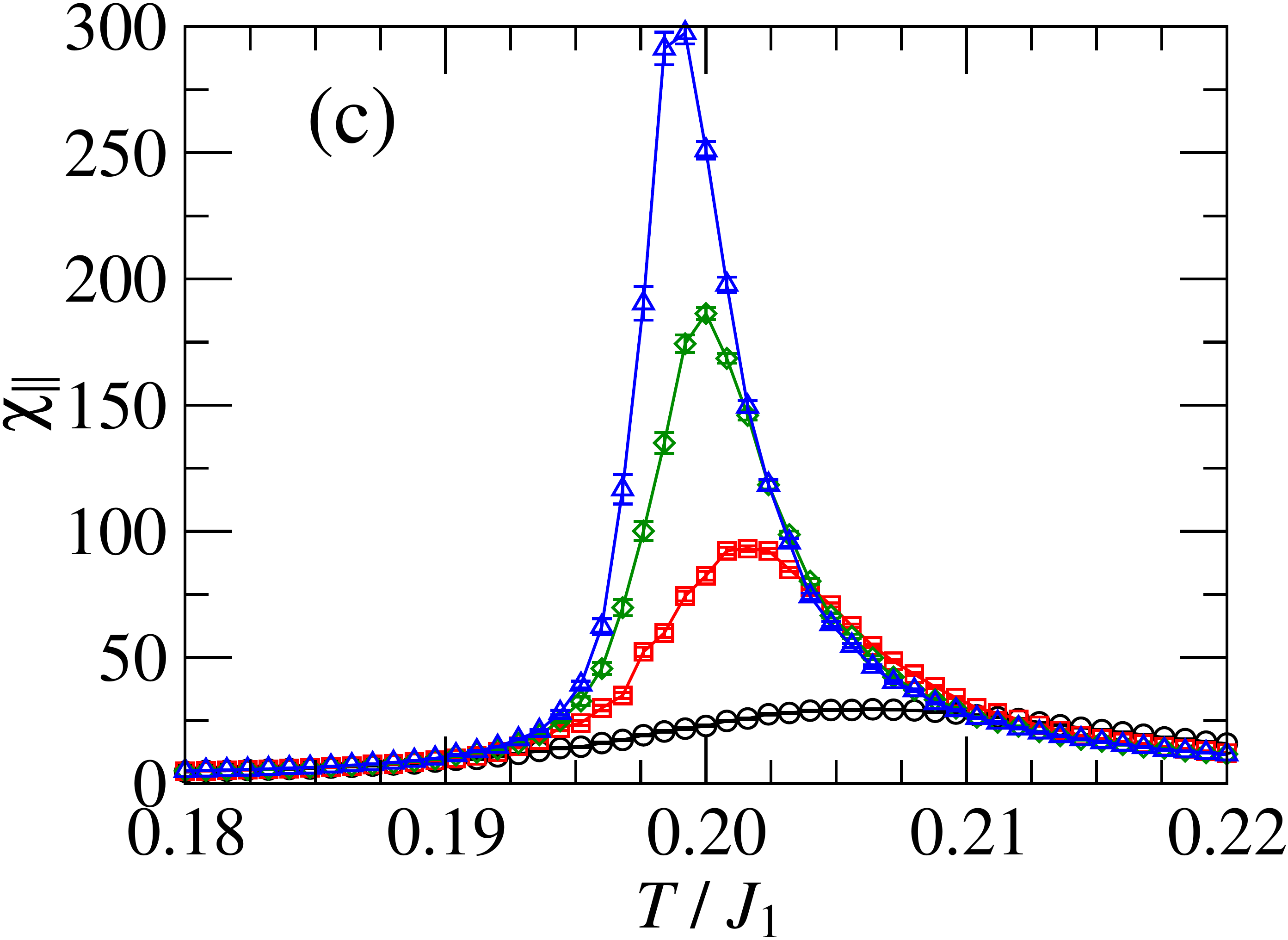}\includegraphics[clip,width=0.5\columnwidth]{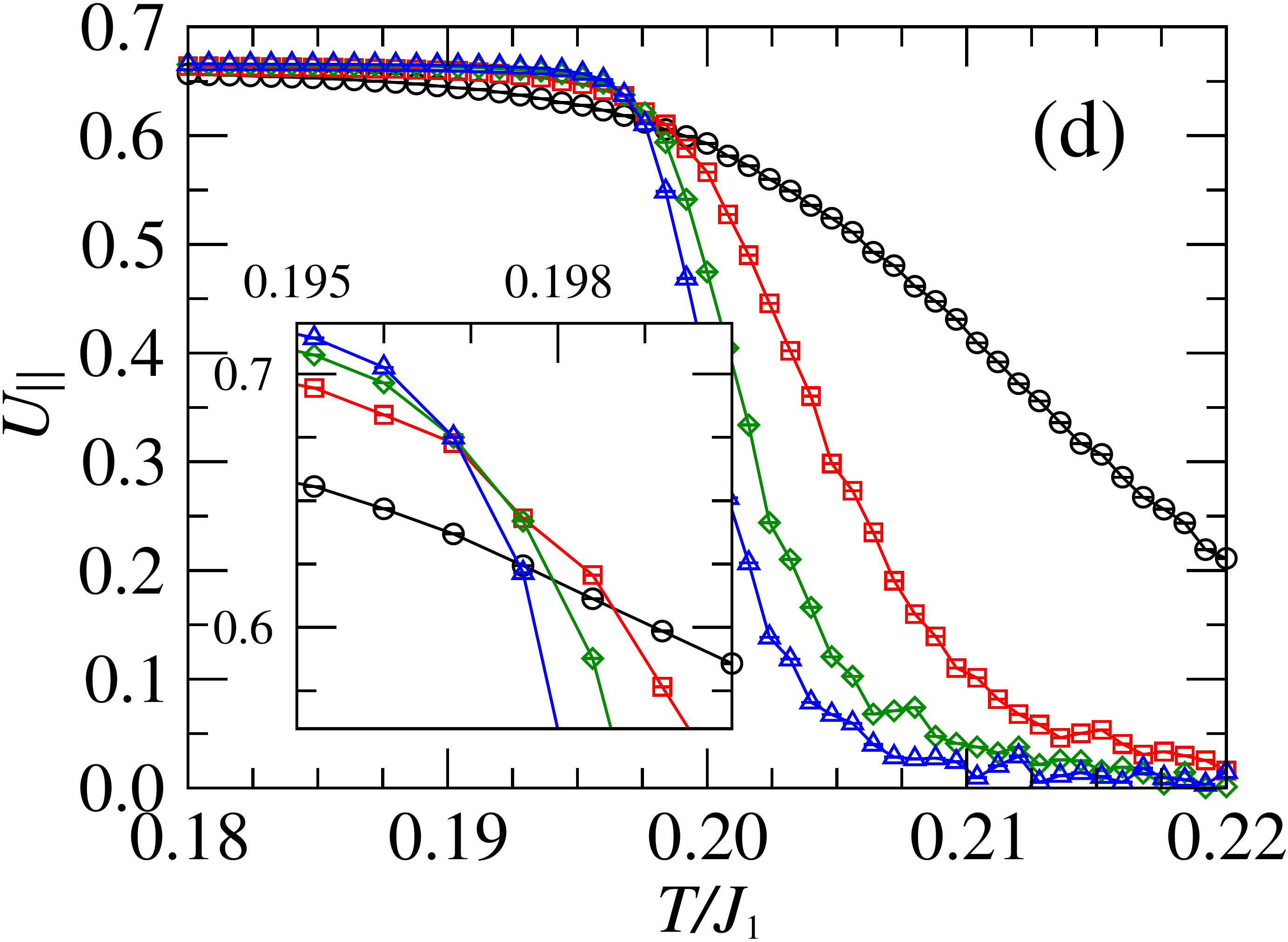}
\par\end{centering}
\caption{\label{fig:cleanPT}The MC results for clean system and $J_{2}=0.55J_{1}$.
(a) The modulus of the nematic order parameter, (b) the specific heat,
(c) the nematic susceptibility, and (d) the Binder cumulant as a function
of the temperature for several system sizes. The inset shows a magnification
around the crossing points signaling the location of the transition
temperature $T_{c}\approx0.197(1)J_{1}$. }
\end{figure}

In Fig.~\ref{fig:cleanPT} we plot the specific heat, the absolute
value of the nematic order parameter, the nematic susceptibility and
Binder cumulant as a function of the temperature for various system
sizes.

\begin{figure}[b]
\begin{centering}
\includegraphics[clip,width=0.5\columnwidth]{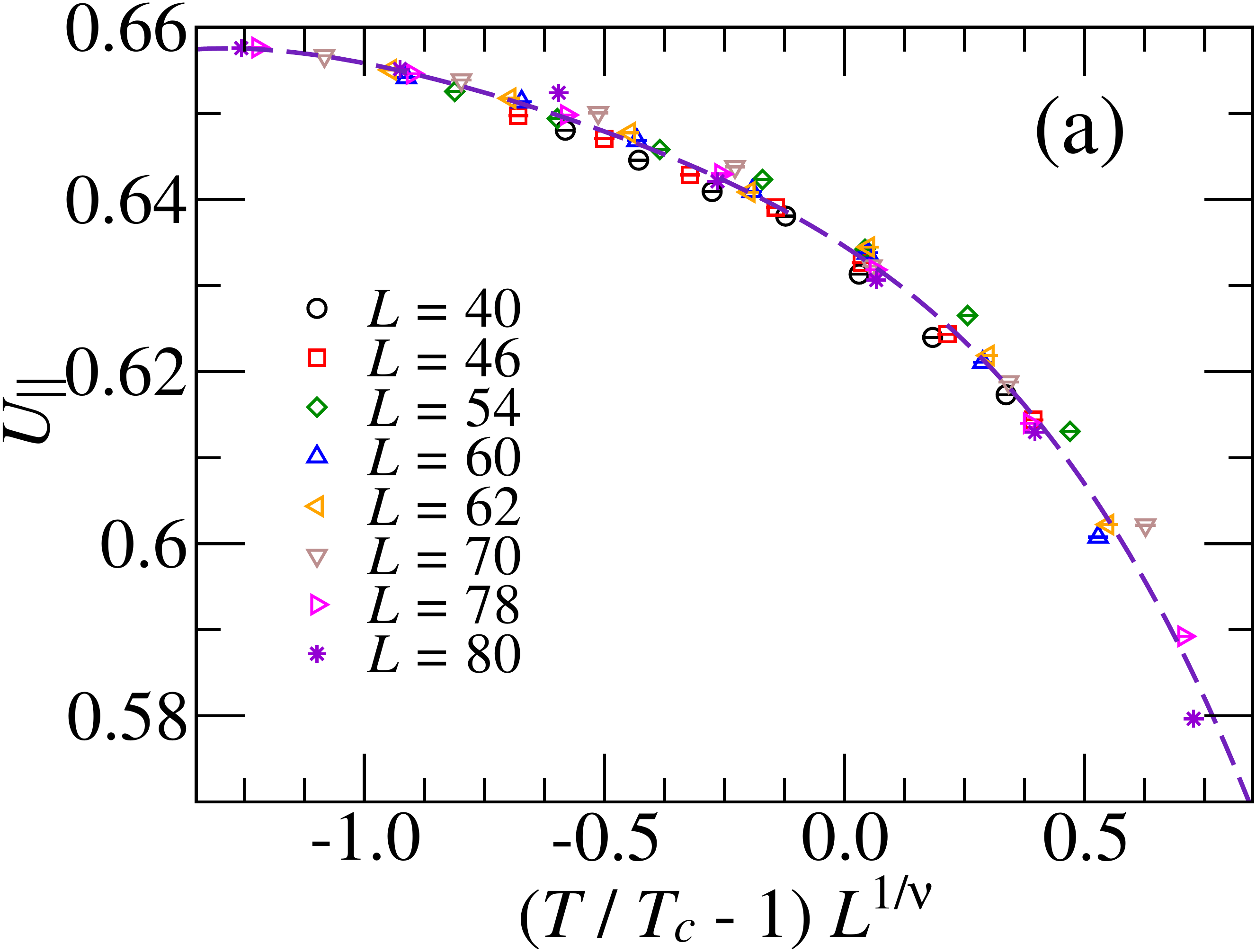}\includegraphics[clip,width=0.5\columnwidth]{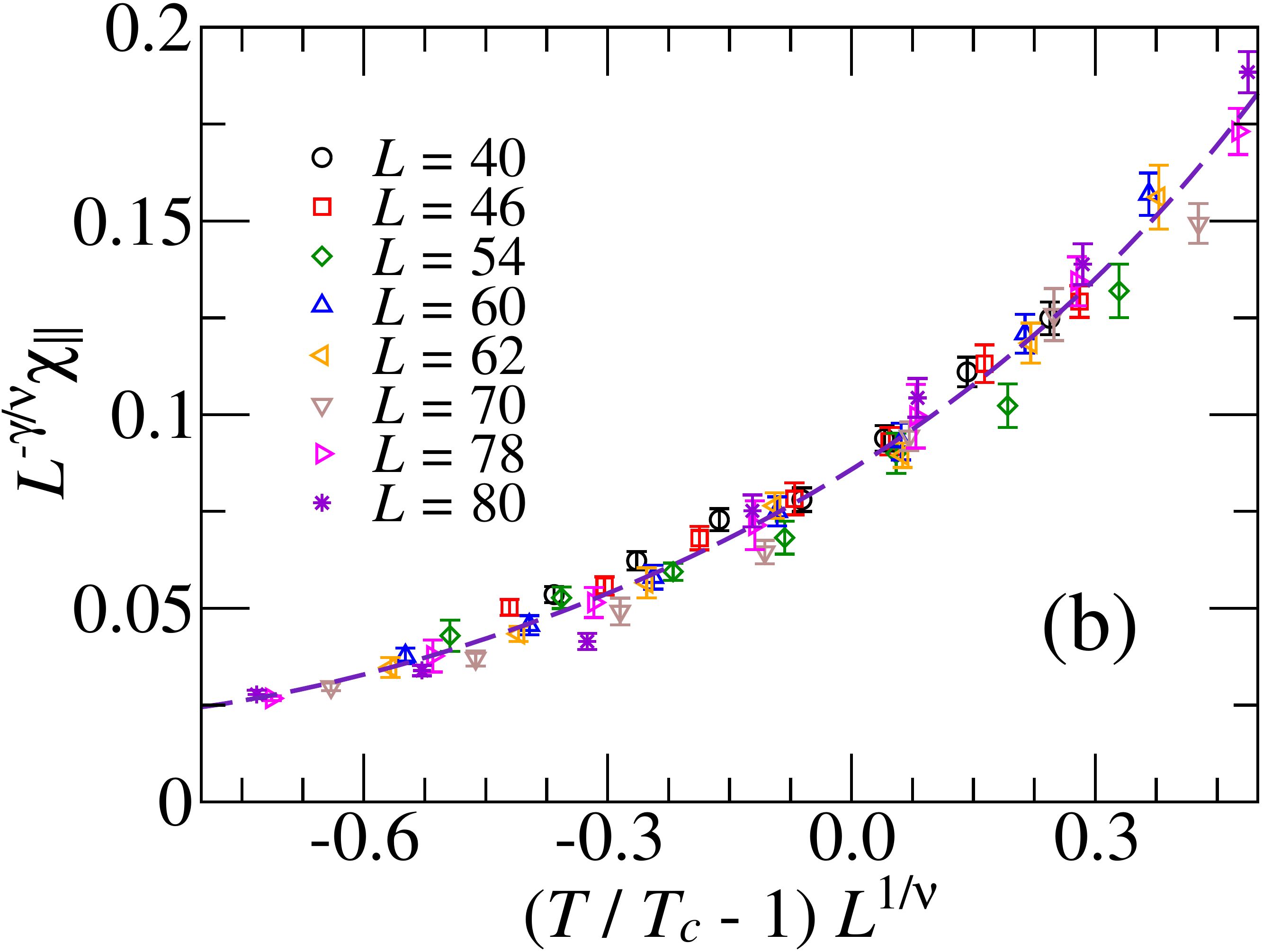}
\par\end{centering}
\caption{\label{fig:cleanPT-FSS}The finite-size scaling replot of (a) the
Binder cumulant and (b) the nematic susceptibility (right) data of
Fig.~\ref{fig:cleanPT}. The lines are simple quartic best fits representing
the corresponding scaling functions.}
\end{figure}

Clearly, these observables exhibit the usual temperature dependence
of a continuous phase transition. In particular, the Binder cumulant
tends to $\frac{2}{3}$ for $T<T_{c}$ and to $0$ for $T>T_{c}$
in the thermodynamic limit. For finite-size systems, extrapolating
the crossings of different sizes (not shown) allows us to estimate
the critical temperature to $T_{c}\approx0.197(1)J_{1}$. The standard
finite-size scaling analysis~\citep{goldenfeld-book,Sandvik-MC}
is shown in Fig.~\ref{fig:cleanPT-FSS} from which we obtain $\nu=1.00(6)$
and $\gamma=1.72(5)$, compatible with the 2D Ising universality class
($\nu=1$ and $\gamma=7/4$) and the earlier estimate reported in
Ref.~\citealp{weber03}.

\subsection{Spatially anisotropic disorder}

In this section we study the effects of dilution of horizontal $J_{1}$
bonds and of horizontal nearest-neighbor site pairs. We do not find
any hint of a phase transition. Instead, the system is always polarized
in the $\mathbf{Q}_{+}$ ($\mathbf{Q}_{-}$) nematic state when of
bond (site) dilution.

We start by reporting the effects of a single $J_{1}$-bond impurity.
We have verified (not shown for the sake of brevity) that, for low
temperatures and small sample sizes, the nematic order parameter \eqref{eq:Nematic-OP}
is always positive in agreement with the predictions of Sec.~\eqref{subsec:single-impurity}.
(Recall that $m_{\parallel}$ averages to zero in the clean system.)
Accordingly, as the temperature is raised and the system size enlarged,
the difference between the $\mathbf{Q}_{\pm}$ states diminishes and
so does the order-parameter mean value. 

\begin{figure}[t]
\begin{centering}
\includegraphics[clip,width=0.5\columnwidth]{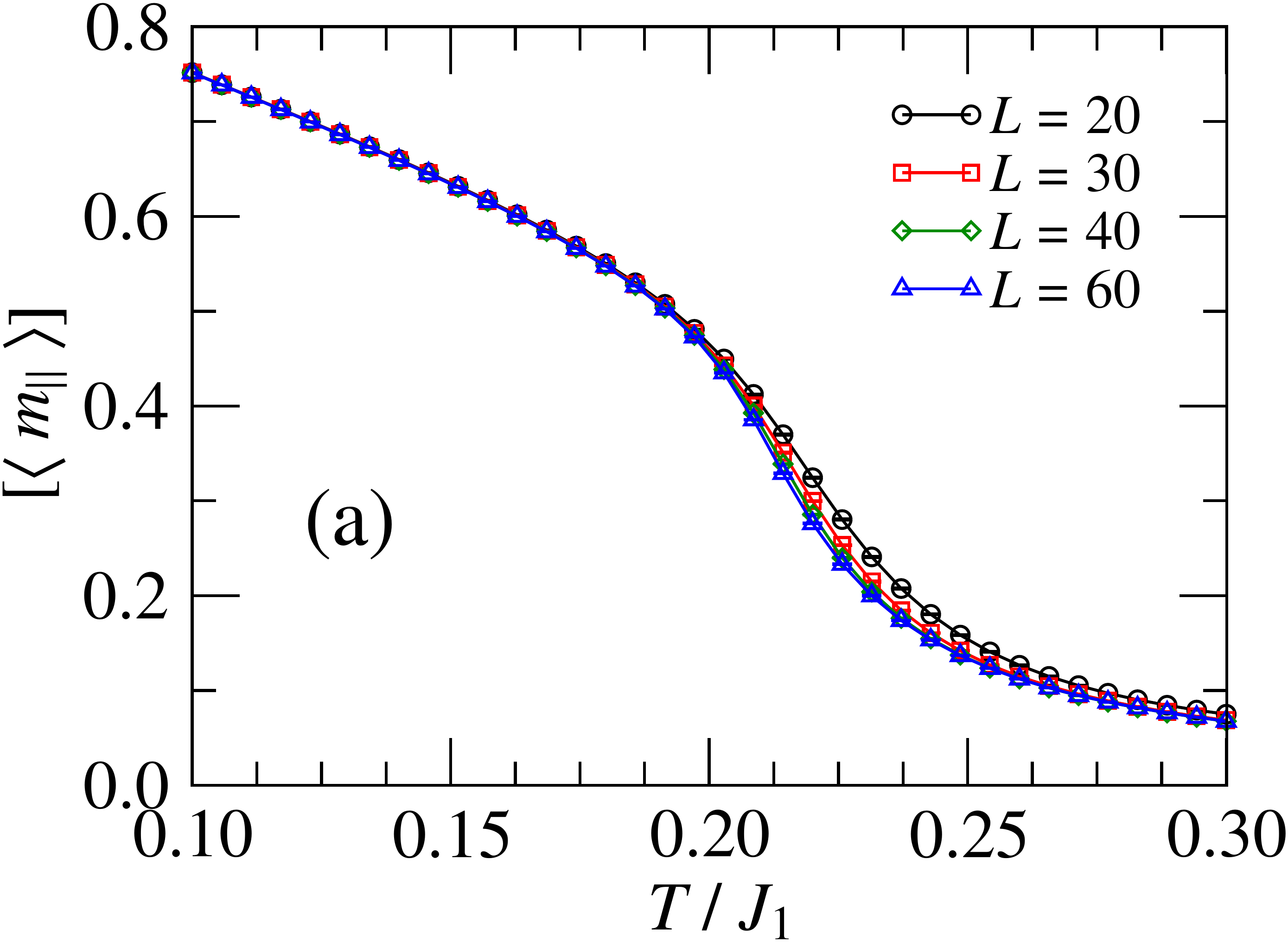}\includegraphics[clip,width=0.5\columnwidth]{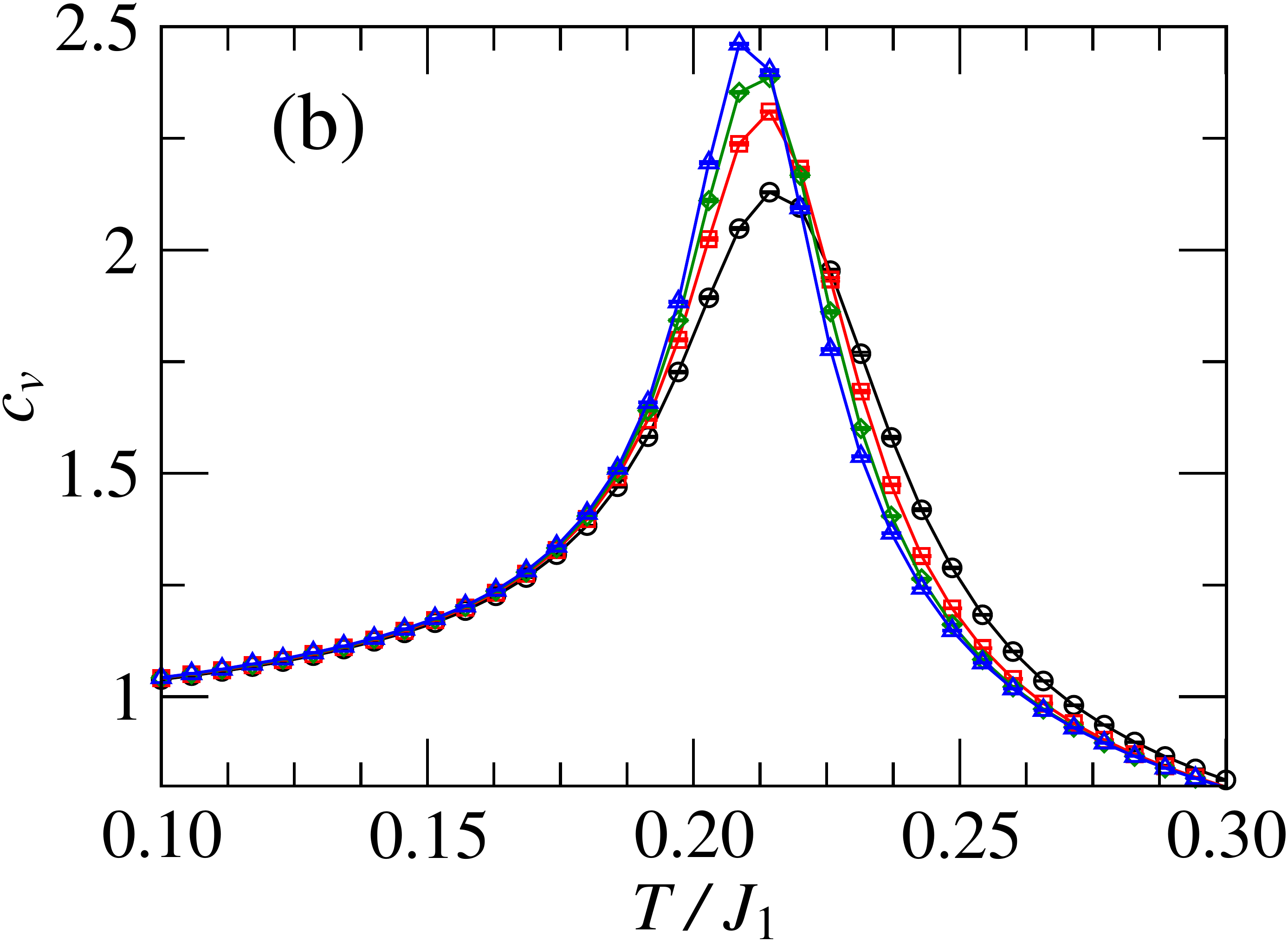}\\
\includegraphics[clip,width=0.5\columnwidth]{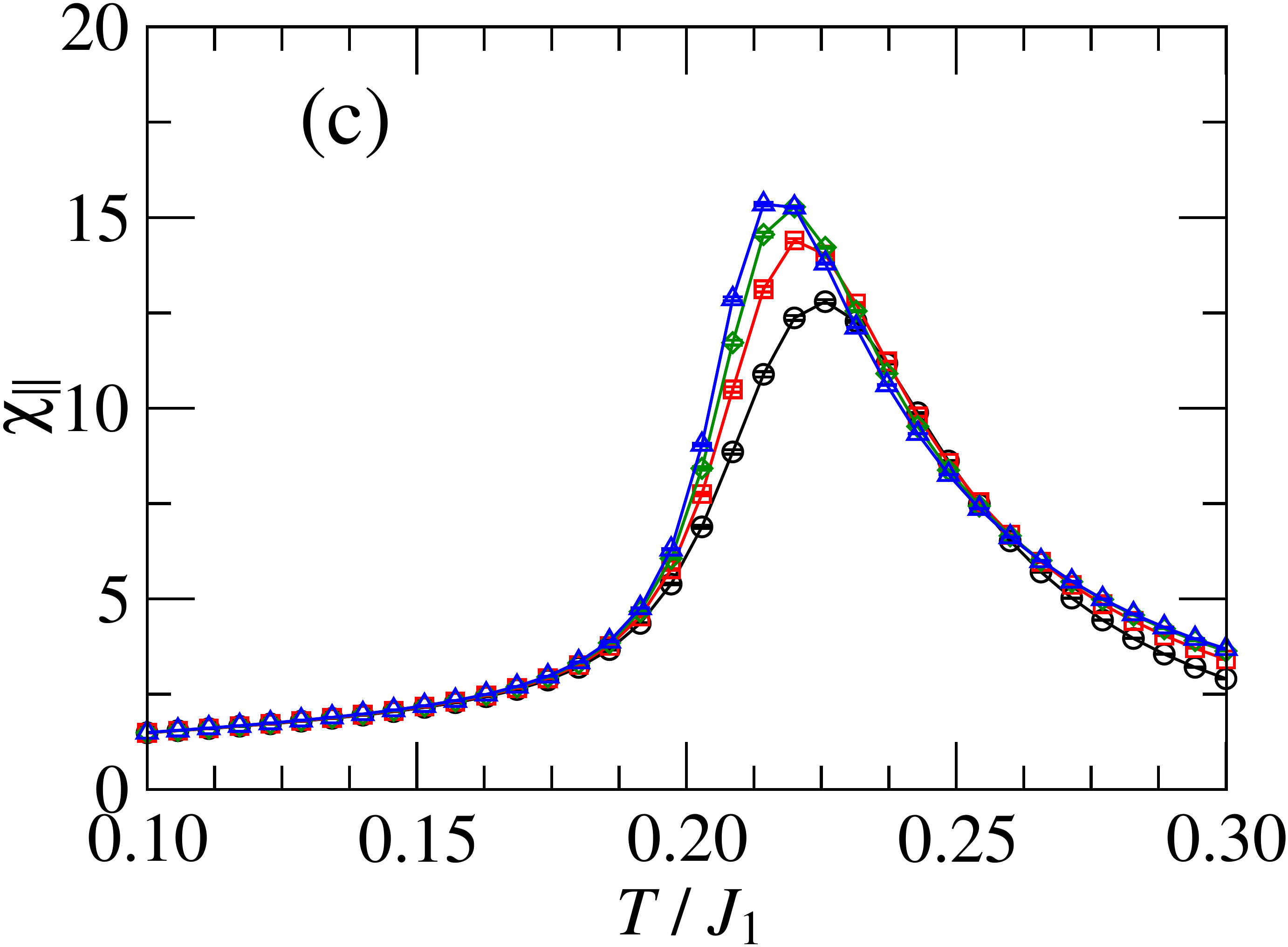}\includegraphics[clip,width=0.5\columnwidth]{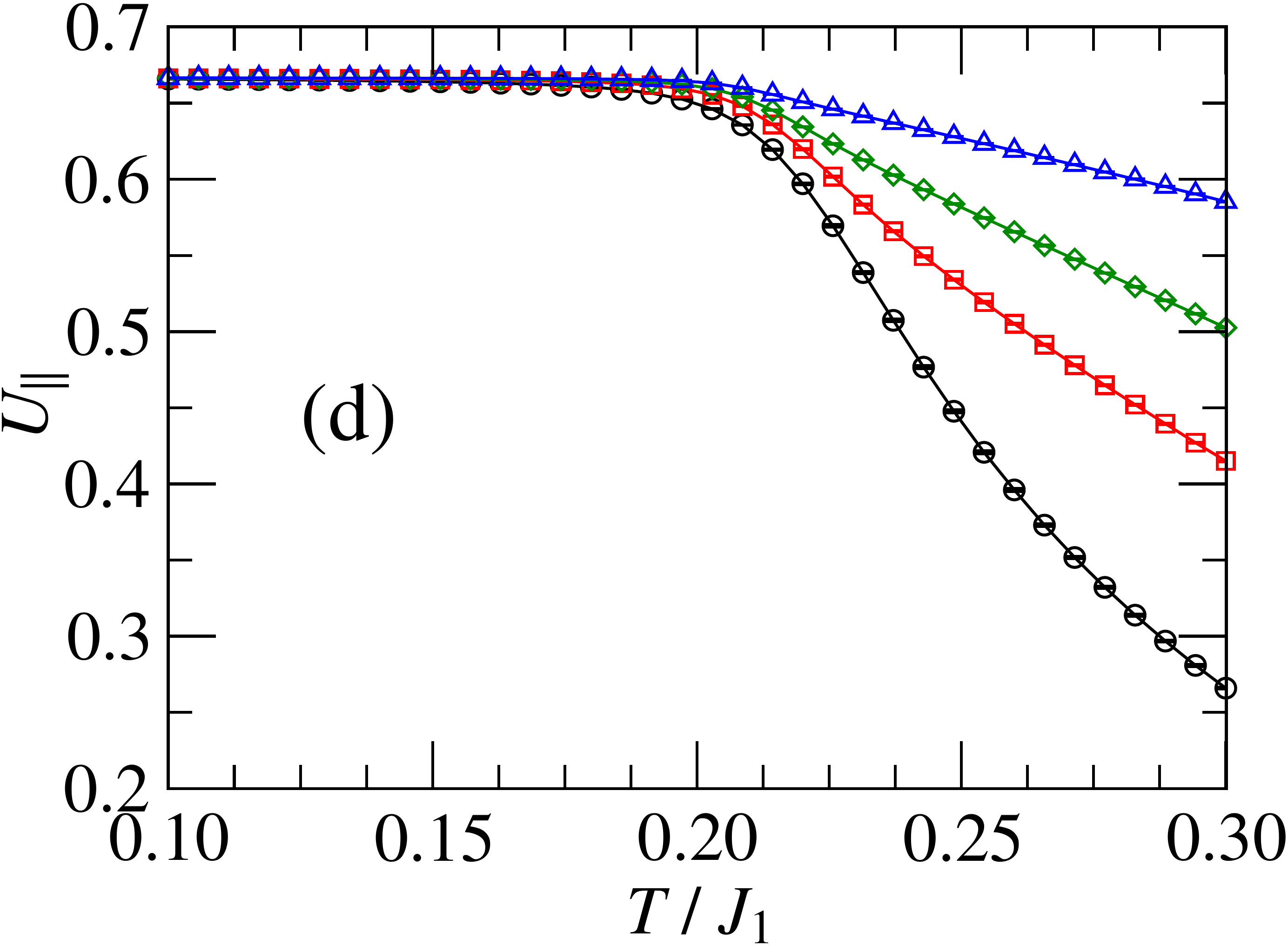}
\par\end{centering}
\caption{\label{fig:hBonds}The MC results for the case of $2\%$ dilution
of horizontal $J_{1}$ bonds where $J_{2}=0.55J_{1}$. (b) The specific
heat, (a) the nematic order parameter and (c) the corresponding susceptibility,
and (d) the Binder cumulant as a function of the temperature for several
system sizes. }
\end{figure}

We now report our study for the case of a finite fraction of impurities.
Figure \ref{fig:hBonds} is equivalent to Fig.~\ref{fig:cleanPT}
but with $2\%$ of horizontal $J_{1}$ bonds missing. Clearly, there
are no indications of singularities as the system size increases.
More importantly, there is no hint of crossings in the Binder cumulant,
implying no phase transition even above the clean critical temperature
$T_{c}\approx0.19J_{1}$. Notice also that the order parameter is
always finite (and positive) for any finite temperature in agreement
with our expectations (see Sec.~\ref{subsec:Nematic-RF}). Finally,
another support of this conclusion is the strong indication of the
Binder cumulant saturating to $\frac{2}{3}$ in the thermodynamic
limit for all temperatures. 

\begin{figure}[b]
\begin{centering}
\includegraphics[clip,width=0.8\columnwidth]{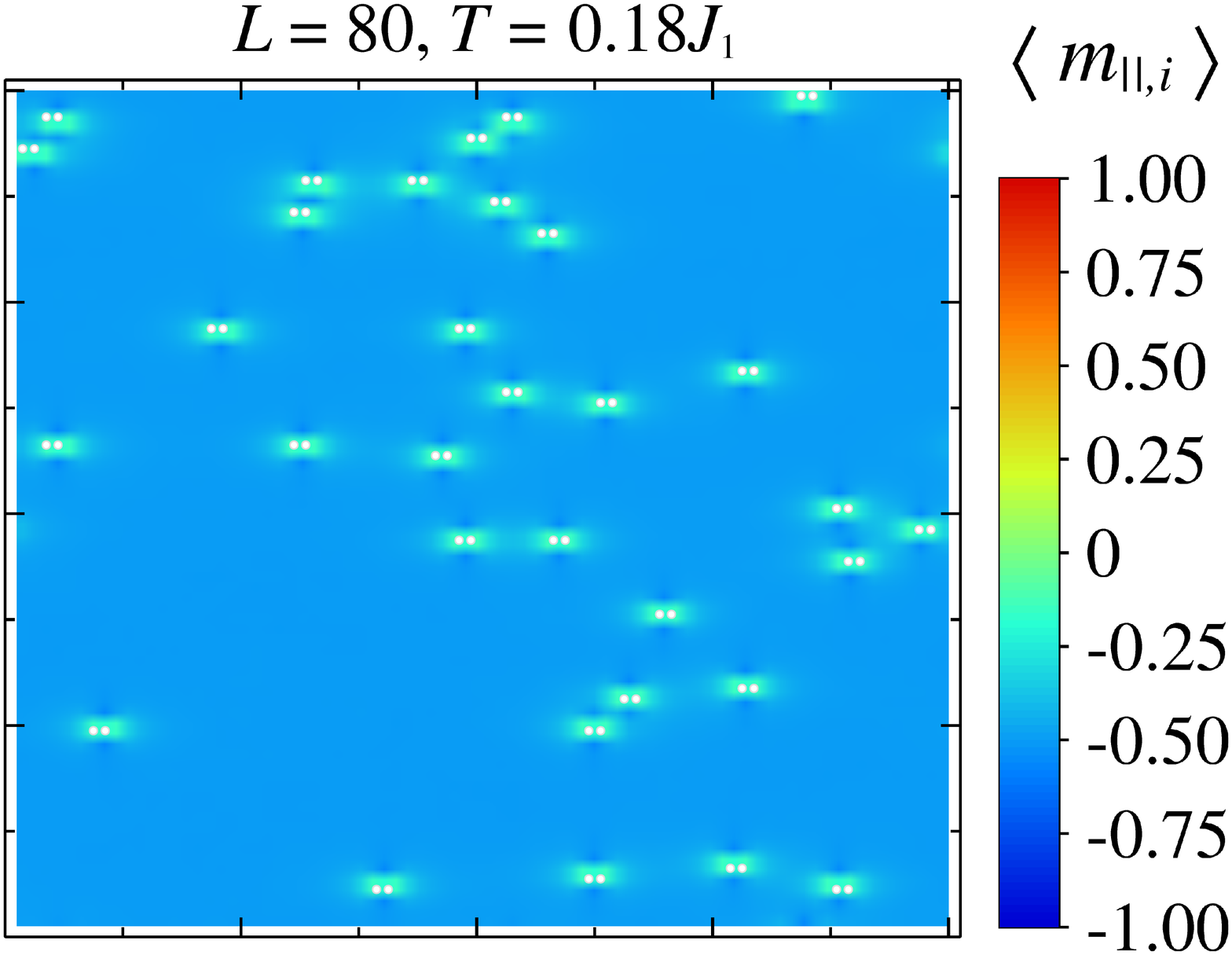}\\
\includegraphics[clip,width=0.8\columnwidth]{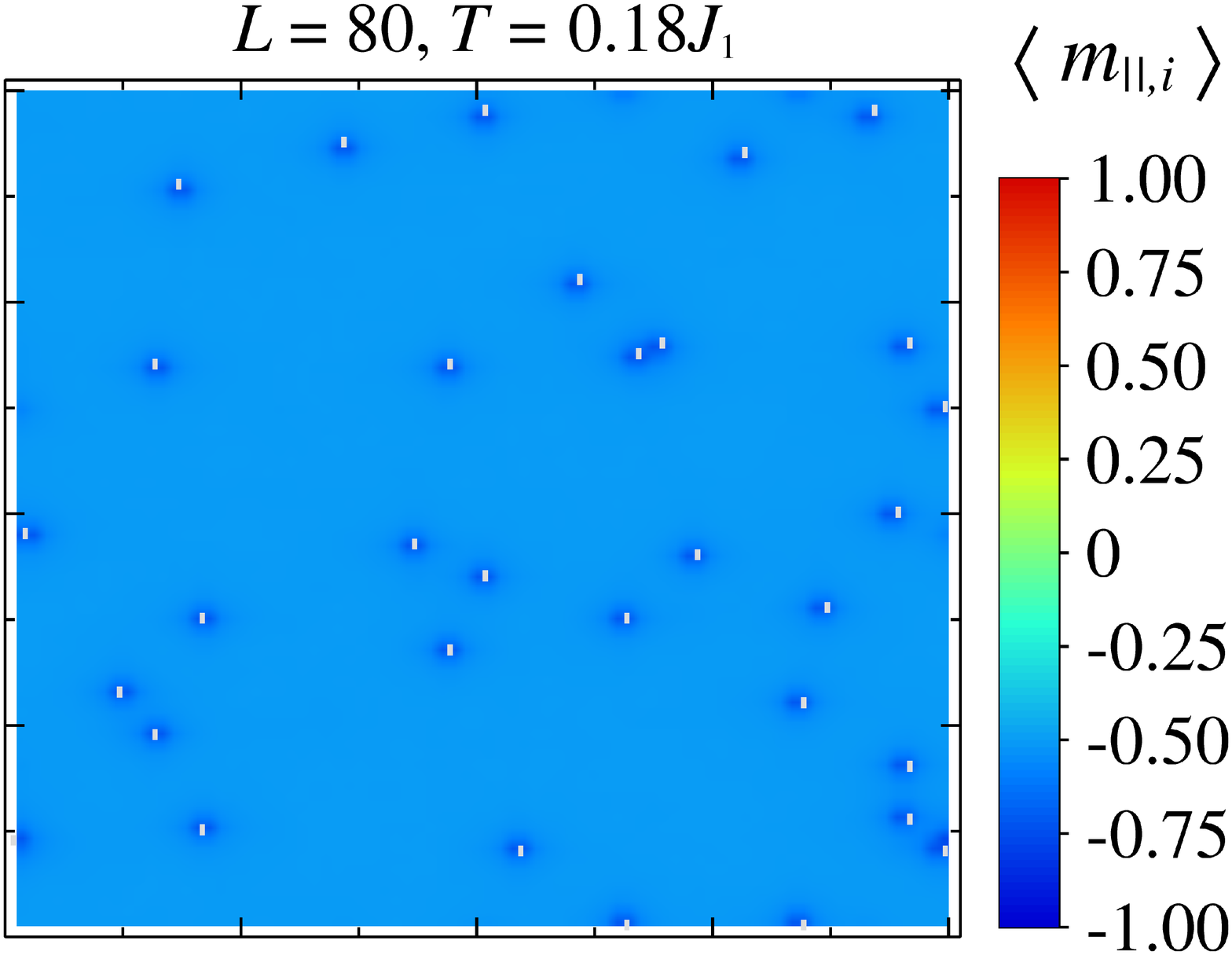}
\par\end{centering}
\caption{\label{fig:Density-HsiteDilution}A typical density plot of the local
nematic order parameter $\left\langle m_{\parallel,i}\right\rangle $
{[}see Eq.~\eqref{eq:Nematic-OP}{]} for a lattice of size $L=80$,
temperature $T=0.18J_{1}$ and $J_{2}=0.55J_{1}$. The light dots
(top) and light ticks (bottom) show the position of the missing sites
($1\%$ on average) and vertical bonds ($0.5\%$ on average), respectively.
Notice that the site impurities always appear in pairs of horizontal
nearest neighbor sites locally favoring the nematic $\mathbf{Q}_{-}$
state. For these particular disorder realizations, the nematic order
parameter averages $\left\langle m_{\parallel}\right\rangle =-0.486(1)$
(site dilution) and $-0.521(1)$ (bond dilution).}
\end{figure}

Finally, we report that a completely similar study was conducted when
$1\%$ of the lattice sites are diluted in such a way that they appear
in pairs of horizontal nearest neighbors. We report that completely
analogous results as in Fig.~\ref{fig:hBonds} were found and, for
that reason, we do not to show them here. Instead, we present a complementary
result: the density plot of the local nematic order parameter for
a typical disorder configuration (see the top panel of Fig.~\ref{fig:Density-HsiteDilution}).
Clearly, the $\mathbf{Q}_{-}$-state is selected. We have verified
that this is the case in all temperatures. For comparison, in the
bottom panel we show a typical density plot for the case of $0.5\%$
of vertical $J_{1}$-bond dilution. Apart from the localized spin
texture around the defects, these plots are essentially the same.

\subsection{Spatially isotropic disorder\label{subsec:Isotropic-disorder}}

We now report on the more subtle case of isotropic disorder in which
the global (statistical) vertical/horizontal real-space symmetry is
preserved. Specifically, we consider the cases of uncorrelated $J_{1}$-bond
dilution. {[}We have also considered site dilution and obtained similar
results but with much larger finite-size effects since $\xi_{\parallel\text{RF}}$
are considerably larger (see Sec.~\ref{subsec:Nematic-RF}).{]}

\begin{figure}[t]
\begin{centering}
\includegraphics[clip,width=0.5\columnwidth]{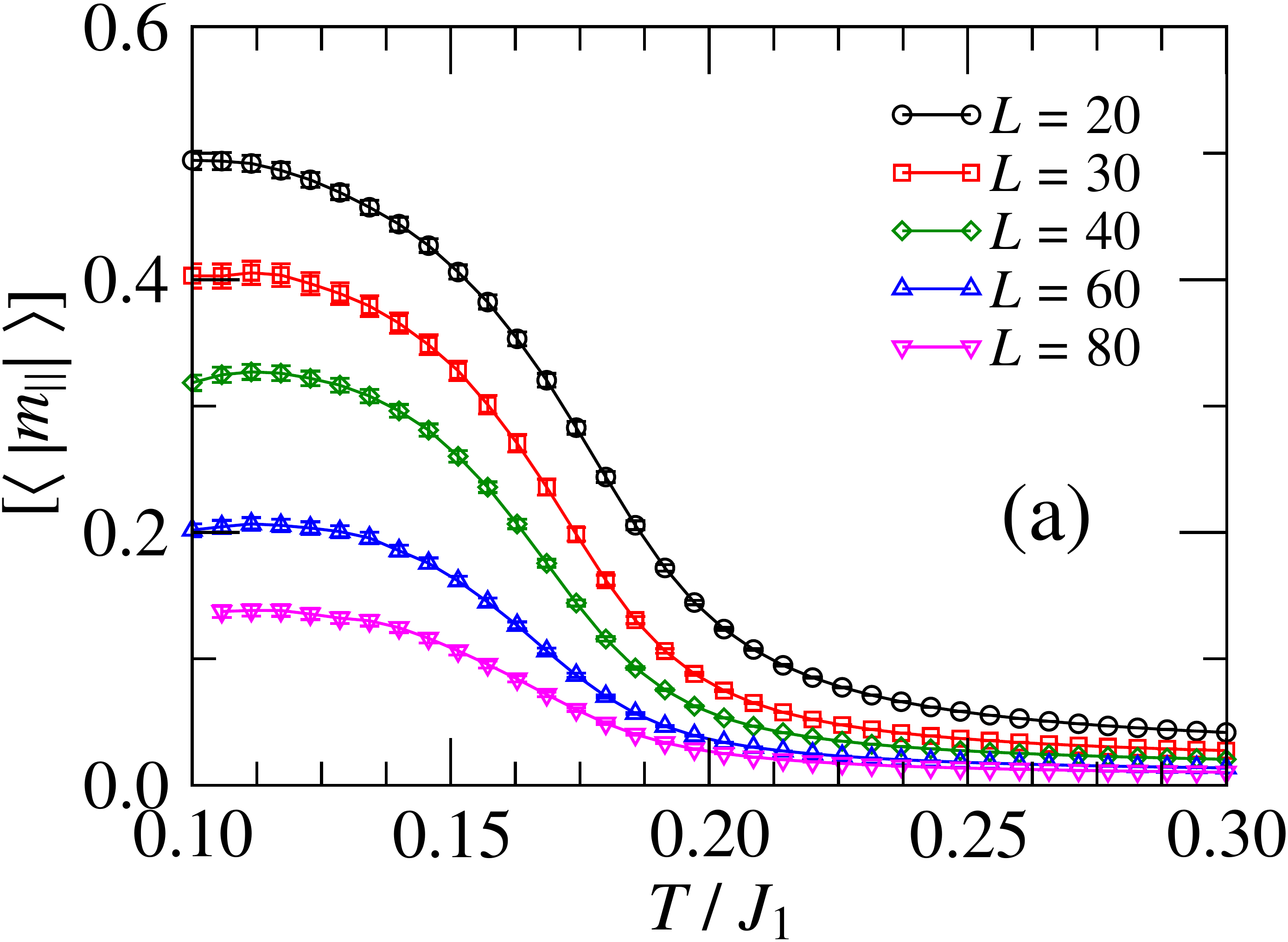}\includegraphics[clip,width=0.5\columnwidth]{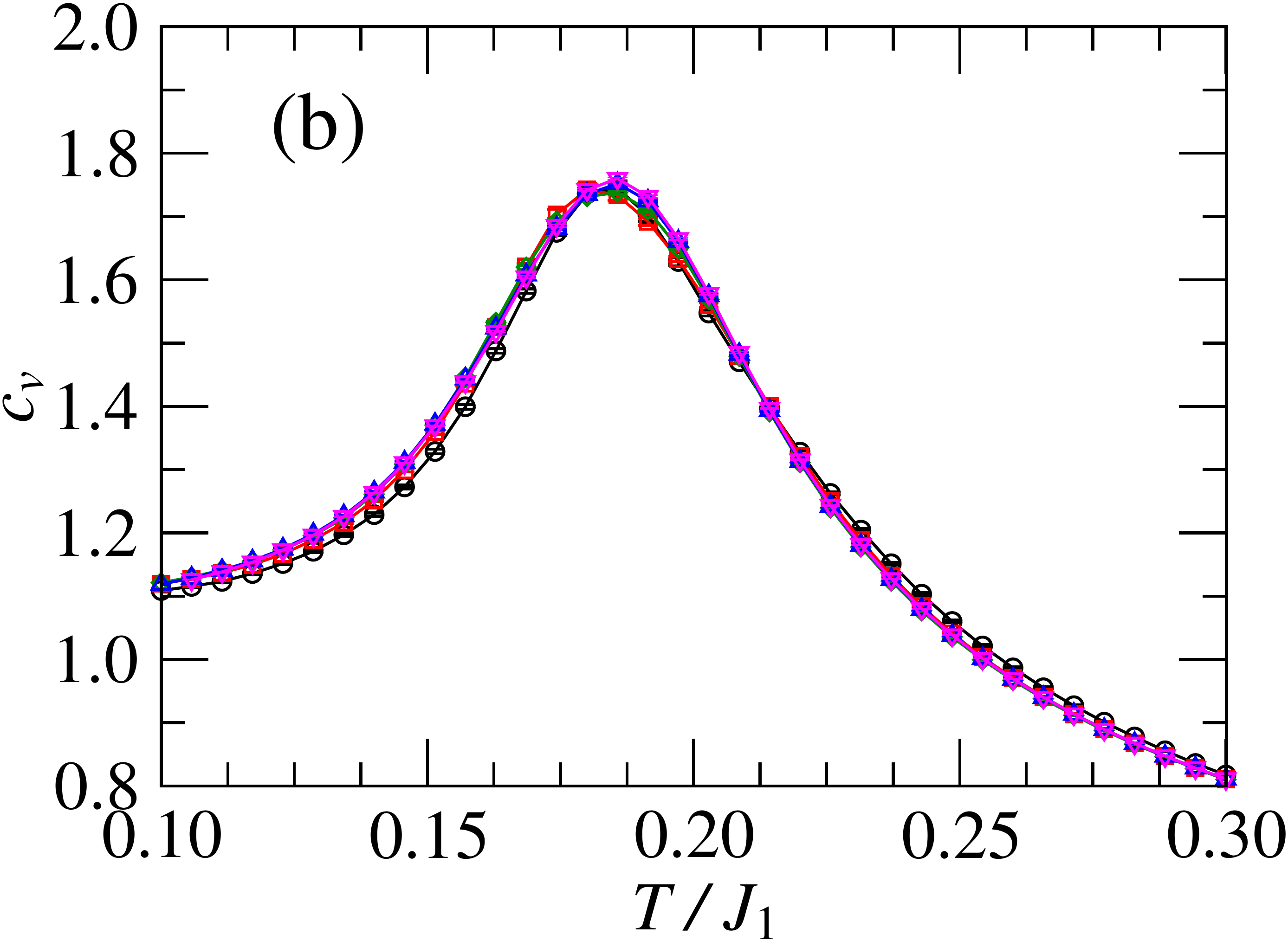}\\
\includegraphics[clip,width=0.5\columnwidth]{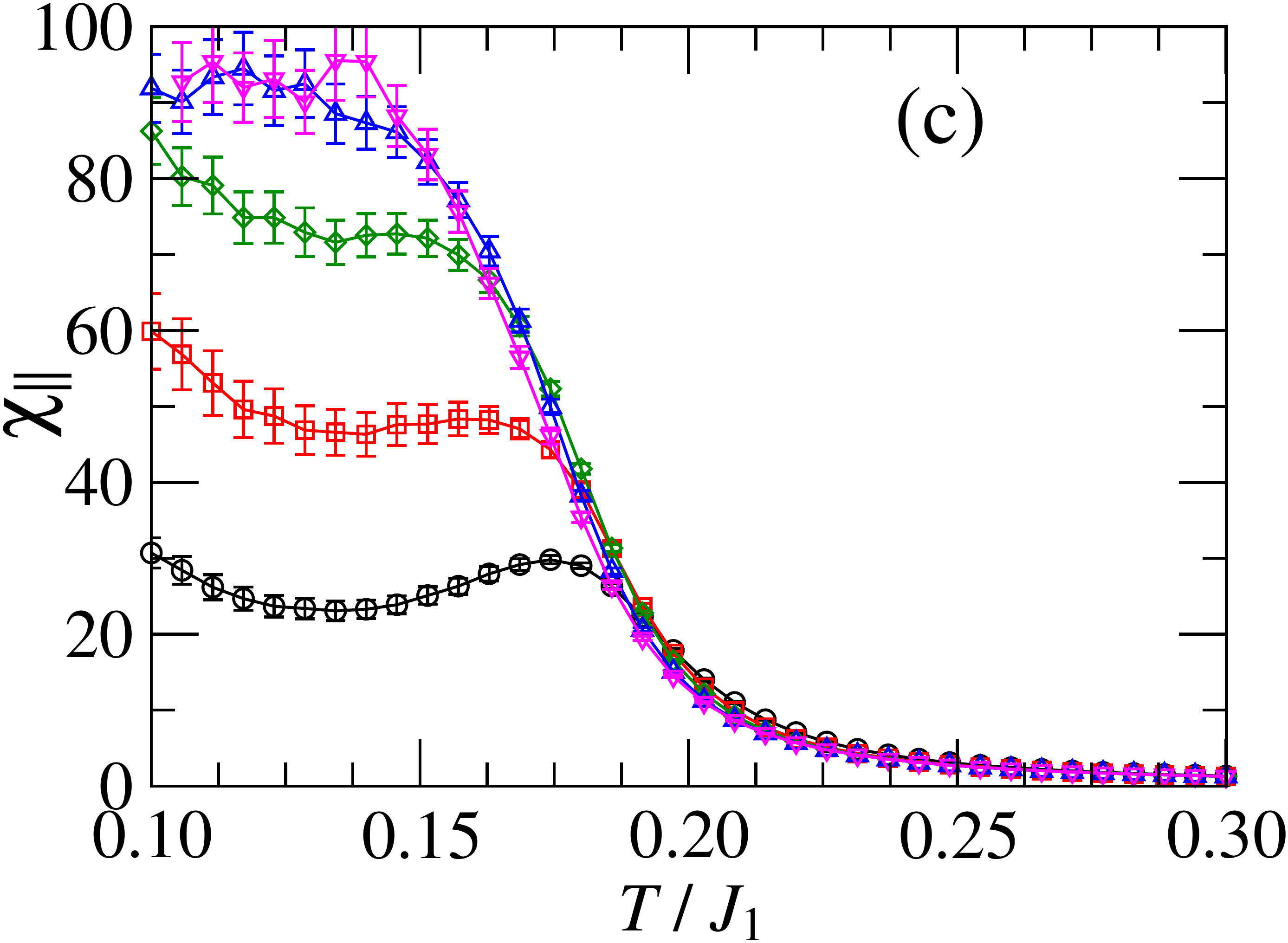}\includegraphics[clip,width=0.5\columnwidth]{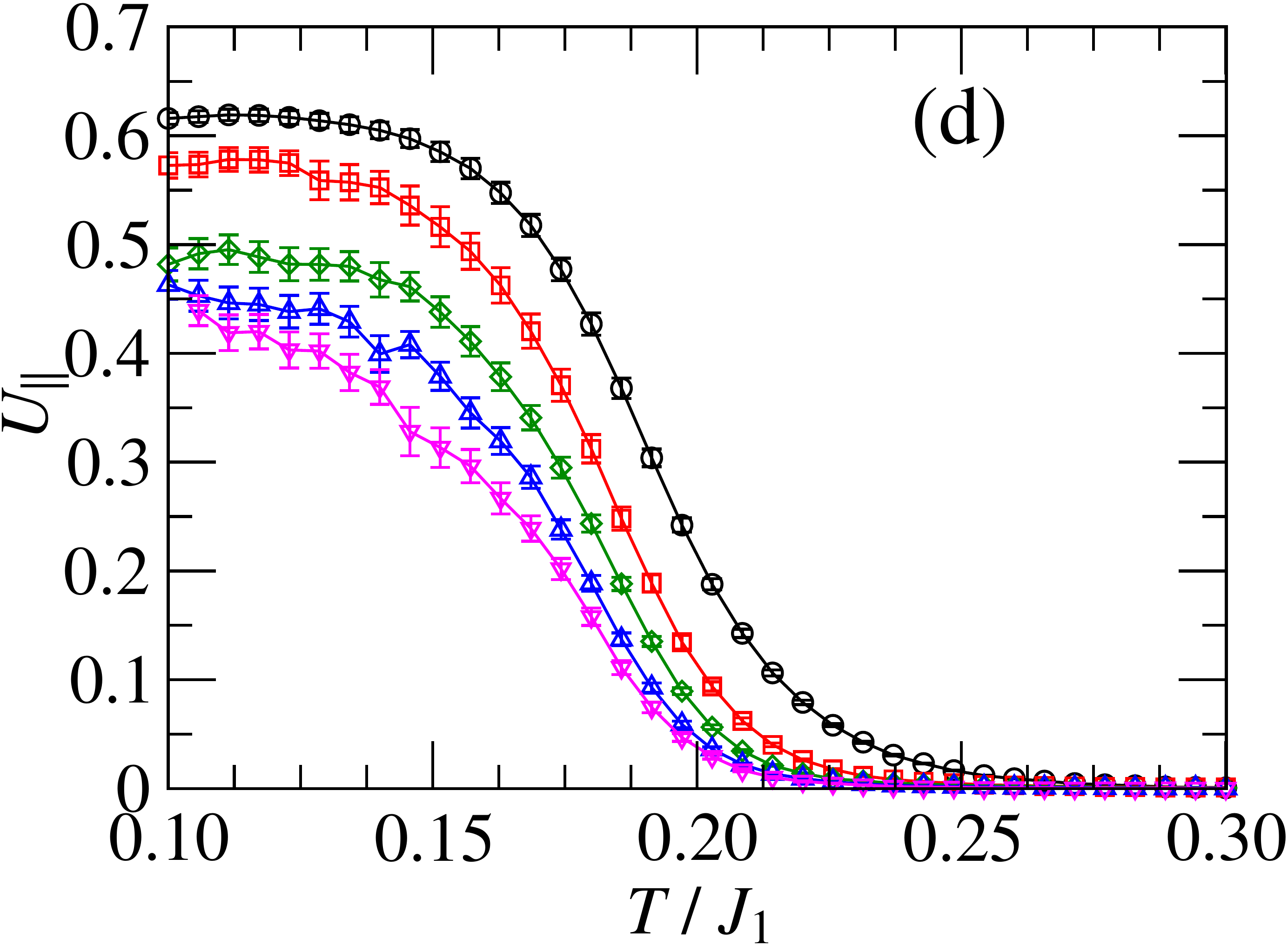}\\
\includegraphics[clip,width=0.5\columnwidth]{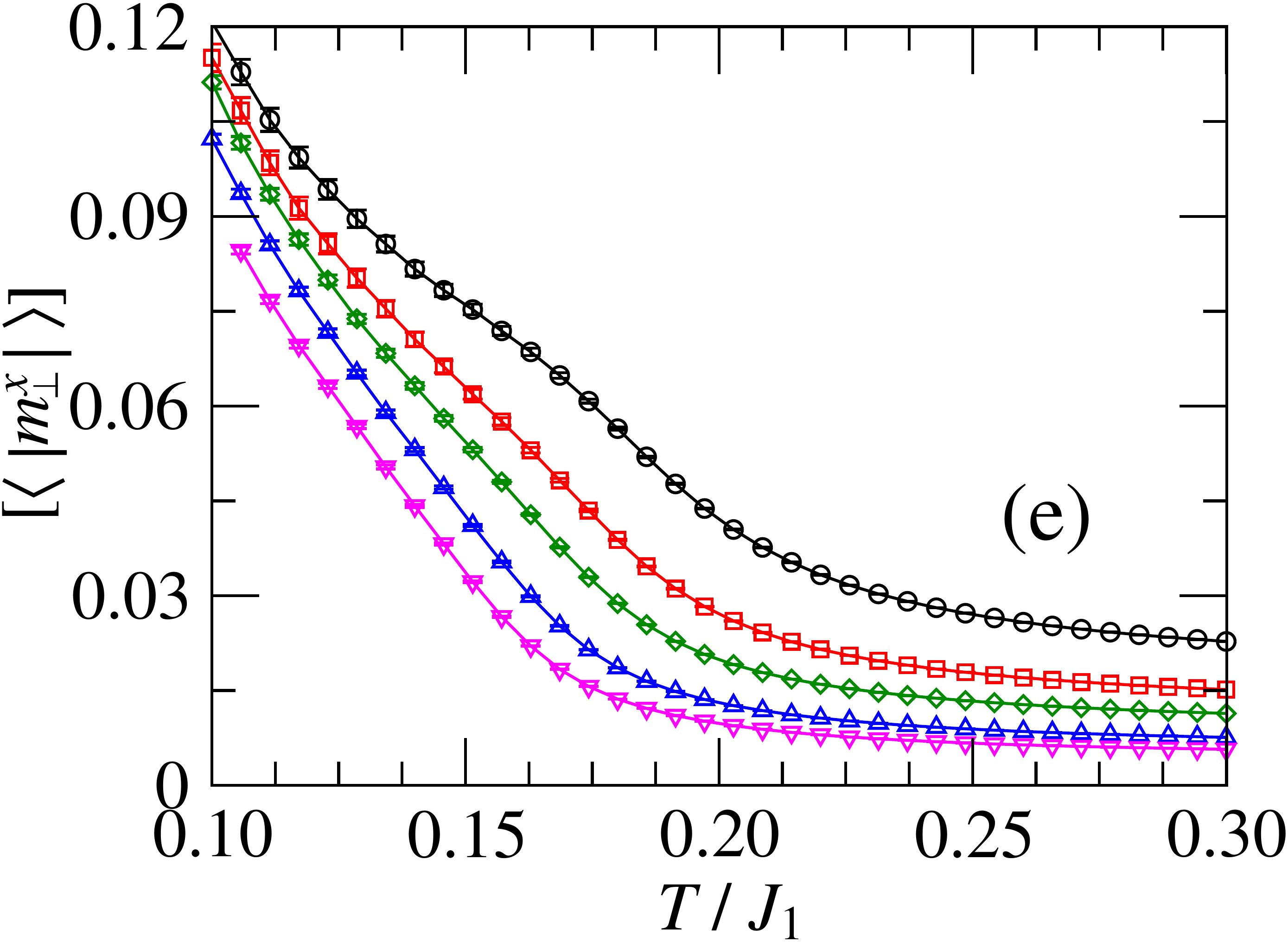}\includegraphics[clip,width=0.5\columnwidth]{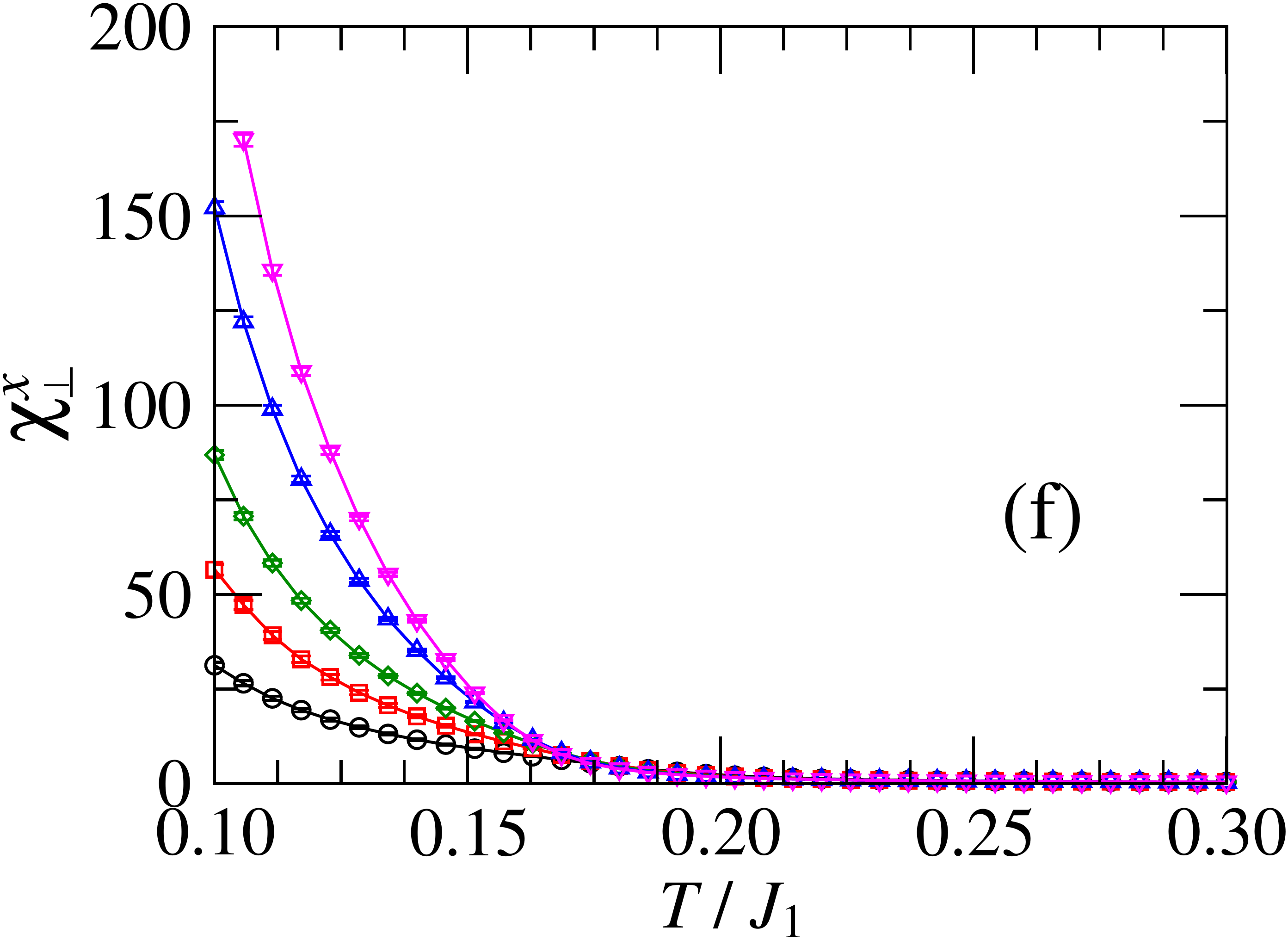}
\par\end{centering}
\caption{\label{fig:hvBonds}The MC results when $2\%$ of $J_{1}$ bonds are
isotropically diluted and $J_{2}=0.55J_{1}$. The (b) specific heat,
the (d) nematic Binder cumulant, the (a) nematic and (e) SVDW order
parameters, and their corresponding susceptibilities {[}(c) and (f),
respectively{]}, all as a function of the temperature and for several
system sizes. {[}We verified that $m_{\perp}^{y,z}$ ($\chi_{\perp}^{y,z}$)
is statistically identical to $m_{\perp}^{x}$ ($\chi_{\perp}^{x}$).{]}}
\end{figure}

In agreement with the arguments of Sec.~\ref{sec:Disorder}, we have
verified that there is no phase transition and the system is always
a paramagnet. This is illustrated in Fig.~\ref{fig:hvBonds} where
the specific heat, the nematic Binder cumulant, order parameter and
susceptibility are plotted as a function of the temperature for samples
of many sizes when $2\%$ of the $J_{1}$ bonds are isotropically
diluted. The SVDW order parameter and susceptibility are also shown.
Notice the striking difference with respect to the clean case. There
is no hint of singularities in any observable. In addition, both the
nematic order parameter and the Binder cumulant decrease with increasing
the system size for all temperatures. This is a strong indication
of a state without nematic long-range order. As expected, the SVDW
order parameter behaves likewise. Notice it vanishes slowly at lower
temperatures since the associated correlation length $\xi_{\perp}$
is exponentially large as $T\rightarrow0$.

\begin{figure}
\begin{centering}
\includegraphics[clip,width=0.5\columnwidth]{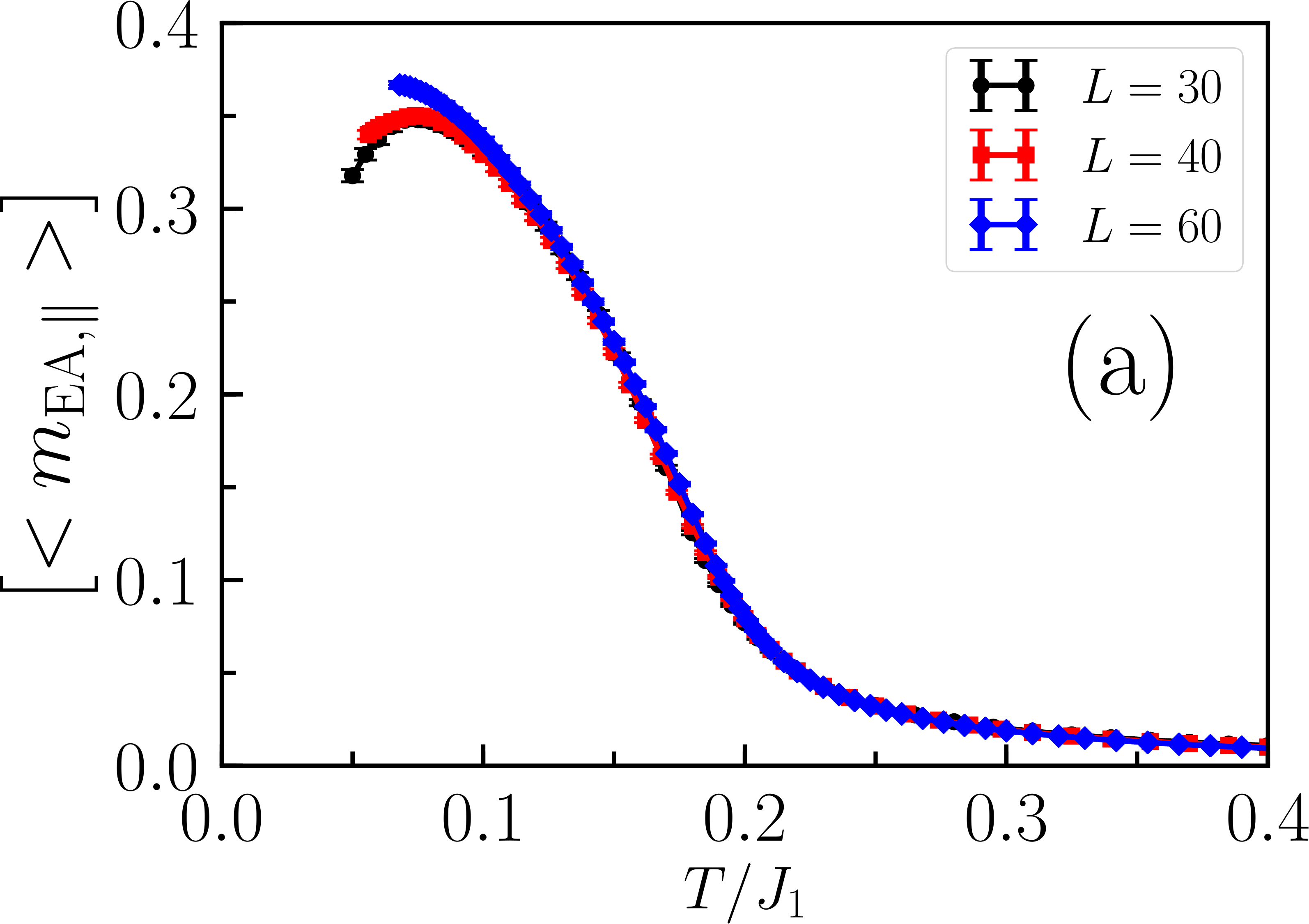}\includegraphics[clip,width=0.5\columnwidth]{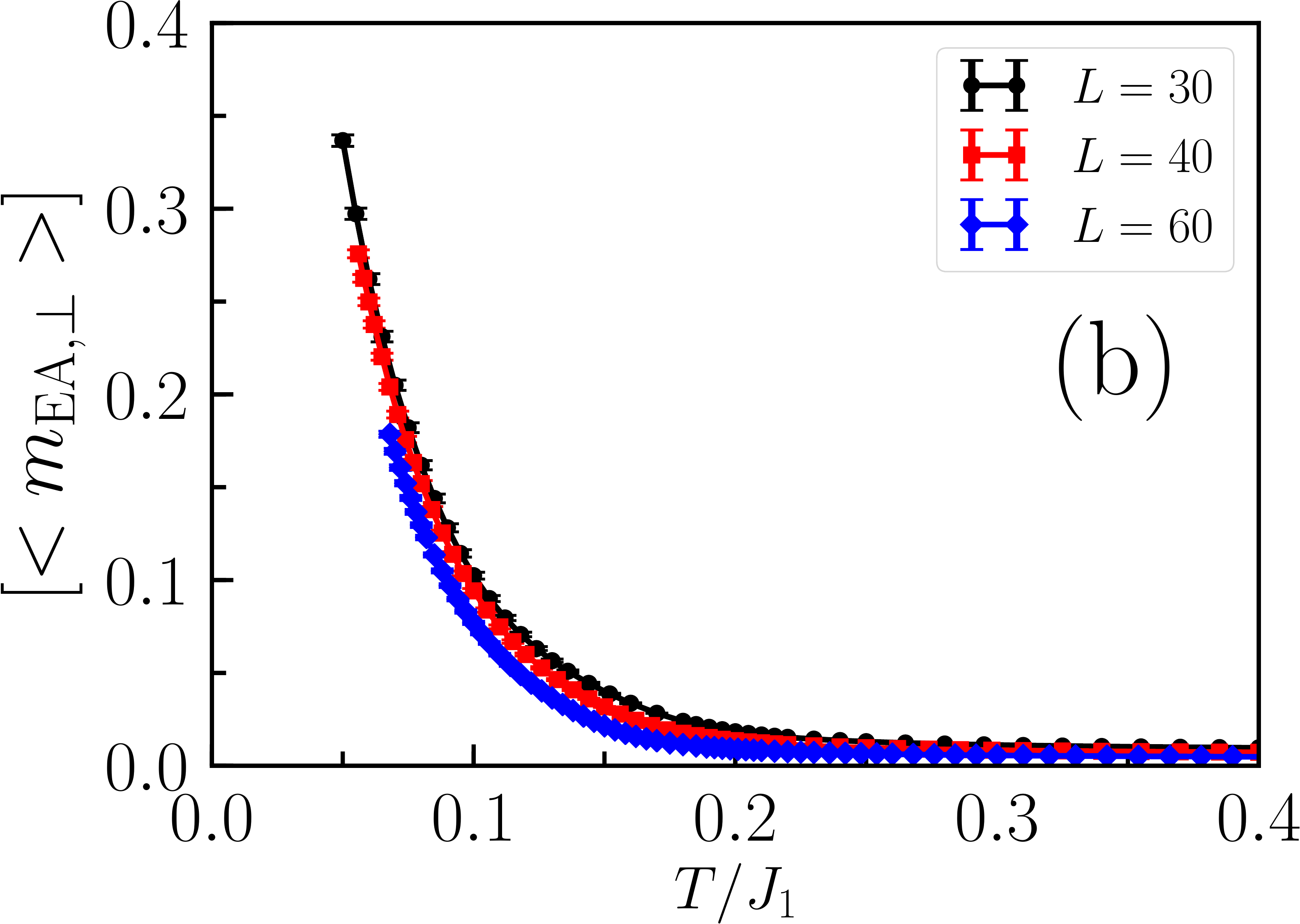}\\
\includegraphics[clip,width=0.5\columnwidth]{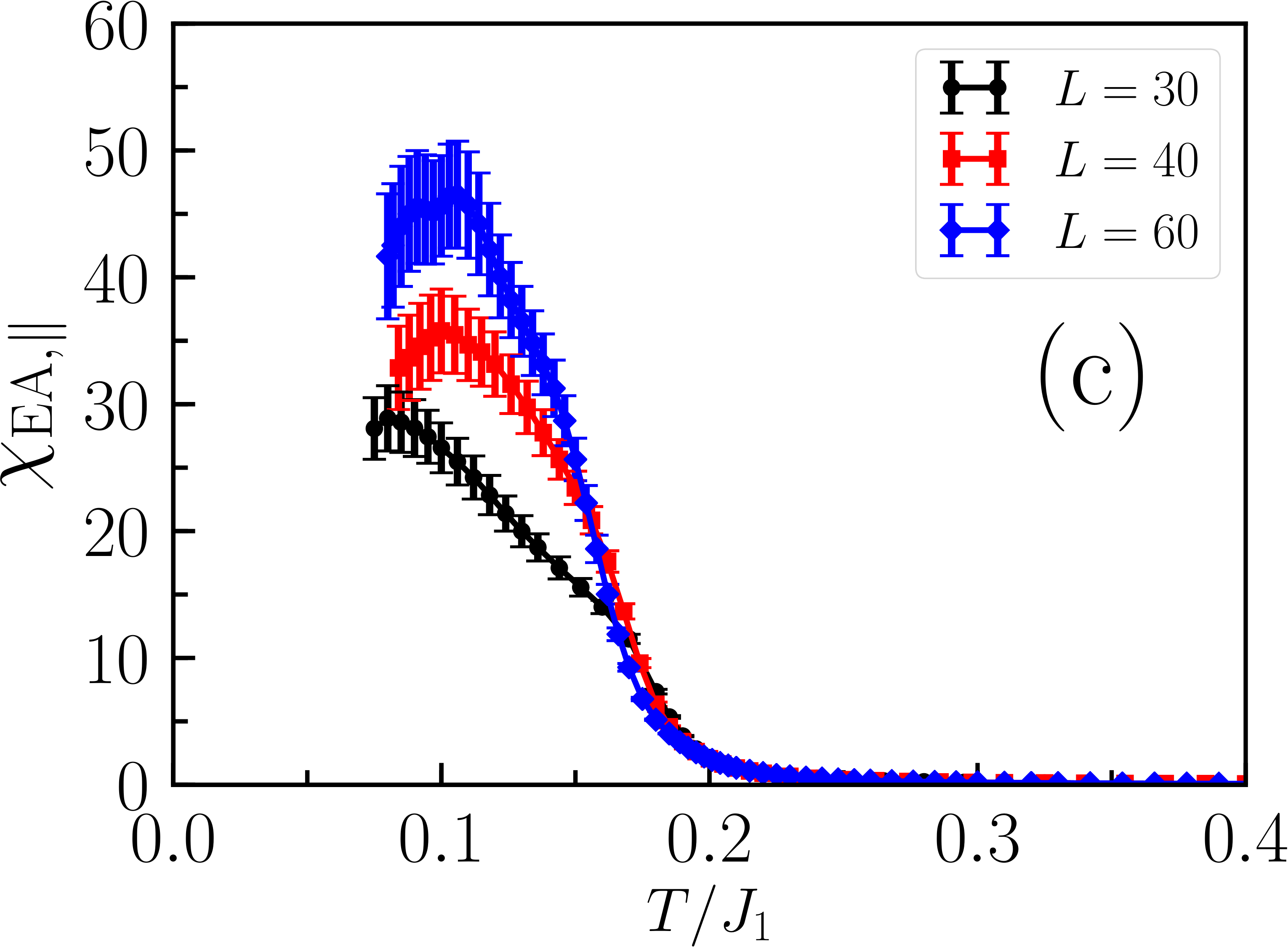}\includegraphics[clip,width=0.5\columnwidth]{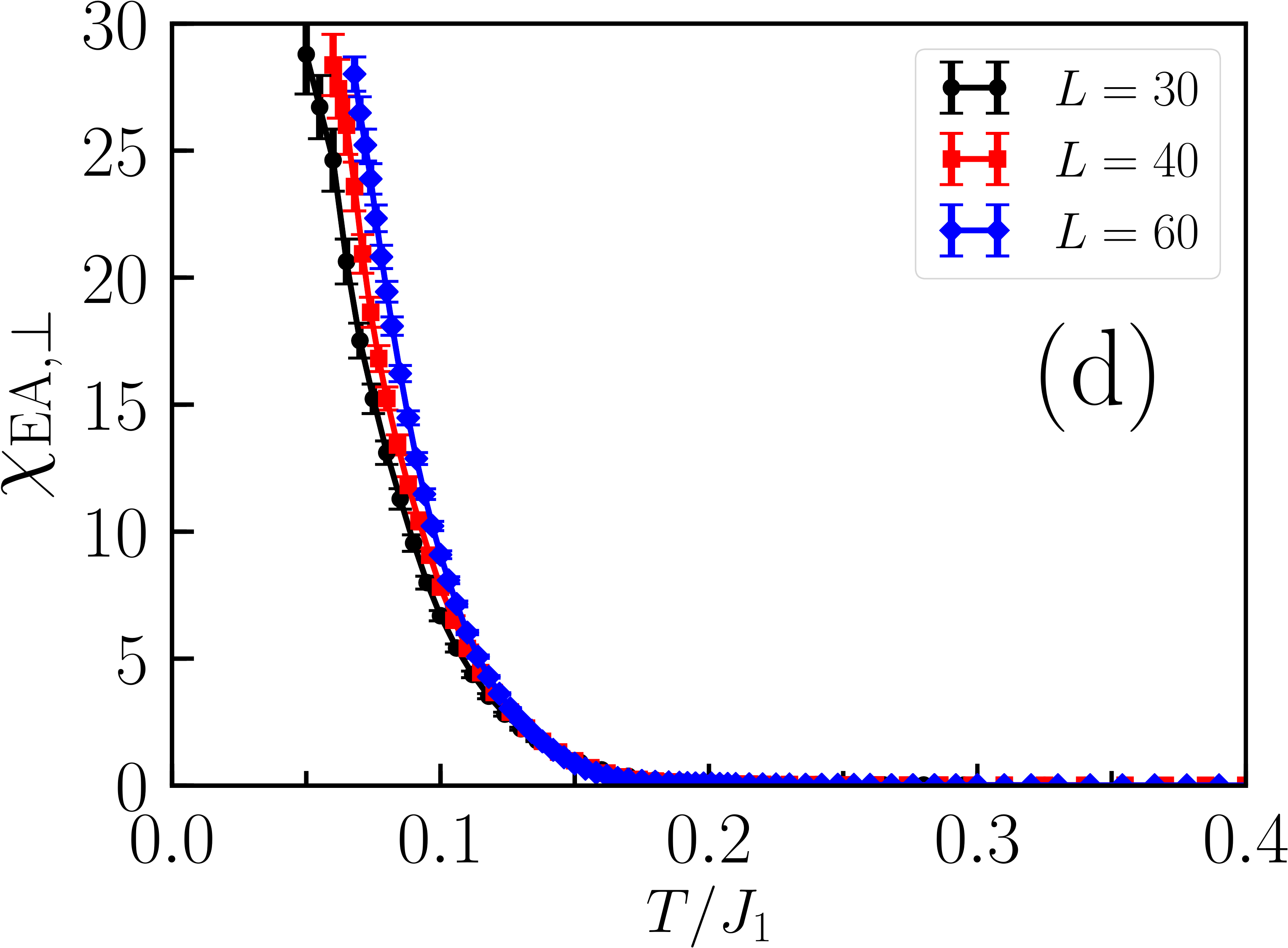}
\par\end{centering}
\caption{The Edwards-Anderson nematic and SVDW order parameters and the corresponding
susceptibilities as a function of the temperature for the same parameters
as in Fig.~\ref{fig:hvBonds}.\label{fig:glass-observables}}
\end{figure}

\begin{figure}
\begin{centering}
\includegraphics[clip,width=0.5\columnwidth]{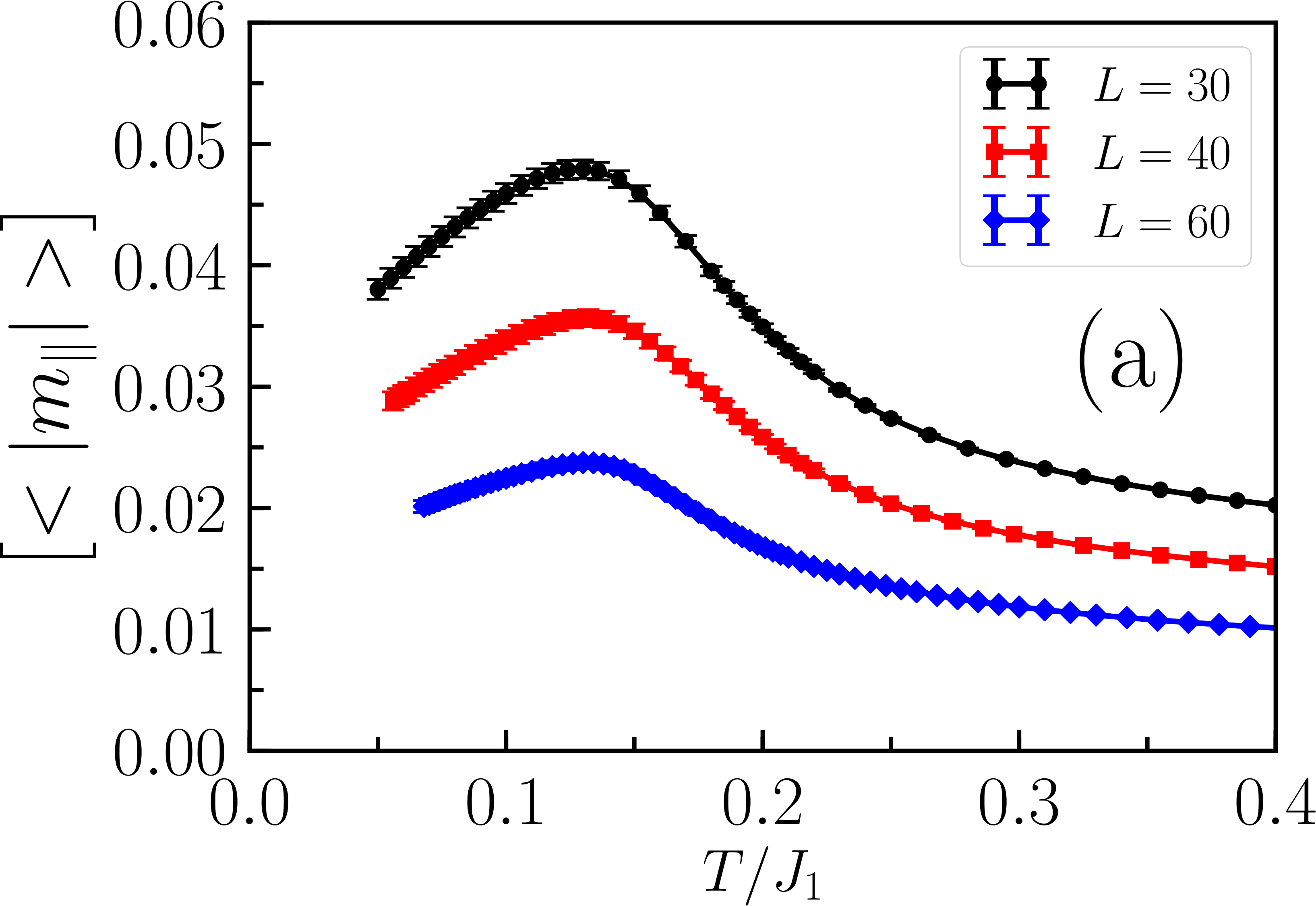}\includegraphics[clip,width=0.5\columnwidth]{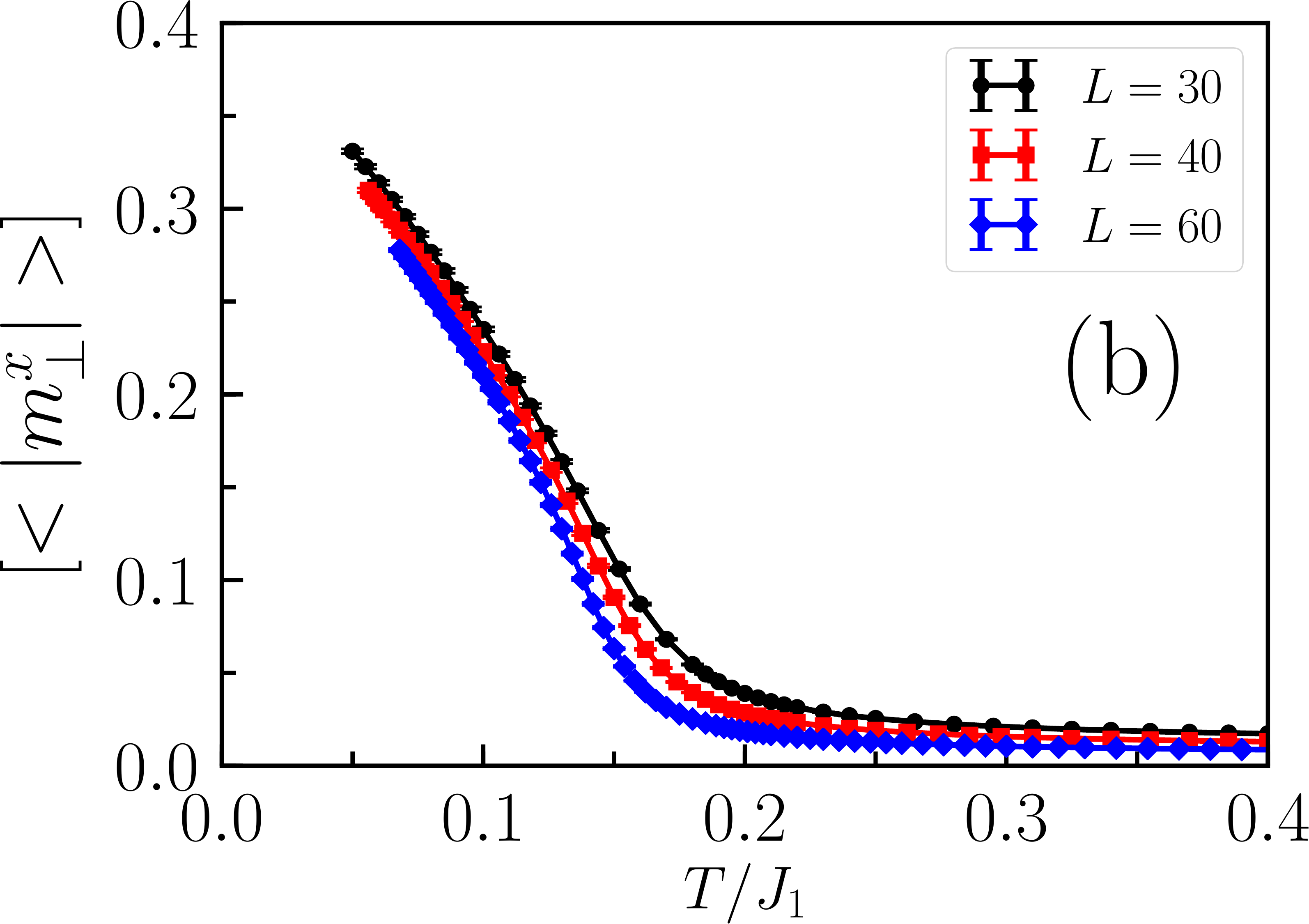}\\
\includegraphics[clip,width=0.5\columnwidth]{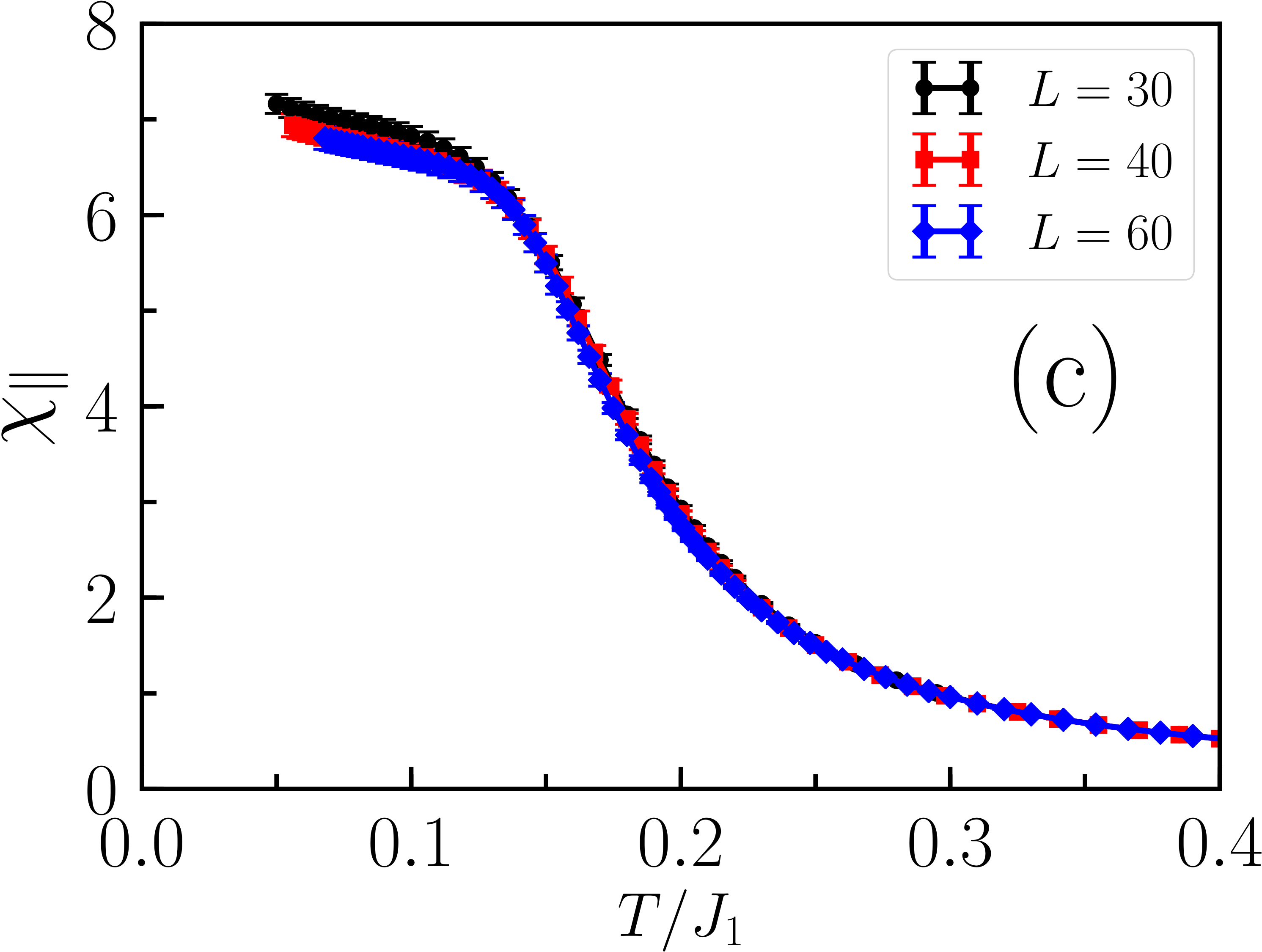}\includegraphics[clip,width=0.5\columnwidth]{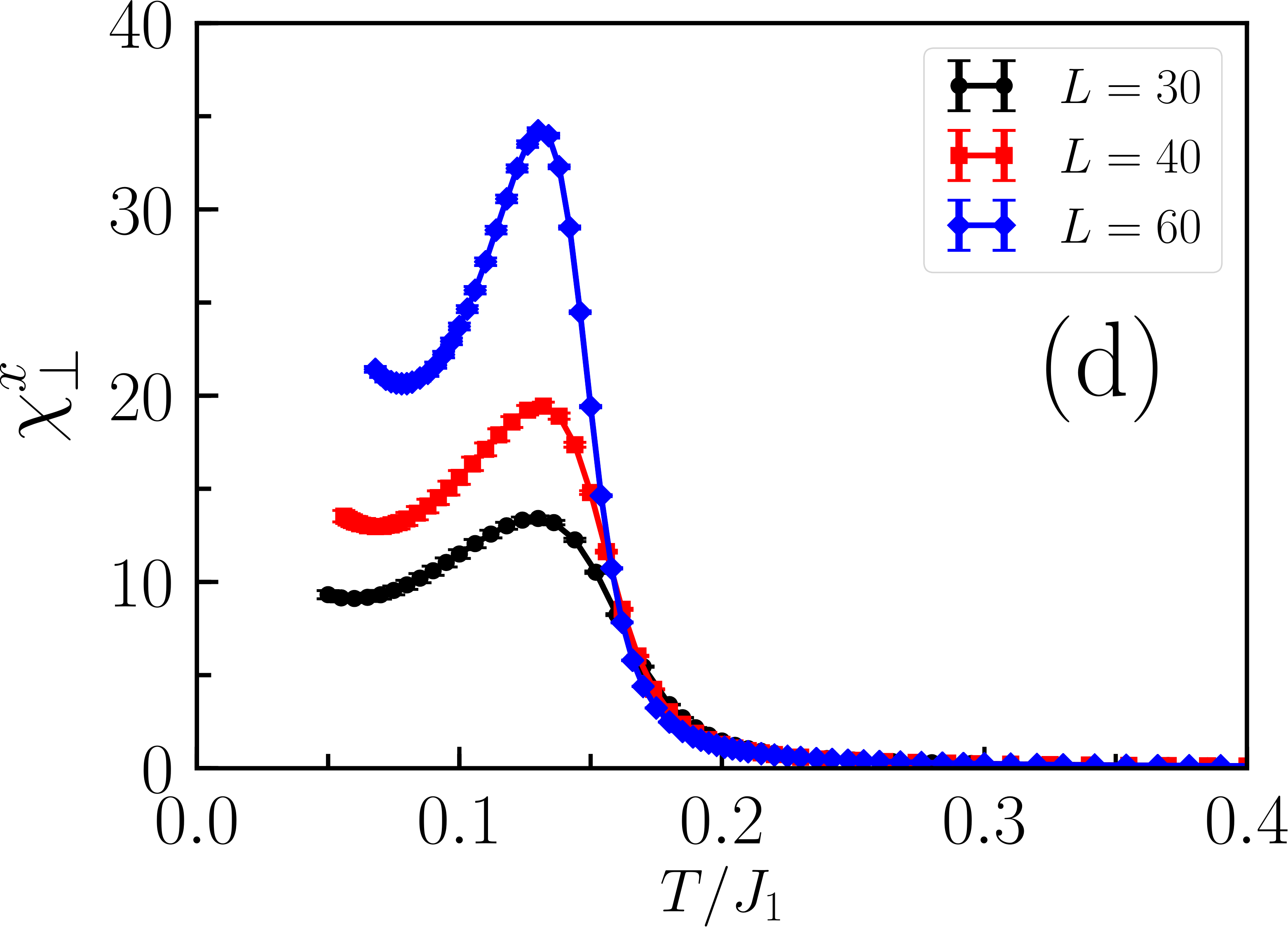}\\
\includegraphics[clip,width=0.5\columnwidth]{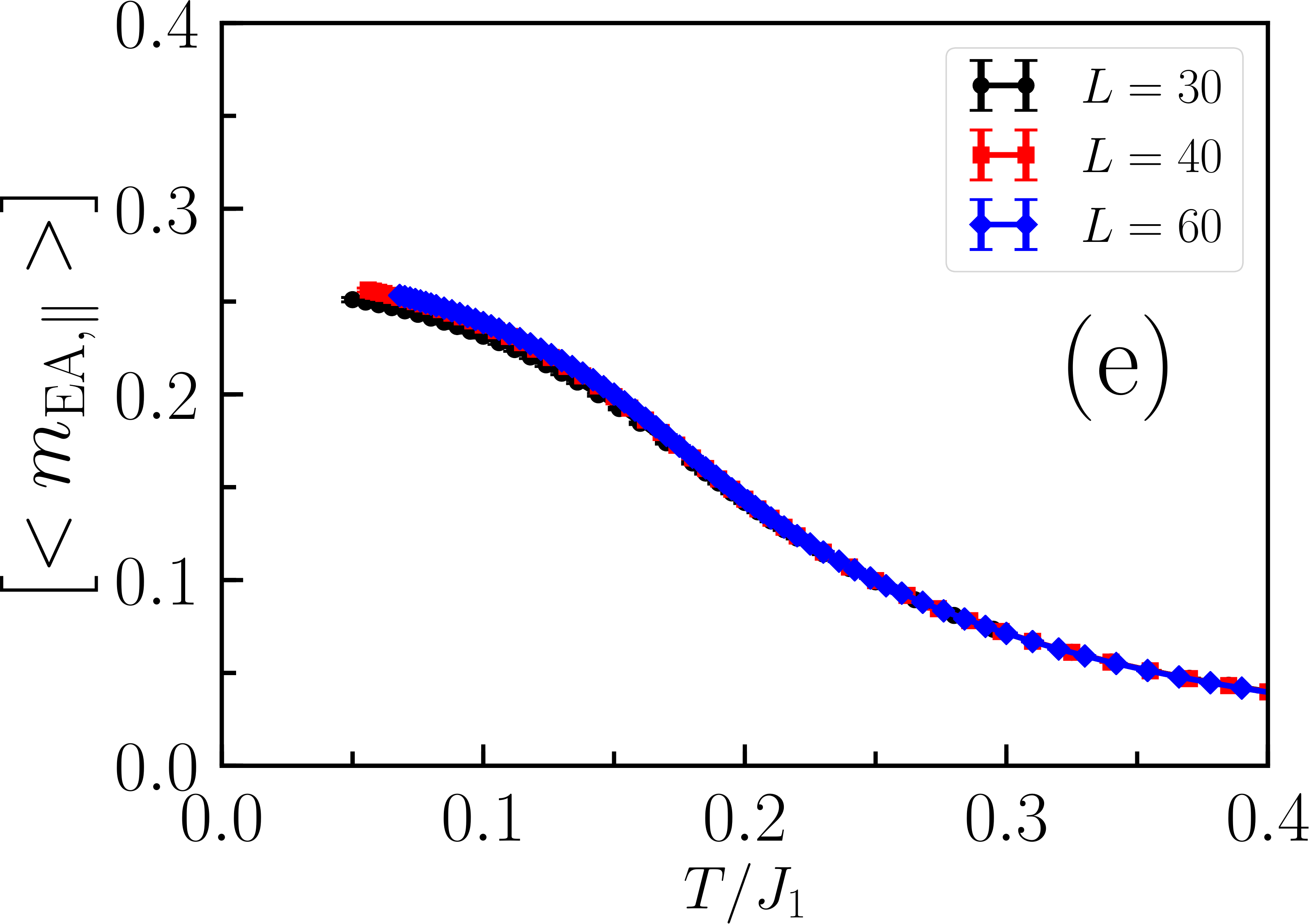}\includegraphics[clip,width=0.5\columnwidth]{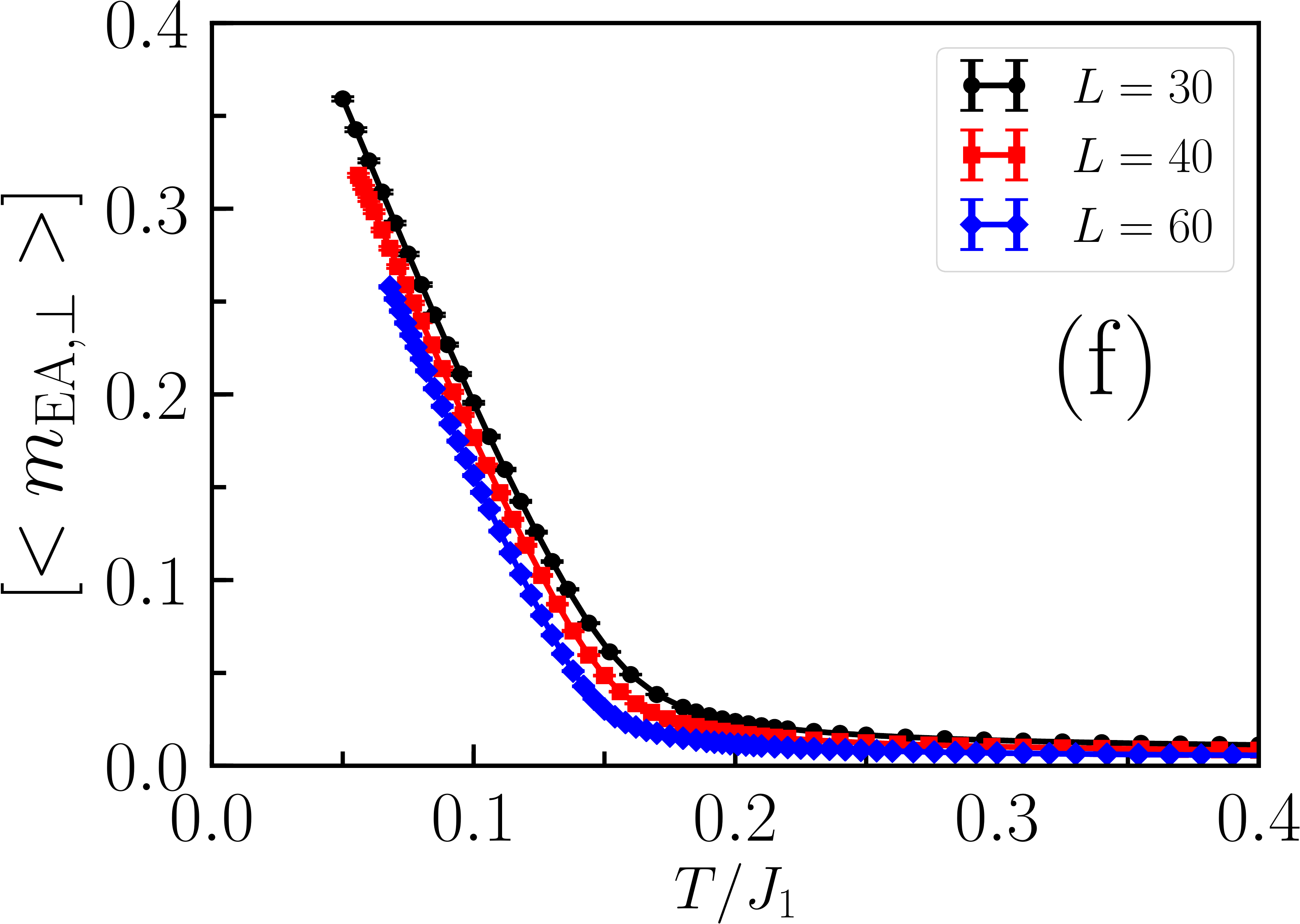}\\
\includegraphics[clip,width=0.5\columnwidth]{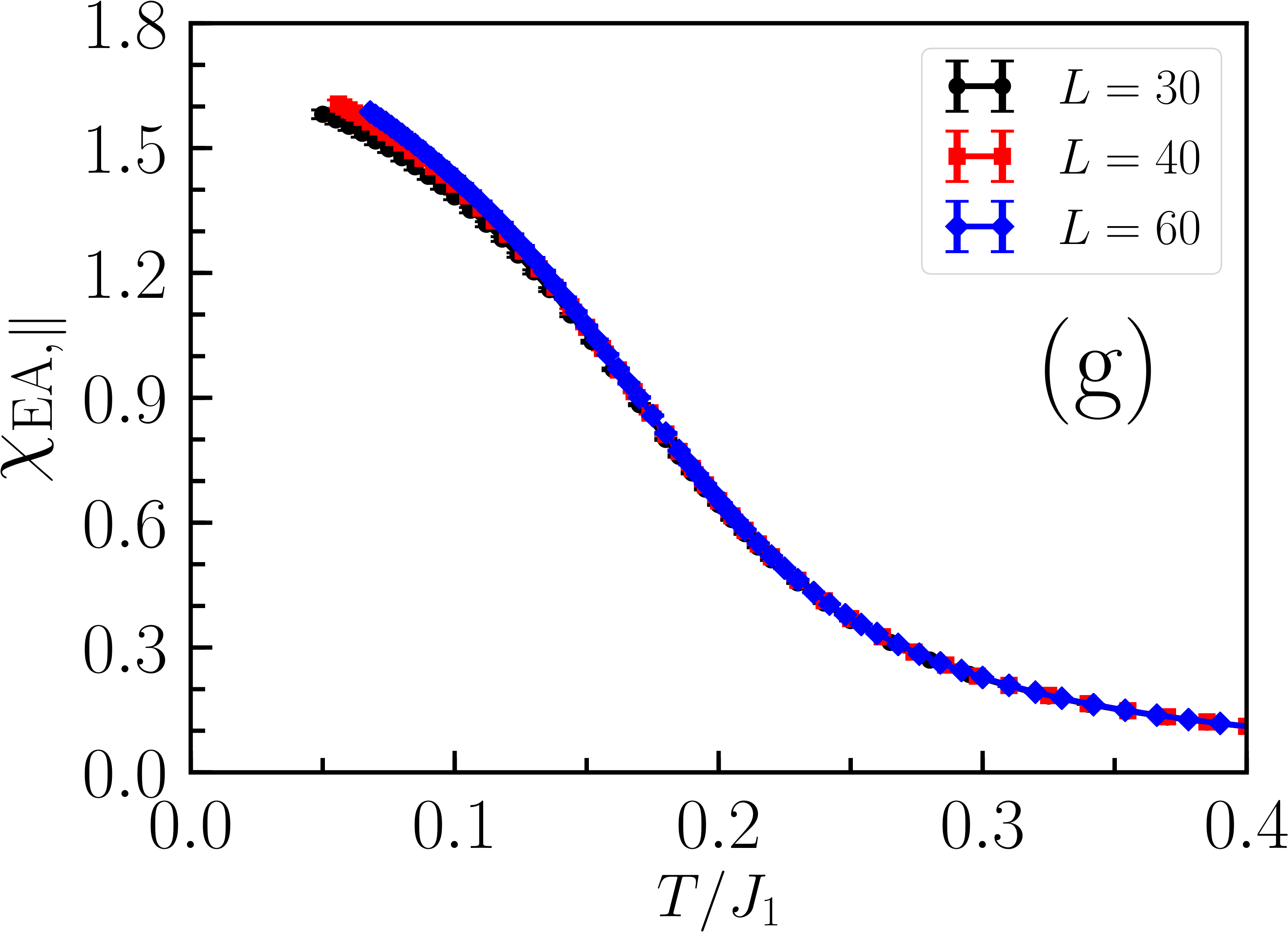}\includegraphics[clip,width=0.5\columnwidth]{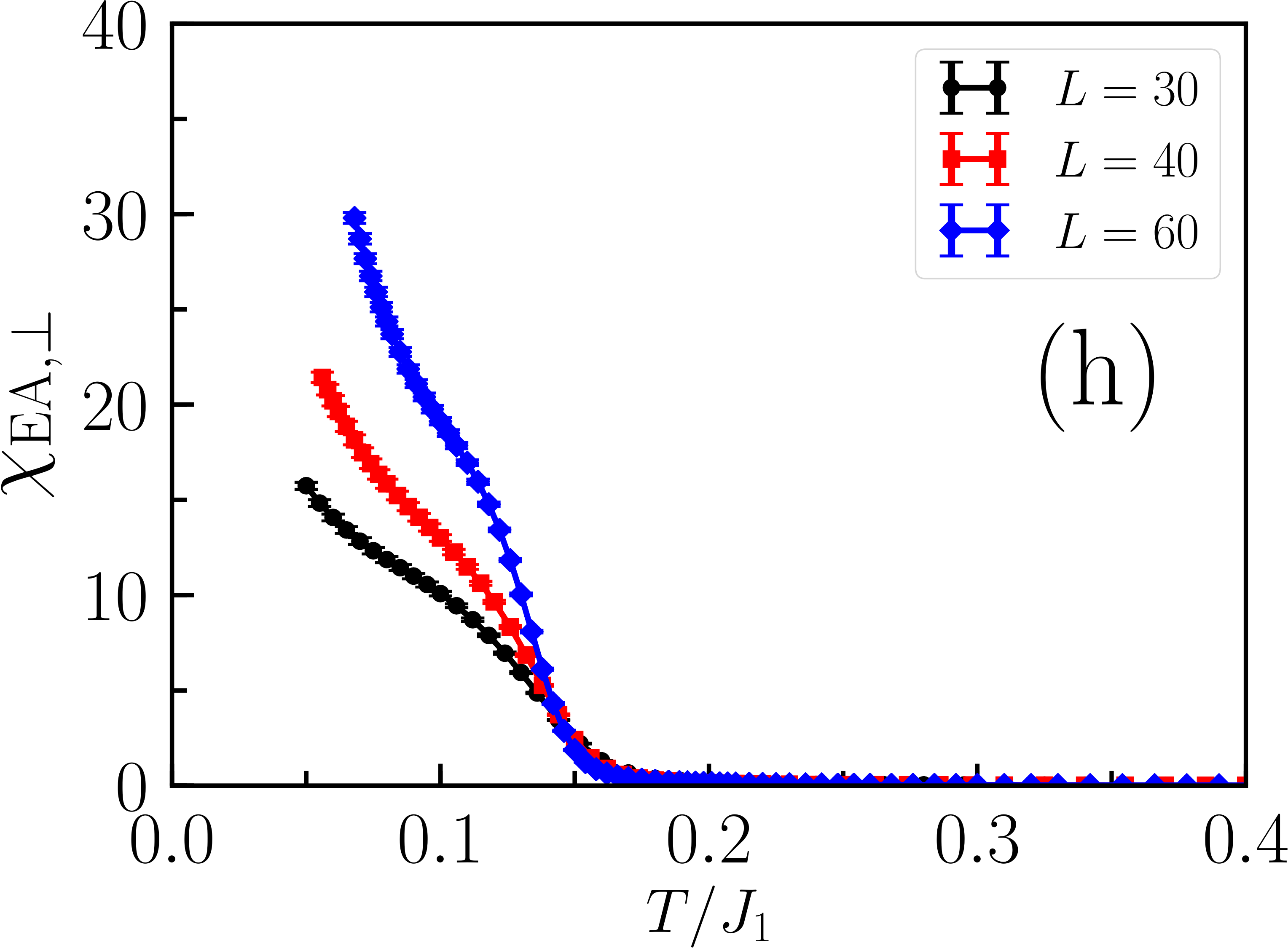}
\par\end{centering}
\caption{The conventional and the Edwards-Anderson nematic and SVDW order parameters
and the corresponding susceptibilities as a function of the temperature
for $10\%$ of isotropically dilution of $J_{1}$ bonds. {[}We verified
that $m_{\perp}^{y,z}$ ($\chi_{\perp}^{y,z}$) is statistically identical
to $m_{\perp}^{x}$ ($\chi_{\perp}^{x}$).{]}\label{fig:glass-susceptibilities}}
\end{figure}

We now present further details on the glassy features of the resulting
paramagnet. In Fig.~\ref{fig:glass-observables}, the nematic Edwards-Anderson
order parameter \eqref{eq:EA-OP} and the corresponding susceptibility
for $2\%$ of diluted $J_{1}$ bonds are plotted. Clearly, $\left[\left\langle m_{\text{EA},\parallel}\right\rangle \right]$
is finite in the entire temperature range studied. Notice that, similar
to the random-field Ising model~\citep{krzakala-etal-prl10}, this
is not a nematic cluster-spin glass but a paramagnet polarized in
that glassy order since $\chi_{\text{EA},\parallel}<\chi_{\parallel}$.
The SVDW Edwards-Anderson order parameter and the corresponding susceptibility
are also plotted. As expected, $\left[\left\langle m_{\perp}\right\rangle \right]$
{[}Fig.~\hyperref[fig:hvBonds]{\ref{fig:hvBonds}(e)}{]} and $\left[\left\langle m_{\text{EA},\perp}\right\rangle \right]$
diminish very slowly as the system size increases. We have studied
the cases of $10\%$ (see Fig.~\ref{fig:glass-susceptibilities})
and $20\%$ (not shown) of $J_{1}$-bond dilution and observed qualitatively
similar results. 

As mentioned in Sec.~\ref{subsec:PD}, the many competing orders
yields this an interesting paramagnet as evident from the non-monotonic
behavior of $\chi_{\perp}$ in Fig.~\hyperref[fig:glass-susceptibilities]{\ref{fig:glass-susceptibilities}(d)}.
In the following, we discuss about the usual and glassy nematic and
SVDW susceptibilities in a qualitative level. A full quantitative
study is beyond the scope of this work. 

Let us start with the nematic susceptibility. In analogy with the
random-field Ising model, one could naively expect $\chi_{\parallel}$
to be round-peaked at a temperature $T_{\parallel}^{*}$ proportional
to the clean paramagnet-nematic transition temperature $T_{c}$ as
this is the relevant energy scale for the nematic fluctuations. The
weaker the disorder, the higher and sharper is the peak (as larger
is the typical size of the nematic clusters $\xi_{\parallel\text{RF}}$).
However, unlike the Ising model, to the left of of this peak the nematic
energy scale $J_{\parallel}$ is temperature dependent and, thus,
we expect $\chi_{\parallel}$ to remain high at lower temperatures
as can be seen in our MC data Figs.~\ref{fig:hvBonds} and \ref{fig:glass-susceptibilities}.
With respect to the Edwards-Anderson nematic susceptibility, also
in analogy with the random-field Ising model, we expect $\chi_{\text{EA},\parallel}$
to be similar to but bounded by $\chi_{\parallel}$, in agreement
with our numerics. 

Let us now discuss the more subtle SVDW susceptibility starting our
analysis from the low-$T$ limit. We expect both $\chi_{\perp}$ and
$\chi_{\text{EA},\perp}$ to diverge as $T\rightarrow0$ since the
associated correlation length $\xi_{\perp}$ diverges.\footnote{As discussed in Sec.~\ref{subsec:Dipolar-RF}, the SVDW order is
perturbatively stable against disorder and thus, $\xi_{\perp}$ is
not bounded by the dipolar random field correlation length $\xi_{\perp\text{RF}}$.} This is only possible because the domain walls are sufficiently thick
at the lowest temperatures occupying most of the bulk. (Thus, the
greater the impurity concentration the greater $\chi_{\perp}$ and
$\chi_{\text{EA},\perp}$.) As temperature increases both $\chi_{\perp}$
and $\chi_{\text{EA},\perp}$ decrease due to (i) thermal fluctuations
and to (ii) the growth of the nematic domains. Beyond $T_{\parallel}^{*}$,
the nematic domains decrease again giving further room to the SVDW
domain wall, and thus, $\chi_{\perp}$ develops inflection points
(or even a local minimum) near $T_{\parallel}^{*}$. The same effect
is expected on $\chi_{\text{EA},\perp}$, however, less intensively
due to the high temperature. 

\begin{figure}[t]
\begin{centering}
\includegraphics[clip,width=0.5\columnwidth]{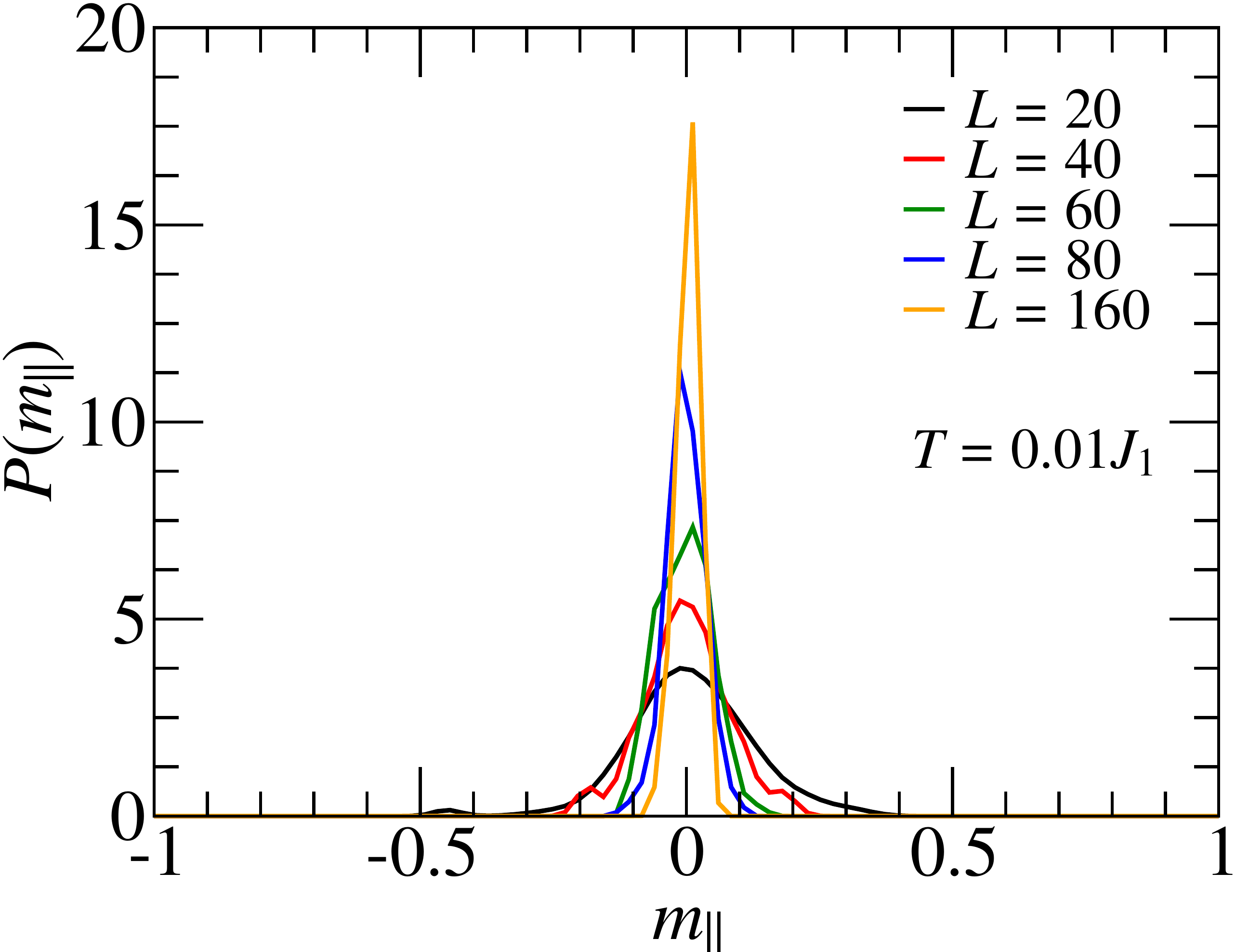}\includegraphics[clip,width=0.5\columnwidth]{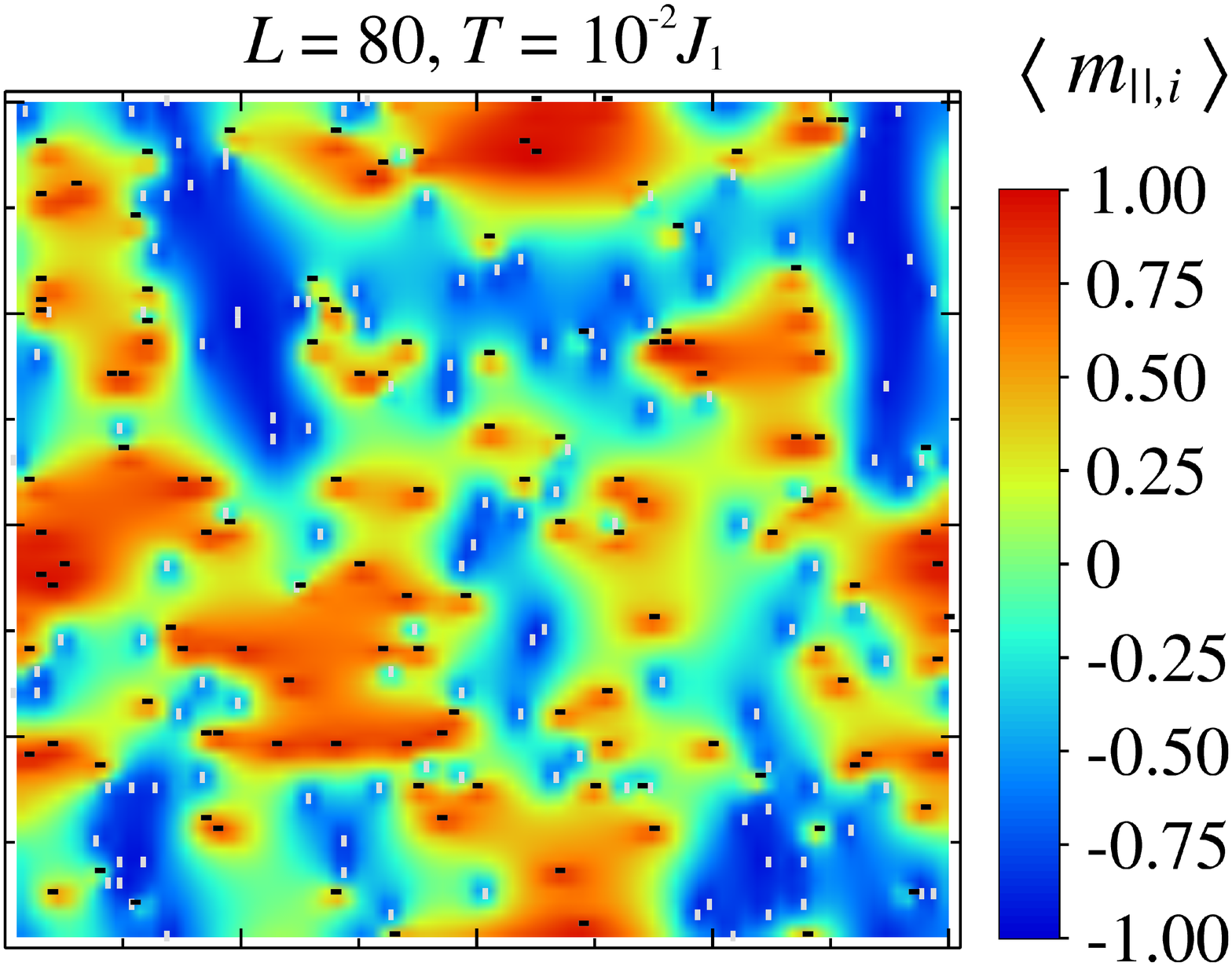}\\
\includegraphics[clip,width=0.5\columnwidth]{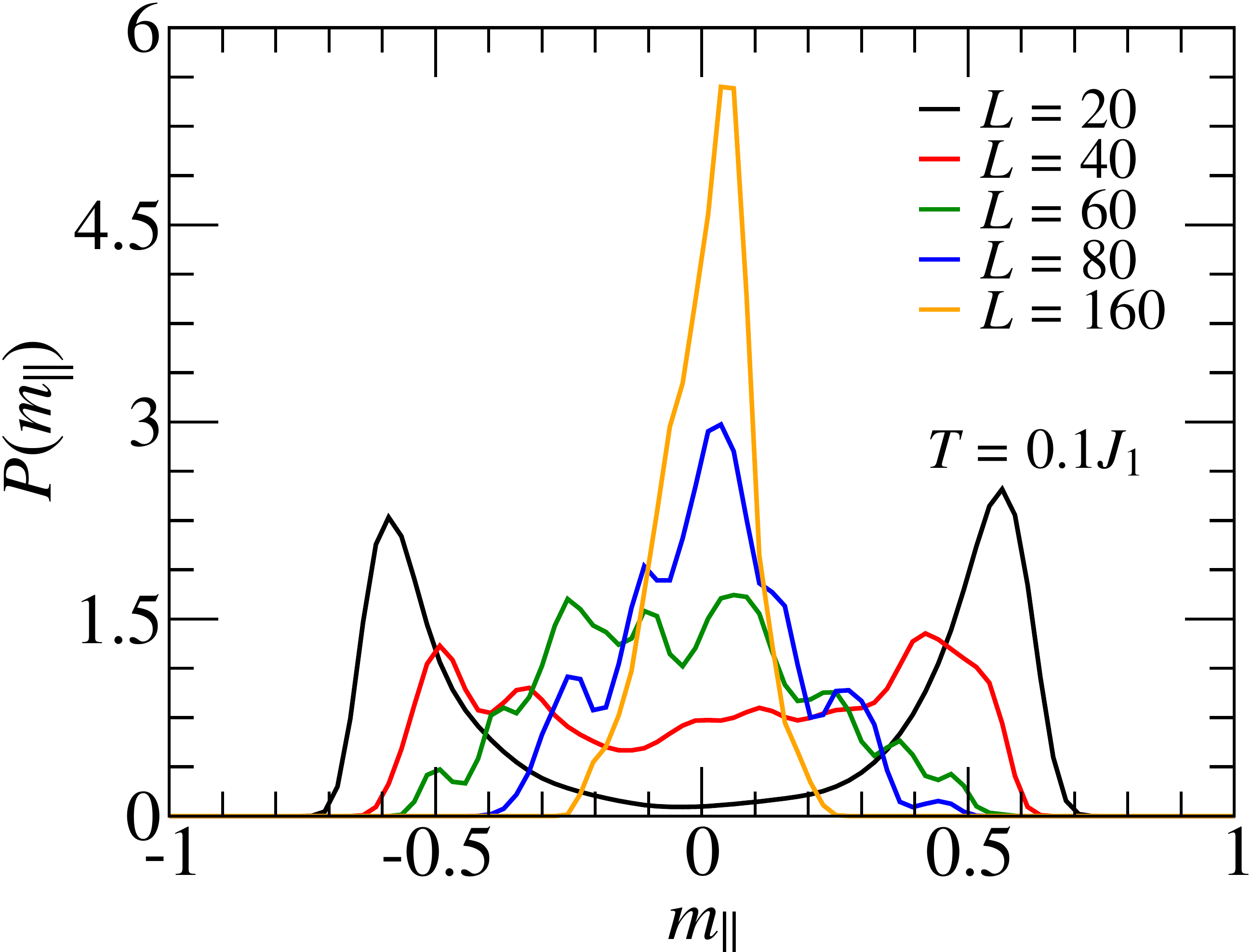}\includegraphics[clip,width=0.5\columnwidth]{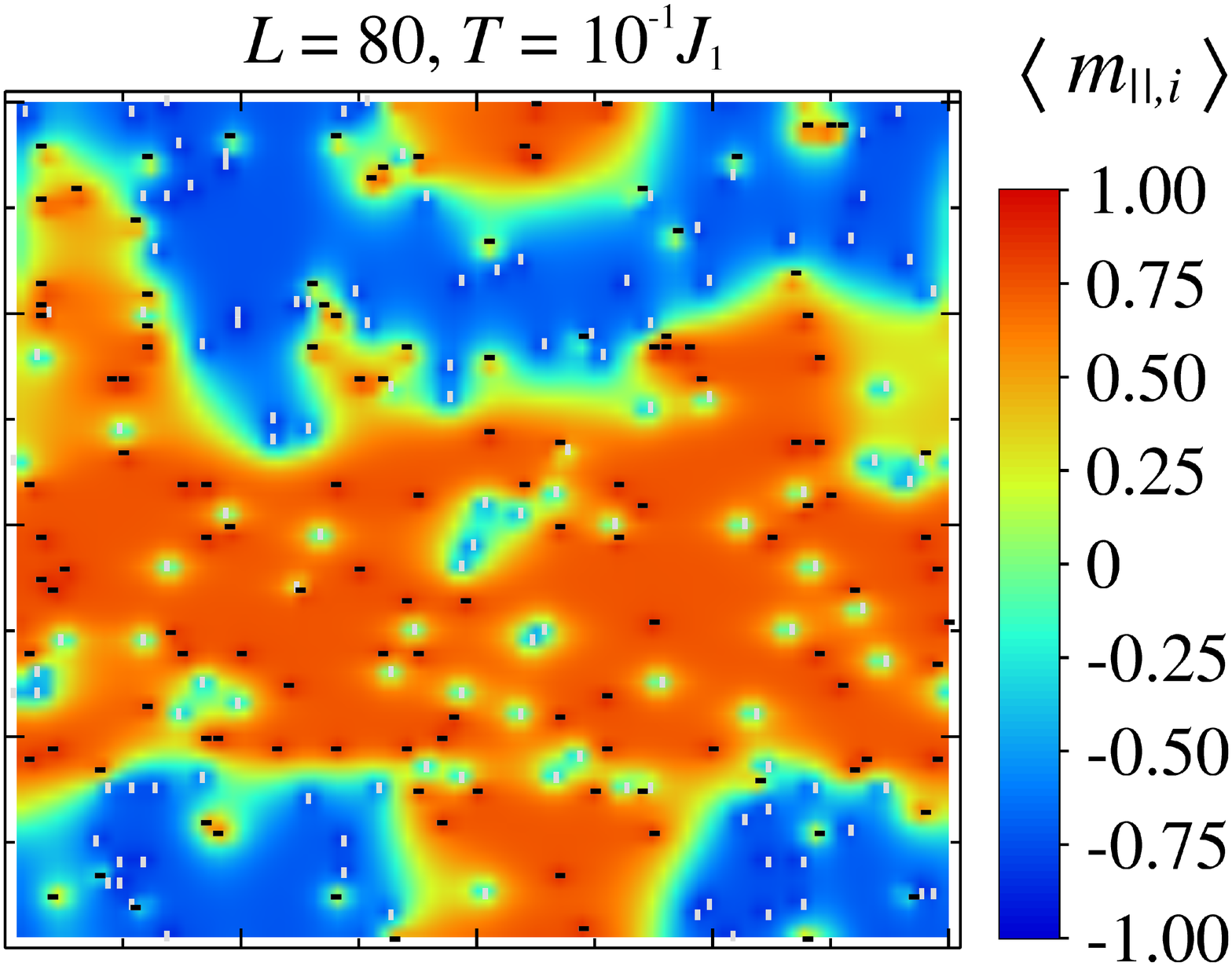}\\
\includegraphics[clip,width=0.5\columnwidth]{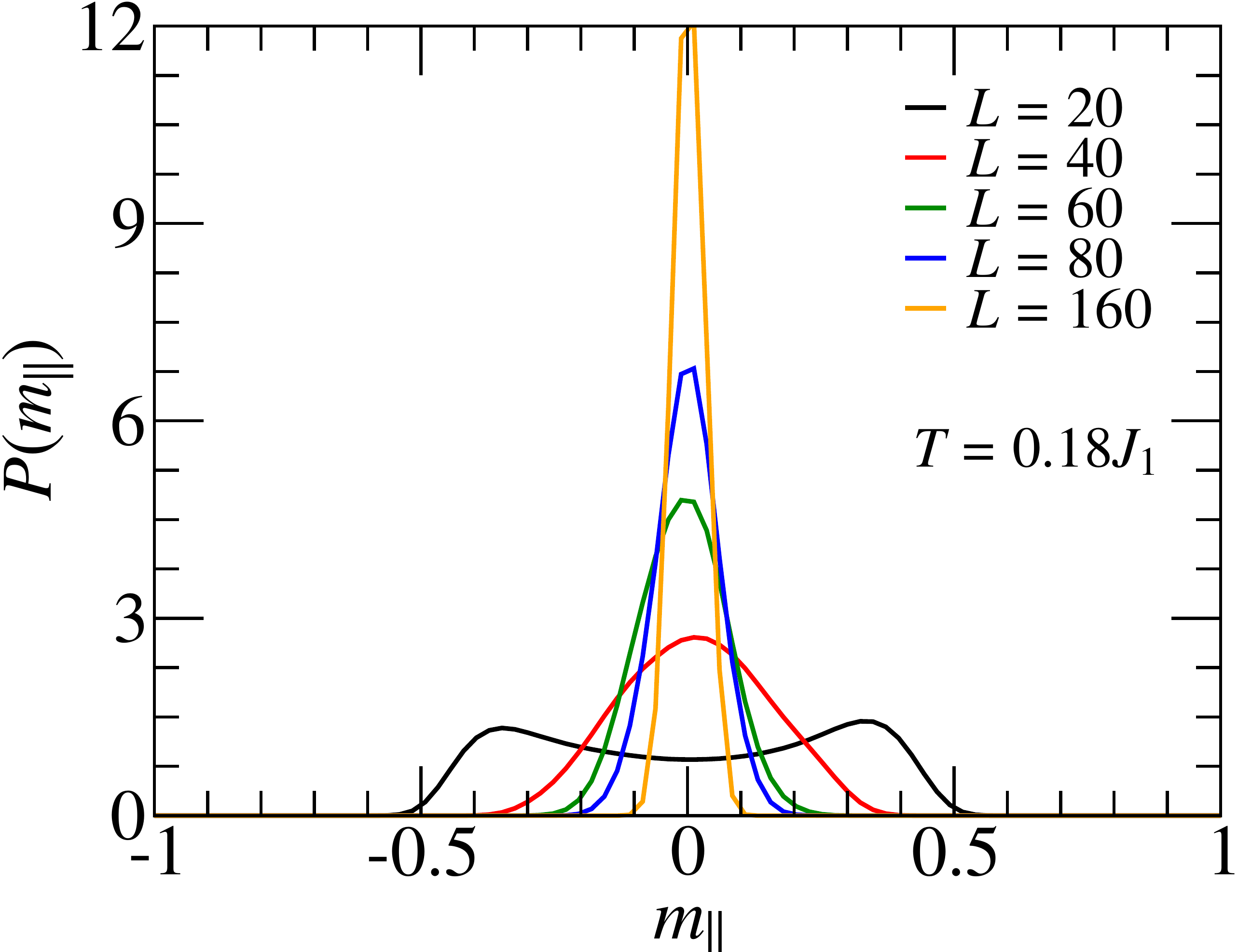}\includegraphics[clip,width=0.5\columnwidth]{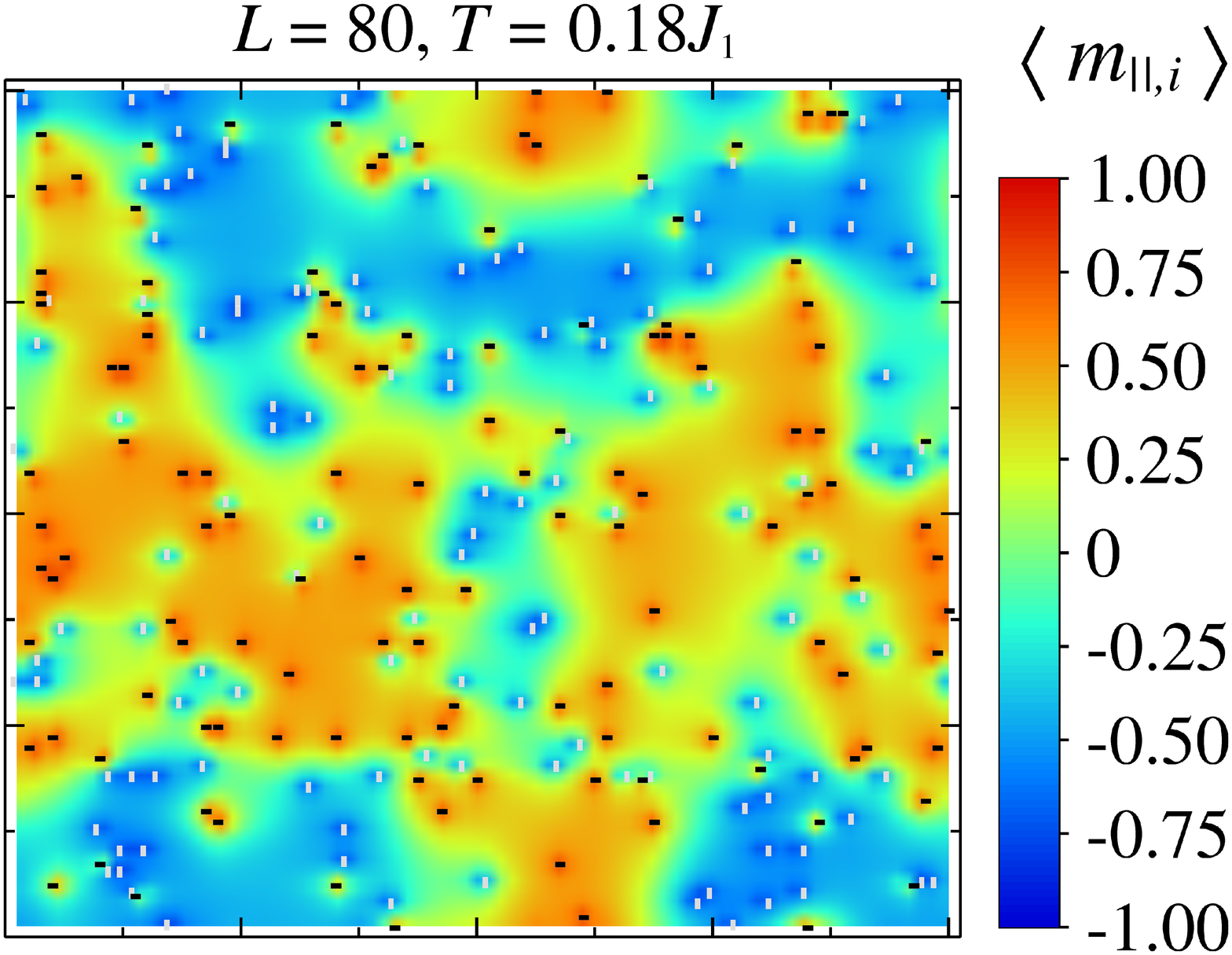}
\par\end{centering}
\caption{\label{fig:hvBonds-Histogram}(Left panels) The nematic order-parameter
normalized histogram $P(m_{\parallel})$ for different system sizes
and temperatures and (right panels) the density of the local nematic
order $\left\langle m_{\parallel,i}\right\rangle $ for a typical
disorder realization. Here, the vertical (light) and horizontal (dark)
ticks represent the missing $J_{1}$ bonds (2\% on average).}
\end{figure}

We end this section by studying more quantitatively the temperature
dependence of the size of nematic domains. In Fig.~\ref{fig:hvBonds-Histogram},
we plot the nematic order parameter distribution $P(m_{\sigma})$
for $2\%$ dilution of $J_{1}$ bonds. It is built as a normalized
histogram of the entire MC time series $m_{\parallel}(t)$ for all
the disorder realizations. We have studied the temperatures $T=10^{-2}J_{1}$,
$10^{-1}J_{1}$ and $0.18J_{1}$. For each temperature, we also show
the density plot of the local nematic order $\left\langle m_{\parallel,i}\right\rangle $
for a typical disorder realization. The density plots show well-defined
domains of static positive (negative) nematic order perfectly coincident
with the regions rich in horizontal (vertical) $J_{1}$-bond dilution.
When the domains are bigger than $L$, $P(m_{\sigma})$ exhibits two
symmetric peaks, a feature observed in the clean (disorder-free) samples
(not shown). Increasing $L$ for fixed impurity density and $T$,
more and more domains fit inside the system and thus $P(m_{\sigma})$
approaches a zero-mean Gaussian of vanishing width $\sim1/L$ as dictates
the central limit theorem. 

\subsection{Disorder on the diagonal couplings\label{subsec:J2-disorder}}

For completeness, we have studied the effects of quenched disorder
only on the next-nearest neighbor $J_{2}$-couplings. We have verified
that the nematic-paramagnet transition is preserved. As discussed
in Sec.~\ref{subsec:Nematic-RF}, this is expected since disorder
on the $J_{2}$-couplings cannot generate random nematic fields. Based
on universality, we then expect this transition to belong to the usual
Ising universality class furnished with logarithmic corrections~\citep{dotsenko-dotsenko-ap83,shalaev-pr94,shankar-prl87}.
Our results (not shown) are compatible with this scenario.~\footnote{Evidently, we cannot exclude other scenarios such as non-universal
disorder-dependent critical exponents since large system sizes (unreachable
for our current numerical resources) are required~\citep{zhu15}.}

\begin{figure}[t]
\begin{centering}
\includegraphics[clip,width=0.7\columnwidth]{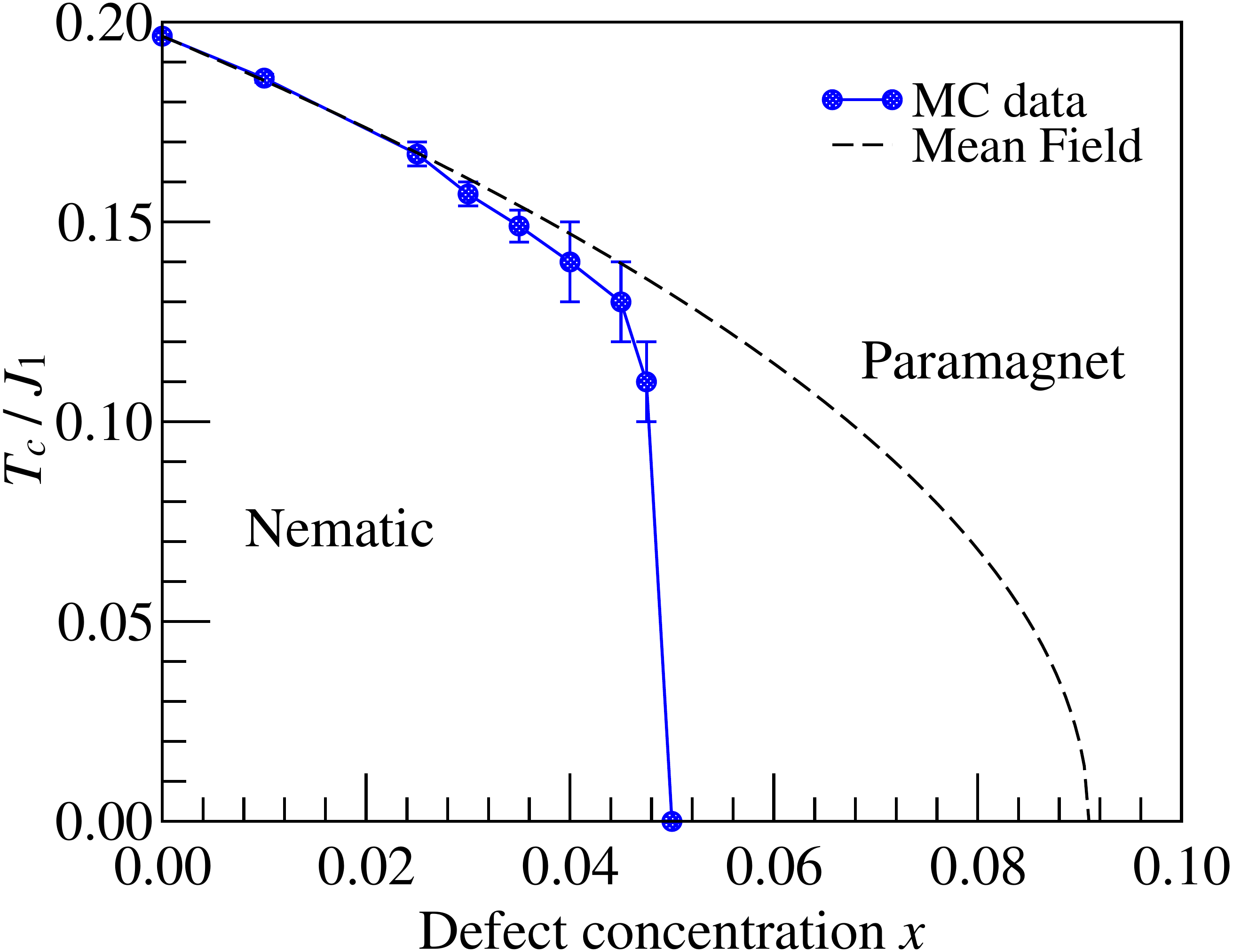}
\par\end{centering}
\caption{Critical temperature $T_{c}$ vs the dilution concentration of $J_{2}$
bonds $x$. The dashed line is the prediction given by mean field
theory $T_{c}=0.62\sqrt{\left(2\left(1-x\right)J_{2}-J_{1}\right)J_{1}}$.
Here, $J_{2}=0.55J_{1}$ and we have studied systems of sizes up to
$L=80$. The MC critical temperature was estimated from the crossings
of the Binder cumulant. The solid line is just a guide for the eyes.\label{fig:PD-J2}}
\end{figure}

In Fig.~\ref{fig:PD-J2} we plot the critical temperature $T_{c}$
as a function of $x$, the density of diluted $J_{2}$ bonds. As before,
we have set $J_{2}=0.55J_{1}$. The critical temperature was obtained
from the crossing of the nematic Binder cumulant \eqref{eq:Binder}.
For lower densities $x\apprle2.5\%$, the critical temperature follows
a simple mean-field prediction as we explain below. At higher densities,
the nematic order is considerably weakened and completely destroyed
above $\approx5\%$.

For $0.5J_{1}\leq J_{2}\apprle0.9J_{1}$, the MC clean critical temperature
obtained in Ref.~\citealp{weber12} is well approximated by $T_{c}^{(\text{clean})}\approx A\sqrt{\left(2J_{2}-J_{1}\right)J_{1}}$
with $A\approx0.62$. In a simple mean-field approximation, we replace
$J_{2}$ by its mean value $J_{2}(1-x)$. The mean-field critical
temperature (shown as a dashed line in Fig.~\ref{fig:PD-J2}) is
thus $T_{c}\approx A\sqrt{\left(2\left(1-x\right)J_{2}-J_{1}\right)J_{1}}$.

\subsection{The spin-vortex-crystal glass state and the spin-vorticity-density-wave
order\label{subsec:GS}}

Finally, for completeness, we briefly study the disorder effects on
the $T=0$ spin configuration obtained from the energy minimization
method. 

\begin{figure}[b]
\begin{centering}
\includegraphics[clip,width=0.7\columnwidth]{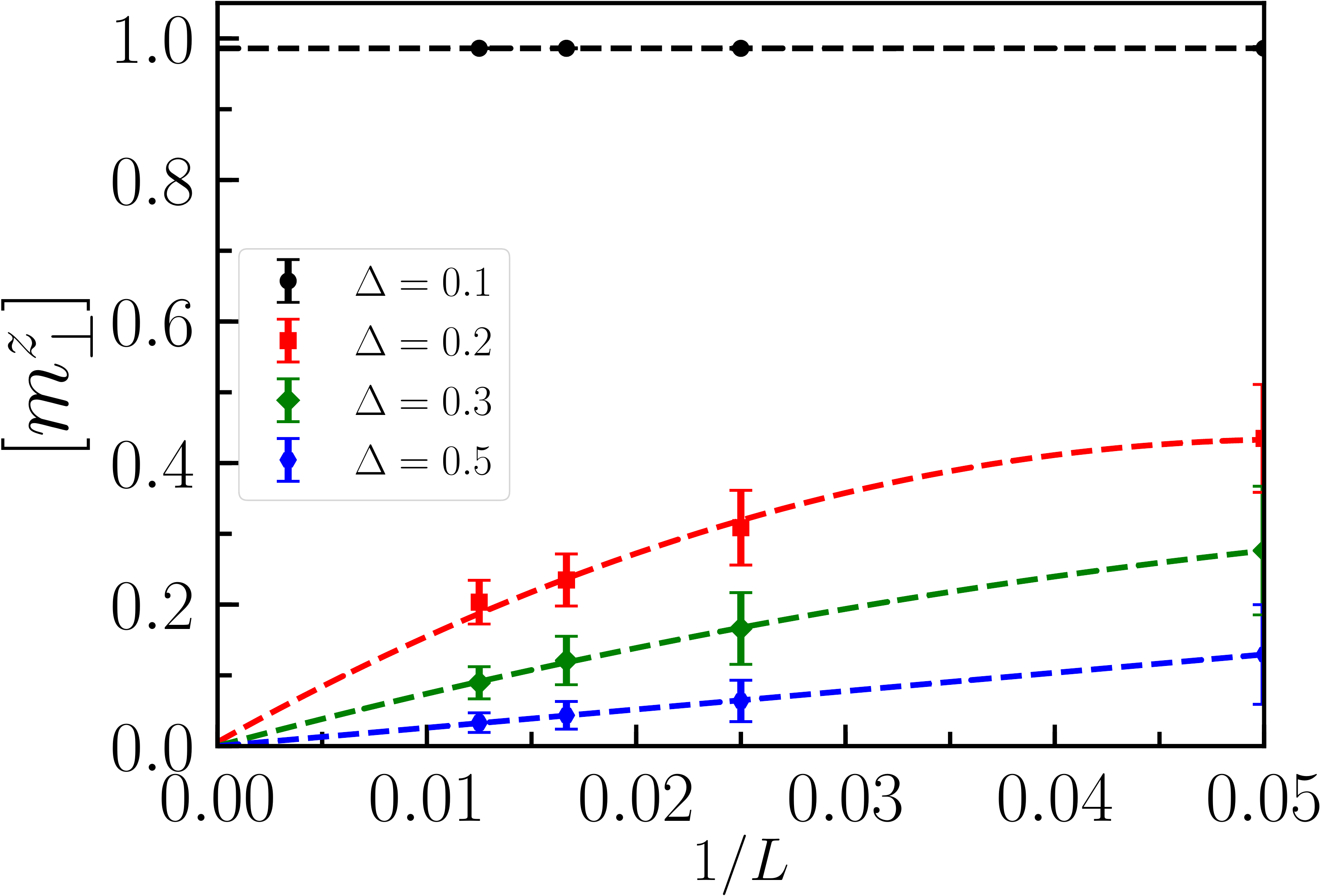}
\par\end{centering}
\caption{The $T=0$ disorder average of the $z$-component of the SVDW order
parameter \eqref{eq:SVDW-OP} as a function of the system size $L$
for various disorder strength $\Delta$ (see text). The spin configuration
is on the $xy$ plane in the weak disorder limit $\Delta\leq\Delta\approx0.2$
and the SVDW order is finite. For stronger disorder $\Delta>\Delta_{c}$,
the spin configuration is no longer coplanar and $\left[m_{\perp}^{z}\right]$
vanishes as $L\rightarrow\infty$. \label{fig:SVDW-GS}}
\end{figure}

As discussed in Sec.~\ref{subsec:Equivalence}, disorder perturbatively
selects the SVC state. Non-perturbative effects, however, destroy
this order (see Sec.~\ref{subsec:Dipolar-RF}). The vestigial SVDW
order, on the order hand, is stable against weak disorder. Our aim
is to confirm this prediction. Turns out that this is not a simple
task because the amount of disorder required to destroy the coplanarity
is not large and thus, large samples are required. We circumvent this
obstacle by adding to the Hamiltonian \eqref{eq:H} a positive biquadratic
term $K\sum_{\left\langle ij\right\rangle }\left(\mathbf{S}_{i}\cdot\mathbf{S}_{j}\right)^{2}$~\citep{henley89,rafael16}
with small $K=0.05$. 

We have considered $J_{1}$-bond disorder such that $J_{1,ij}=1+\Delta_{ij}$
with $\Delta_{ij}$ being either $\Delta$ or $-\Delta$ with equal
probability. Thus, $\Delta$ parameterizes the disorder strength.
The next-nearest neighbor interaction is kept fixed at $J_{2}=0.55$.
In Fig.~\ref{fig:SVDW-GS} we plot the SVDW axial order parameter
\eqref{eq:SVDW-OP} projected onto the ($z$) direction perpendicular
to the ($xy$) coplanar spin configuration. Clearly, the SVDW order
is finite for weak disorder $\Delta\leq\Delta_{c}\approx0.2$. On
the other hand, the SVDW order vanishes in the thermodynamic limit
for $\Delta>\Delta_{c}$. We have verified (not shown) that the spin
configuration is no longer coplanar in this regime. It is plausible
that the destruction of the SVDW order coincides with the loss of
coplanarity (and not before).

\section{\label{sec:conclusion}Conclusions and outlook}

In this section we summarize our results and discuss on their implications.
Combining symmetry arguments, numerical minimization, and large-scale
Monte Carlo simulations, we have revisited the square lattice $J_{1}$-$J_{2}$
classical Heisenberg model and have shown that long-range order at
any finite temperature and in the frustrated order-by-disorder regime
$2J_{2}>J_{1}$ is perturbatively unstable against any finite density
of generic bond and/or site disorder due to random-field effects.
At zero temperature and low density of impurities, the state is a
coplanar spin-vortex-crystal glass with finite spin-vorticity-density-wave
vestigial order. 

\subsection{Equivalence between site and bond disorder}

It is well-known that a single bond defect and a site impurity have
quite different effects as they select different states out of the
ground-state manifold (stripe and spin-vortex crystal, respectively).
We have shown that two nearest-neighbor site vacancies mimic a $J_{1}$-bond
defect and, likewise, two site-sharing non-collinear $J_{1}$-bond
defects mimic a site vacancy. Thus, we have established the equivalence
between the effects of site and bond disorder when a finite density
of impurities is considered. They both can select one of the stripe
states or the SVC state, and induce random dipolar SVC fields and
random nematic fields. 

Interestingly, disorder on the $J_{2}$ bonds has a quite different
effect. It does not lift the ground-state degeneracy and, thus, cannot
induce any sort of symmetry-breaking random term. It can only induce
random mass. Thus, long-range nematic order is stable against weak
$J_{2}$-bond disorder.

We recall that in XY easy-plane pyrochlores, a related order-by-disorder
frustrated system, the effects of site and bond disorder were numerically
verified to be equivalent~\citep{andrade18}. Finally, our equivalence
extends the results of Ref.~\citealp{santanu20} (which states that
generic non-collinear coplanar order is perturbatively unstable against
bond disorder in $d\leq2$) to site disorder as well.

\subsection{Random fields, random easy-axes and transverse dipolar fields}

In the ObD regime, both the perturbatively selected states (the collinear
nematic at finite temperatures and the anti-collinear spin-vortex
crystal at $T=0$) break a real-space symmetry. As a consequence,
disorder generates random conjugate fields which have non-perturbative
effects at $d\leq2$. For the discrete symmetry stripe/nematic order,
the generation of the random field is readily understood. Disorder
locally breaks the same symmetry lifting the degeneracy between the
equivalent ordered states. The order-parameter discrete symmetry character
ensures that the generated field is short ranged, i.e., it is local.
Notice that these arguments are not tied to the ObD mechanism. They
apply to any phase where a real-space symmetry is spontaneously broken.
An important example is the destruction of the stripe phase in the
$J_{1}$-$J_{2}$ frustrated Ising magnets~\citep{fernandez-epl88}. 

Fundamentally, the random easy axes discussed in the \hyperref[sec:easy-axes]{Appendix}
and the nematic random fields have the same common origin: impurities
locally lifting the ground-state degeneracy. In the latter case, the
disorder completely lifts the $\text{Z}_{2}$ order-parameter degeneracy,
and thus, can be recast as a random field. In the former, the order
parameter $\text{Z}_{3}\otimes\text{Z}_{2}$ symmetry is broken down
to $\text{Z}_{2}$ and, thus, has the effects of random easy axes.

This symmetry-based strategy for determining the disorder effects
in discrete-symmetry orders is very appealing due to its simplicity.
However, it is not totally clear whether it is relevant or useful
to the case of inversion-symmetry-breaking order since this symmetry
is not broken by a single $J_{1}$-bond defect (or a pair of nearest-neighbor
vacancies). Instead, it locally releases frustration which acts as
a local transverse field. Goldstone modes then communicate this perturbation
as a slowly decaying dipolar field $\sim r^{1-d}$. Complicating even
further this scenario, the effects on the associated spin-vorticity-density-wave
vestigial order is much milder since the parent SVC ``handness'' is
preserved (disorder acts as a weak SVDW $s$-wave field which decays
$\sim r^{-2d}$). While in $d\leq2$ SVC order is completely destroyed,
the SVDW order is perturbatively stable. Clearly, it is desirable
to study the disorder effects on other non-collinear ground states.
We have also verified that in the regime of strong disorder, the spin
configuration becomes non-coplanar. Describing and understanding how
disorder destroys coplanarity is a task left for future research.

Recently, it was noted that local correlations on the disorder variables
can prevent random mass and, consequently, Griffiths singularities
on a phase transition~\citep{hoyos-etal-epl11}. More recently~\citep{kunwar18},
it was shown that local correlations on the disordered variables can
also prevent random stripe fields as well. This is also possible in
the model Hamiltonian \eqref{eq:H} if, for instance, two vacancies
are forbidden to be nearest neighbors. Thus, there will be no generation
of local nematic fields and of transverse dipolar fields. In sum,
this class of local correlation in the disorder variables can prevent
the generation of random fields ensuring the phase transition and
the low-temperature long-range nematic order. 

In experiments, bond defects induced by chemical doping distorting
the local lattice may induce similar vertical and horizontal local
bond defects. This can either prevent or greatly diminish the amplitude
of the random fields and, therefore, the phase transition may either
be preserved or appear to be preserved.\footnote{An analogous result was obtained for the case of random mass~\citep{getelina-etal-prb16}.}
Finally, we mention the recently synthesized Sr$_{2}$CuTe$_{1-x}$W$_{x}$O$_{6}$
compound which is a quasi-two-dimensional spin-1/2 magnet modeled
by the Hamiltonian \eqref{eq:H}~\citep{mustonen-etal-prb18}. Disorder
in this compound is of random-couplings type~\citep{katukuri-etal-prl20}.
Interestingly, the couplings appear equally disordered in the plaquette.
In other words, it realizes the correlated disorder mechanism above
mentioned. In agreement with our results, the nematic phase is quite
robust against weak disorder. It is worthy noting that our results
also explains the numerical findings of Ref.~\citealp{hong-etal-prl21}.

\subsection{Quantum mechanical effects}

We have established that the classical (finite-$T$) paramagnet-nematic
phase transition is precluded by generic quenched disorder (i.e.,
disorder which includes random $J_{1}$ bonds). Since quantum fluctuations
are expected to play no role on the critical behavior of this finite-$T$
transition, it is very plausible that our result directly applies
to quantum mechanical systems as well. 

We have also shown that the resulting paramagnet is polarized in the
glassy nematic order for all temperatures, i.e., the system is broken
into domains of local nematic order. How is this paramagnet changed
by quantum fluctuations? Deep in the frustrated regime $J_{2}\gg J_{1}$
where the semi-classical approach is expected to be valid, the naive
expectation is that the local nematic domains are not melted by quantum
fluctuations since they also select the stripe state via the order-by-(quantum)disorder
mechanism~\citep{henley89}.\footnote{This is certainly valid even beyond the semi-classical approximation
in a temperature window ranging from higher temperatures down to slightly
below the clean critical temperature.} We also expect a weaker SVC and SVDW response at lower temperatures
because the size of the nematic domains will not vanish as $T\rightarrow0$. 

Our results are thus relevant for a wide range of materials such as
(K or Rb)MoOPO$_{4}$Cl~\citep{ishikawa-etal-prb17}, Li$_{2}$VO(Si
or Ge)O$_{4}$~\citep{melzi-etal-prb01,rosner-etal-prb03}, and abVO(PO$_{4}$)$_{2}$
(with ab = BaCd, SrZn, BaZn, Pb$_{2}$, PbZn)~\citep{nath-etal-prb08,kaul-etal-jmmm04,tsirlin-rosner-prb09,tsirlin-etal-prb10}.
They are modeled by the Hamiltonian \eqref{eq:H} and disorder effects
can be introduced by chemical substitution.

Around the antiferromagnetic-stripe quantum phase transition $0.4\apprle J_{2}/J_{1}\apprle0.6$,
the system does not display long-range order and its description lies
beyond the reach of the semi-classical approach~\citep{jiang12,gong14,morita-etal-jpsj15,wang18}.
The effects of disorder on the phases appearing in that interval is
an important question which has received considerable attention recently~\citep{kawamura18,kawamura19,liu18b,kimchi18,baek-etal-prb20,liu-guo-sandvik-prb20}. 
\begin{acknowledgments}
We thank Pedro Consoli, Santanu Dey, Matthias Vojta, André Vieira,
Eduardo Miranda and Rafael Fernandes for useful discussions and collaborations
on related topics. We also thank N. P. Teodosio for collaboration
at an early stage of this project. This work was supported by the
CAPES \textendash{} Finance Code 001, by FAPESP Grants No. 2015/23849-7,
No. 2016/10826-1, No. 2019/17026-9, and No. 2019/17645-0, by CNPq
Grants No. 307548/2015-5, No. 406399/2018-2, and No. 302994/2019-0,
by the Deutsche Forschungsgemeinschaft (DFG) through SFB 1143 (project
id 247310070), the W\"urzburg-Dresden Cluster of Excellence ct.qmat
(EXC 2147, Project ID No. 390858490), and by the IMPRS for Chemistry
and Physics of Quantum Materials at MPI-CPfS.
\end{acknowledgments}

\appendix

\section{Random easy-axes\label{sec:easy-axes}}

Several recent works~\citep{zhu17,andrade18,parker18,kunwar18,santanu20}
have discussed specific examples of frustrated magnetic systems in
which the effects of interaction-symmetry-preserving disorder can
be understood in terms of effective random-fields~\citep{aharony78,fernandez-epl88,fyodorov-shender-jpc91}.
Here, we show that effective random axes~\citep{fischer,aharony-jmmm83}
are generated by site and bond disorder in easy-plane pyrochlores
providing a clear physical interpretation of the numerical data. The
important issue is that these new terms have much more dramatic effects
such as precluding any phase transition at $d\leq2$, or even stabilizing
a cluster-spin glass phase at $d\geq3$.

Our strategy is straightforward. As discussed in Secs.~\eqref{subsec:single-impurity}
and \eqref{subsec:Dipolar-RF} we simply analyze how a single impurity
lifts the ground-state degeneracy.

\begin{figure}[b]
\begin{centering}
\includegraphics[clip,width=0.6\columnwidth]{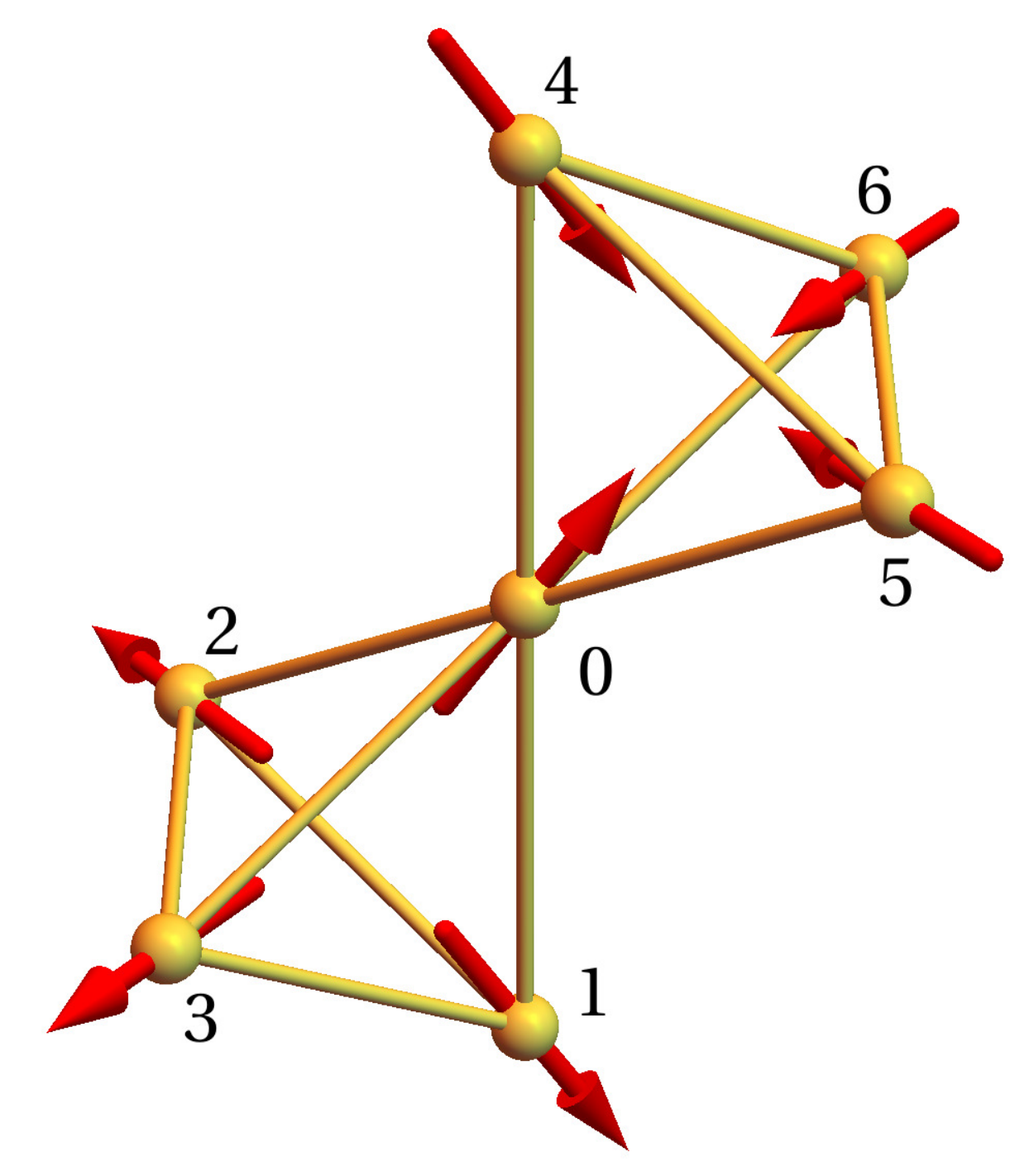}
\par\end{centering}
\caption{The sketch of the six nearest-neighbors sites in a pyrochlore lattice.
The local $z$ axes are also shown as red arrows. For clarity, the
local $x$ and $y$ axes are not shown.\label{fig:pyrochlore}}
\end{figure}

The easy-plane pyrochlores are XY frustrated magnets interacting via
diagonal and off-diagonal (spin-orbit induced) interactions which
can be modeled by the system Hamiltonian~\citep{ross11} 
\begin{equation}
\mathcal{H}=-\sum_{\left\langle jk\right\rangle }\left[J_{jk}^{xx}S_{j}^{x}S_{k}^{x}+J_{jk}^{yy}S_{j}^{y}S_{k}^{y}+J_{jk}^{xy}\left(S_{j}^{x}S_{k}^{y}+S_{j}^{y}S_{k}^{x}\right)\right],\label{eq:Hxy}
\end{equation}
 where the sum runs over pairs of nearest-neighbor sites on a cubic
pyrochlore lattice (see Fig.~\ref{fig:pyrochlore}), the coupling
constants are 
\begin{equation}
J_{jk}^{xx(yy)}=J_{jk}\left(1\mp\alpha_{jk}\mbox{cos}\gamma_{jk}\right),\ J_{jk}^{xy}=J_{jk}\alpha_{jk}\mbox{sin}\gamma_{jk},\label{eq:exchange}
\end{equation}
 the classical spins $\mathbf{S}_{j}=\left(\cos\theta_{j},\sin\theta_{j}\right)$
are 2-component unity vectors, $J_{jk}$ parameterizes the local energy
scale, and $\alpha_{jk}$ parameterizes the relative strength of the
local spin-orbit coupling. The direction-dependent angles are $\gamma_{01}=\gamma_{23}=\gamma_{04}=\gamma_{56}=0$,
$\gamma_{02}=\gamma_{13}=\gamma_{05}=\gamma_{46}=2\pi/3$, and $\gamma_{03}=\gamma_{12}=\gamma_{06}=\gamma_{45}=-2\pi/3$.

In the clean limit ($J_{jk}=J$ and $\alpha_{jk}=\alpha$) and for
$-2<\alpha<2$, the ground state manifold is O(2) accidentally degenerate
being a ferromagnetic state (in the local reference frame) with spins
pointing along any direction $\theta_{j}=\theta$ in the XY plane.
The ground-state energy $E_{0}=-3JN$ is $\alpha$ independent. As
can be expected, the order-by-(thermal/quantum-)disorder mechanism
lifts the O(2) ground-state degeneracy down to $\text{Z}_{6}$: for
$0<\alpha<2$ ($-2<\alpha<0$) the dubbed $\psi_{2}$ ($\psi_{3}$)
state is selected which corresponds to spins pointing along one of
the $\cos\left(\frac{\pi}{3}n\right)\hat{x}+\sin\left(\frac{\pi}{3}n\right)\hat{y}$
{[}$\cos\left(\frac{\pi}{3}n+\frac{\pi}{6}\right)\hat{x}+\sin\left(\frac{\pi}{3}n+\frac{\pi}{6}\right)\hat{y}${]}
directions, with $n=0,\dots,5$, in the local reference frame~\citep{zhitomirsky12,savary12b}.

The bare interaction symmetry is actually a simple $\text{Z}_{2}$
one ($\theta_{j}\rightarrow\theta_{j}+\pi$). This becomes evident
when rewriting the local interaction energy as 
\begin{equation}
{\cal H}_{jk}=-J_{jk}\left(\cos\left(\theta_{j}-\theta_{k}\right)-\alpha_{jk}\cos\left(\theta_{j}+\theta_{k}+\gamma_{jk}\right)\right).\label{eq:local-Hxy}
\end{equation}
 The symmetry is enhanced to $\text{Z}_{6}$ when there is a real-space
equivalence between the three axes connecting the site pairs (0,1),
(0,2), and (0,3), i.e., ($\theta_{j}\rightarrow\theta_{j}+\frac{\pi}{3}$
and ${\cal H}_{01}\rightarrow{\cal H}_{02}\rightarrow{\cal H}_{03}\rightarrow{\cal H}_{01}$).
Not totally surprisingly, thermal and/or quantum fluctuations lift
the O(2) ground-state degeneracy down to $\text{Z}_{6}$ in the clean
system. Likewise, quenched disorder lifts down to a $\text{Z}_{2}$
manifold as shown explicitly in the following. Consider a single bond
defect between sites $j$ and $k$ such that $J_{jk}=J+\delta J_{jk}$
and, for simplicity, $\alpha_{jk}=\alpha$. Thus, the ground-state
degenerate energy $E_{0}$ is changed by 
\begin{equation}
\delta E=\delta J_{jk}\left(\alpha\cos\left(2\theta+\gamma_{jk}\right)-1\right).\label{eq:dE-pyrochlore}
\end{equation}
 Hence, a single bond defect locally lifts the accidental O(2) ground-state
manifold down to a $\text{Z}_{2}$ manifold, the same one of the interactions.
Consequently, the generated term in the field theory is a random axis.
For $\delta J_{jk}\alpha$ negative (positive), the preferred axis
is in the direction $-\frac{\gamma_{jk}}{2}$ $\left(\frac{\pi-\gamma_{jk}}{2}\right)$
which is one of the three axis of the $\psi_{2}$ ($\psi_{3}$) set
of states.

A single-site defect, on the other hand, does not lift the degeneracy
between the three different axes and, therefore, does not generate
any easy-axis term. Two nearest-neighbor site defects, on the other
hand, do break this symmetry generating the random easy-axis term.
Thus, the equivalence between the effects of site and bond disorder
is restored when a finite density of impurities is considered.

What is the resulting state? In the absence of disorder and in the
ObD regime, the $\psi_{2,3}$ fluctuations at low temperatures can
be modeled by a short-ranged $\text{Z}_{6}$ planar clock model. In
the presence of weak disorder, this model must be furnished with local
random axes. The Imry-Ma criterion~\citep{imry_ma} directly applies
for this case of random axes. Thus, the lower critical dimension is
$d_{c}^{-}=2$. Since $d=3$, then long-range $\psi_{2,3}$ order
is stable against weak disorder. For strong disorder, the system thus
breaks into local domains. Very likely, the effective $\alpha$ throughout
the system has zero-mean (or larger variance compared to its mean
value), and thus, the domains of $\psi_{2}$ and $\psi_{3}$ types
are equally (or almost equally) probable. In other words, there are
six equally probable random easy axes.

There is an important difference with respect to the nematic case
discussed in Sec.~\ref{subsec:Nematic-RF}. Here, the domains have
two equally probable states favored by the local easy axis. Thus,
at sufficiently high temperatures, the local $\psi_{2(3)}$ order
is vanishing. In addition, there are an exponentially large number
of equivalent ground states.\footnote{Evidently, there is an exponentially weak effective coupling between
these domains which are unimportant at finite temperatures.} For these reasons, it is plausible to expect a transition to a spin-cluster
glass phase at low temperatures. All these conclusions were confirmed
numerically in Refs.~\citealp{sarkar17,andrade18}.

Finally, notice that the same disorder-induced random-axes term appears
in the two-dimensional analog of the model \eqref{eq:Hxy} on the
triangular lattice with the three different $\gamma_{ij}$-angles
being related to the interactions along the $0$, $\pm\frac{2\pi}{3}$
directions. In $d=d_{c}^{-}=2$, however, long-range order is precluded
by any amount of disorder. For zero-mean disorder $\left[\delta J_{jk}\right]=0$,
stripe domains on the six directions $n\frac{\pi}{3}$ ($n=0,\,1,\dots,5$)
are expected.\footnote{We find interesting to imagine that, if there was order-by-disorder
in this system selecting, say, the even-$n$ domains, then, the odd-$n$
domains would be penalized. The symmetry would be restore only at
stronger disorder.} A glassy state, however, may be melted by thermal fluctuations as
the lower critical dimension of the problem may be $5/2$~\citep{maiorano-parisi-pnas18}.
Quantum fluctuations, on the other hand, may not be sufficiently strong
to destroy the glassy order (which is compatible with the results
of Ref.~\citealp{zhu17}). The understanding of the disorder effects
as inducing the formation of domains of different stripe orders were
previously reported in Ref.~\citealp{parker18}. 

Although we have focused on easy-plane pyrochlores, our arguments
also apply other similar models interacting via strong anisotropic
interactions such as Heisenberg-Kitaev model~\citep{price-perkins-prl12,av14}.
In those cases, the ground state exhibits $\text{Z}_{6}$ degeneracy
which is locally lifted by disorder. 

\bibliography{obd}

\end{document}